\documentclass[12pt,journal,a4paper,onecolumn]{IEEEtran}

\usepackage{graphicx}
\usepackage{amsmath,epsfig,amssymb, amstext, verbatim,amsopn,cite,subfigure,multirow,multicol,lipsum}
\usepackage{balance}
\usepackage[all]{xy}
\usepackage{url}
\usepackage{amsfonts}
\usepackage{epsfig}
\usepackage{slashbox}
\usepackage{setspace}
\usepackage{stmaryrd}
\usepackage{psfrag}	
\usepackage{multirow}
\usepackage{float}
\allowdisplaybreaks
\newcommand{\Equal}{\hspace{-0.7mm}=\hspace{-0.7mm}}
\newcommand{\Add}{\hspace{-0.7mm}+\hspace{-0.7mm}}
\newcommand{\Minus}{\hspace{-0.7mm}-\hspace{-0.7mm}}

\newtheorem{theo}{Theorem}
\newtheorem{lem}{Lemma}
\newtheorem{remk}{Remark}
\newtheorem{defin}{Definition}
\newtheorem{Prop}{Proposition}

\begin{document}

\title{Achievable Rate Region of the Bidirectional Buffer-Aided Relay Channel with Block Fading}

\author{Vahid Jamali$^\dag$, Nikola Zlatanov$^\ddag$, Aissa Ikhlef$^\ddag$, and Robert Schober$^\dag$ \\
\IEEEauthorblockA{$^\dag$ Friedrich-Alexander University (FAU), Erlangen, Germany \\
 $^\ddag$ University of British Columbia (UBC), Vancouver, Canada}\vspace{-2cm}
\thanks{This paper has been accepted for presentation in part at IEEE Globecom  2013 and EUSIPCO 2013.}}

\maketitle

\begin{abstract}
The bidirectional relay channel, in which two users  communicate  with  each  other 
through a relay node, is a simple but fundamental and practical network architecture. In  this  paper,  we  consider the block fading bidirectional relay channel and propose efficient transmission strategies that exploit the block fading property of the channel. Thereby, we consider a decode-and-forward relay and assume that a direct link between the two users is not present. Our aim is to characterize the long-term achievable rate region and to develop protocols which achieve all points of the obtained rate region. Specifically, in the bidirectional relay channel, there exist six possible transmission
modes: four point-to-point modes (user 1-to-relay, user 2-to-relay,
relay-to-user  1,  relay-to-user  2),  a  multiple-access  mode  (both
users  to  the  relay),  and  a  broadcast  mode  (the  relay  to  both
users). Most existing protocols assume a fixed schedule for using a
subset of the aforementioned transmission modes. Motivated by this limitation, we
develop protocols which are not restricted to adhere to a predefined schedule for
using the transmission modes. In fact, based on the instantaneous  channel  state  information  (CSI)  of  the
involved links, the proposed protocol selects the optimal transmission mode in each time slot to maximize the long-term achievable rate region. Thereby, we consider two different types of transmit power constraints: 1) a joint long-term  power constraint for all nodes, and 2) a fixed transmit power for each node. Furthermore, to enable the use of a non-predefined schedule for transmission mode selection, the relay has to be equipped with two buffers for storage of the information received from both users. As data buffering increases the end-to-end delay, we consider both delay-unconstrained and delay-constrained transmission in the paper.    Numerical results confirm the superiority of the proposed buffer-aided protocols compared to  existing bidirectional relaying protocols. 
\end{abstract}

\begin{keywords}
Bidirectional transmission, rate region, adaptive mode selection, power allocation, buffer-aided relaying.
\end{keywords}

\begin{figure}
\centering
\psfrag{U1}[c][c][0.75]{$\text{User 1}$}
\psfrag{U2}[c][c][0.75]{$\text{User 2}$}
\psfrag{R}[c][c][0.75]{$\text{Relay}$}
\psfrag{M1}[c][c][0.75]{$\mathcal{M}_1:\quad$}
\psfrag{M2}[c][c][0.75]{$\mathcal{M}_2:\quad$}
\psfrag{M3}[c][c][0.75]{$\mathcal{M}_3:\quad$}
\psfrag{M4}[c][c][0.75]{$\mathcal{M}_4:\quad$}
\psfrag{M5}[c][c][0.75]{$\mathcal{M}_5:\quad$}
\psfrag{M6}[c][c][0.75]{$\mathcal{M}_6:\quad$}
\includegraphics[width=0.7 \linewidth]{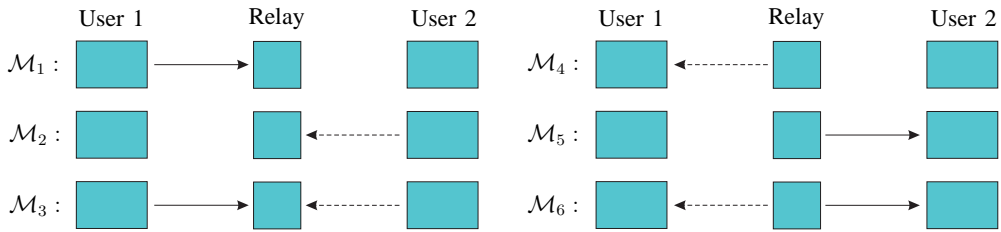}
\caption{The six possible transmission modes in the considered bidirectional relay channel.}
\label{FigModes}
\end{figure}

\section{Introduction} \label{Sec I (Intro)}
In a bidirectional relay channel, two users exchange information via a relay node\footnote[1]{We assume that there is no direct link between the users, and thus, they communicate with
each other only through the relay node.}. The communication in several practical applications such as satellite communication and cellular communication with a base station can be modeled via the bidirectional relay channel. Different protocols have been proposed
for this channel under the practical half-duplex constraint,
i.e., a node cannot transmit and receive at the same time and
in the same frequency band. Most of the existing protocols utilize a subset of the available transmission modes illustrated in Fig. \ref{FigModes}. In particular, there are four point-to-point modes
(mode $\mathcal{M}_1$: user 1-to-relay, mode $\mathcal{M}_2$: user 2-to-relay, mode $\mathcal{M}_4$: relay-to-user 1, mode $\mathcal{M}_5$: relay-to-user 2), a multiple-access mode (mode $\mathcal{M}_3$: both users to  relay),  and a broadcast mode (mode $\mathcal{M}_6$: relay to both users). The capacity
region of each of these six transmission modes is known for decode-and-forward processing at the relay \cite{Cover}, \cite{BocheIT}. Given this knowledge, there has been a significant research effort to obtain achievable rate regions for the bidirectional relay channel \cite{Tarokh,BocheIT,BochePIMRC,PopovskiICC,6ModeMIMO}.  The simplest protocol is the traditional two-way relaying protocol in which the transmission
is accomplished via four successive point-to-point phases, i.e., by using modes $\mathcal{M}_1,\mathcal{M}_5,\mathcal{M}_2,$ and $\mathcal{M}_4$ \cite{6ModeMIMO}. In  contrast,  the  time  division  broadcast  (TDBC)  protocol
exploits the broadcast capability of the wireless medium and utilizes the broadcast mode $\mathcal{M}_6$ along with the point-to-point modes $\mathcal{M}_1$ and $\mathcal{M}_2$ \cite{TDBC}. Thereby, the relay broadcasts a superimposed codeword, carrying information for both
user 1 and user 2, such that each user is able to recover its
intended information by self-interference cancellation. Another
existing  protocol  is  the  multiple-access  broadcast  (MABC)
protocol in which multiple-access mode $\mathcal{M}_3$ is used in addition to broadcast mode $\mathcal{M}_6$
\cite{MABC}.  In  the  multiple-access  phase,  both  user  1  and  user  2
simultaneously transmit to the relay which is able to decode
both messages. A hybrid broadcast (HBC) protocol was proposed in \cite{Tarokh} as a
generalization of the MABC and TDBC protocols, i.e., modes $\mathcal{M}_1,\mathcal{M}_2,\mathcal{M}_3,$ and $\mathcal{M}_6$ are used. Recently, in \cite{6ModeMIMO}, a protocol was proposed  which utilizes all six available transmission modes. For a comprehensive overview of the available protocols for the bidirectional relay channel, we refer to \cite{Tarokh,BochePIMRC,6ModeMIMO}, and references therein.

Due to mobility of the nodes and environmental effects such as path loss, shadowing, and multipath fading, the fluctuations of the transmitted signal vary with time. A general strategy to combat the time-varying nature of the channel is to utilize dynamic resource allocation such as adaptive power allocation and/or adaptive rate and coding scheme selection at the transmitter based on instantaneous
channel state information (CSI) \cite{Goldsmith,Tse}. For instance, Goldsmith and Varaiya \cite{Goldsmith} obtained the ergodic capacity of a  point-to-point Gaussian channel with fading using water-filling power allocation in time and adapting the rate to the instantaneous channel capacity. For general (multi-hop) relay channels, the instantaneous  capacity is limited by the minimum of the capacities of the involved links/hops (i.e., the bottleneck capacity). To overcome this limitation, adaptive power allocation can be used to allocate more power to the weaker channels such that the bottleneck capacity increases \cite{HostPower}. Moreover, for one-way relaying, in \cite{Poor}, a buffer was used at the relay to enable the relay to receive for a fixed number of time slots before retransmitting the information to the destination. The buffering capability allows the protocol in \cite{Poor} to overcome the limitation imposed by the instantaneous bottleneck capacity. In particular, assuming an infinite-size buffer, the achievable rate depends only on the ergodic capacities of the links and not on the instantaneous capacities. Building upon the idea of using relays with buffers, recently, adaptive link selection was proposed for one-way relaying in \cite{NikolaJSAC}. Thereby, based on the instantaneous CSI, in each time slot,
either the source-relay or the relay-destination link is selected
for transmission. Motivated by the performance gains reported in \cite{NikolaJSAC}, we extend the idea of adaptive link selection from one-way relaying to bidirectional relaying where, based on the instantaneous CSI, the optimal \textit{transmission mode} is selected in each time slot. Furthermore, we characterize the achievable rate region of  the block fading half-duplex bidirectional relay channel with adaptive mode selection and develop a protocol which achieves the points on the boundary surface of the rate region.

In order to be able to apply adaptive mode selection, the relay has to be equipped with two buffers for storage of the information received from each of the users. We note that the advantages of adaptive mode selection come at the expense of an increased end-to-end delay caused by data buffering. Therefore, we consider both delay-unconstrained transmission and delay-constrained transmission. Delay-unconstrained transmission provides a performance upper bound in terms of the achievable rate region for the delay-constrained case. We formulate an optimization problem to obtain the boundary points of the achievable rate region for delay-unconstrained transmission. To this end, we provide a useful condition for the queues of the buffers which significantly simplifies the problem formulation. For the considered bidirectional relay channel, the selected transmission mode and the corresponding transmission rates in each time slot are variables with a degree of freedom and have to be optimized to obtain the boundary surface of the rate region. Moreover, depending on the type of power constraint imposed, the  powers  of  the  nodes  might also be  optimized  and   constitute  degrees  of  freedom. In this paper, we consider two different transmit power constraints: 1) a joint long-term power constraint for all nodes, and 2) a fixed transmit power for each node. The first power constraint is interesting theoretically since it leads to the largest feasible set for the optimization variables for a given total transmit power and hence, results in the largest achievable rate region. The second power constraint is more practical and simpler from an implementation point of view. The solution of the formulated optimization problem for each of the two power constraints results in a protocol which specifies the optimal transmission strategy, i.e., the optimal transmission mode, the optimal transmission rates, and the optimal transmit powers  based on the statistical and instantaneous CSI of the involved links. For the case of delay-constrained  transmission, we propose a heuristic but efficient modification of the proposed protocols which takes into account the effect of finite-size buffers at the relay  and consequently limits the end-to-end delay. Numerical results reveal that the proposed protocols  with  adaptive mode selection outperform the available protocols \cite{TDBC,MABC,Tarokh,BochePIMRC,PopovskiICC,6ModeMIMO} considerably even in cases where only a small delay is permitted.

We note that adaptive mode selection for the bidirectional relay channel was also considered in \cite{PopovskiLetter} for a sum-rate maximization problem. However, the selection policy in \cite{PopovskiLetter} does not use all possible modes but  only selects from two point-to-point modes and the broadcast mode, and assumes that the transmit powers of all three  nodes  are  fixed  and  identical. Therefore, the rates achievable with the protocol in \cite{PopovskiLetter} lie in the interior of the  achievable rate region of the proposed protocols.

The  remainder  of  this  paper  is  organized  as  follows.  In
Section  II,  the  system  model and problem formulation are  presented.  In  Section  III, assuming a joint long-term  power constraint for all nodes, the achievable rate regions for both delay-constrained and delay-unconstrained transmission are investigated and the corresponding relaying protocols are provided. In  Section  IV,
the achievable rate regions and the corresponding relaying protocols for the case of a fixed transmit power for each node are presented. Numerical results are provided
in Section V, and conclusions are drawn in Section VI.

\section{System Model and Problem Formulation}\label{SysMod}
In this section, we introduce the considered channel model, review
 the achievable rates of the six possible transmission modes in each time slot, and discuss the degrees of freedom for optimization. Furthermore, we formulate an optimization problem for characterization of the long-term achievable rate region.
\begin{figure}
\centering
\psfrag{W1}[c][c][0.8]{$W_{12}(i)$}
\psfrag{W2}[c][c][0.8]{$W_{21}(i)$}
\psfrag{W2p}[c][c][0.8]{$\widehat{W}_{21}(i)$}
\psfrag{W1p}[c][c][0.8]{$\widehat{W}_{12}(i)$}
\psfrag{W2r}[c][c][0.8]{$\widetilde{W}_{21}(i)$}
\psfrag{W1r}[c][c][0.8]{$\widetilde{W}_{12}(i)$}
\psfrag{X1}[c][c][0.8]{$X_1(i)$}
\psfrag{X2}[c][c][0.8]{$X_2(i)$}
\psfrag{Xr}[c][c][0.8]{$X_r(i)$}
\psfrag{Y1}[c][c][0.8]{$Y_1(i)$}
\psfrag{Y2}[c][c][0.8]{$Y_2(i)$}
\psfrag{Yr}[c][c][0.8]{$Y_r(i)$}
\psfrag{U1}[l][c][1]{\text{User 1}}
\psfrag{U2}[l][c][1]{\text{User 2}}
\psfrag{R}[l][c][1]{\text{Relay}}
\psfrag{E1}[c][c][1]{\text{Encoder}}
\psfrag{E2}[c][c][1]{\text{Encoder}}
\psfrag{Er}[c][c][1]{\text{Encoder}}
\psfrag{D1}[c][c][1]{\text{Decoder}}
\psfrag{D2}[c][c][1]{\text{Decoder}}
\psfrag{Dr}[c][c][1]{\text{Decoder}}
\psfrag{B1}[c][c][1]{$B_1$}
\psfrag{B2}[c][c][1]{$B_2$}
\psfrag{h1}[c][c][0.8]{$h_1(i)$}
\psfrag{h2}[c][c][0.8]{$h_2(i)$}
\psfrag{Z1}[c][c][0.8]{$Z_1(i)$}
\psfrag{Z2}[c][c][0.8]{$Z_2(i)$}
\psfrag{Zr}[c][c][0.8]{$Z_r(i)$}
\includegraphics[width=0.9\linewidth]{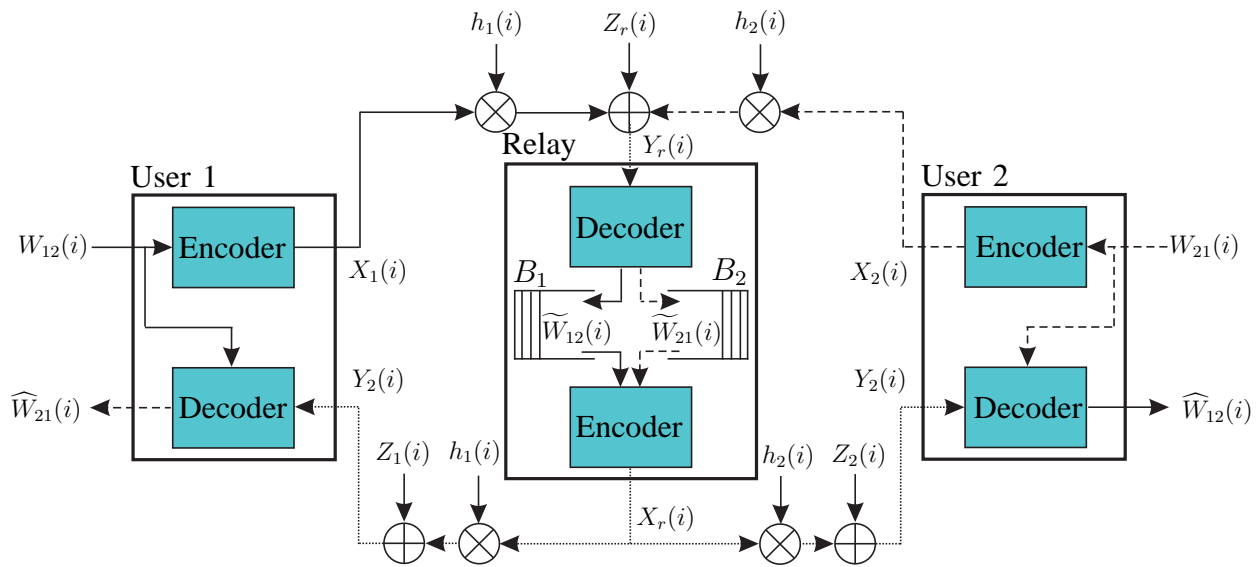}
\caption{Three-node bidirectional relay channel consisting of two users and a decode-and-forward relay.}
\label{FigSysMod}
\end{figure}

\subsection{Channel Model}
We consider the bidirectional relay channel in which user 1 and user 2
exchange information with the help of a decode-and-forward relay node as shown
in Fig. \ref{FigSysMod}.  We assume that there is no direct link between user
1 and user 2, and thus, user 1 and user 2 communicate with
each other only through the relay node. We assume that all
three nodes in the network are half-duplex. Furthermore, we
assume that time is divided into slots of equal length and that
each node transmits codewords which span one time slot. We assume
that the user-to-relay and relay-to-user channels are impaired
by additive white Gaussian noise (AWGN) with unit variance and block fading, i.e., the channel
coefficients are constant during one time slot and change from
one time slot to the next. Moreover, in each time slot, the
channel coefficients are assumed to be reciprocal such that the
user 1-to-relay and the user 2-to-relay channels are identical to
the relay-to-user 1 and relay-to-user 2 channels, respectively. The channel reciprocity assumption is valid for time-division-duplex (TDD) systems where the user-to-relay and relay-to-user links utilize the same frequency band. The received  codewords for the considered channel can be modelled as 
\begin{IEEEeqnarray}{lll}\label{Gaussian}
  Y_1(i) &= h_1(i)X_r(i)+Z_1(i), &\text{if user 1 receives} \IEEEeqnarraynumspace\IEEEyesnumber\IEEEyessubnumber \\
   Y_2(i) &= h_2(i)X_r(i)+Z_2(i), &\text{if user 2 receives}\IEEEyessubnumber \\
   Y_r(i) &= h_1(i)X_1(i) + h_2(i)X_2(i)+Z_r(i), \quad &\text{if the relay receives}\IEEEyessubnumber
\end{IEEEeqnarray}
where $X_j(i),Y_j(i),$ and $Z_j(i),\,\,j\in\{1,2,r\}$, denote the transmitted codeword of node $j$, the received codeword at node $j$, and the noise at node $j$ in the $i$-th time slot, respectively. Furthermore, $h_1(i)$ and $h_2(i)$ denote the channel coefficients between user 1 and the relay and between user 2 and the relay in the $i$-th time slot, respectively.  The squares  of the channel coefficient amplitudes in the $i$-th time slot are denoted by $S_1(i)=|h_1(i)|^2$ and $S_2(i)=|h_2(i)|^2$. Moreover, $S_1(i)$ and $S_2(i)$   are assumed to be ergodic and stationary random processes with means $\Omega_1=E\{S_1\}$ and  $\Omega_2=E\,\{S_2\}$, respectively\footnote[1]{In this paper, we drop time index $i$ in expectations for notational simplicity.}, where $E\{\cdot\}$ denotes expectation.  We also assume that $S_1(i)$ and $S_2(i)$ have continuous probability density functions. Moreover,  since the noise is AWGN, in order to achieve the channel capacity in each mode, the nodes transmit
Gaussian distributed codewords, i.e., $X_j(i)$ is comprised of symbols which are zero-mean rotationally invariant complex  Gaussian random variables with variance $P_j(i), \,\,j\in\{1,2,r\},\,\forall i$. $P_j(i)$ also represents the transmit power of node $j$ in the $i$-th time slot. For notational convenience, we use the definition $C(x)\triangleq \log_2(1+x)$ throughout the paper.

\subsection{Transmission Modes and Their Achievable  Rates}

In the considered bidirectional relay channel, six transmission modes are possible, cf. Fig. \ref{FigModes}. Let  $R_{jj'}(i)\geq 0, \,\,j,j'\in\{1,2,r\}$, denote the transmission rate from node $j$ to node $j'$ in the $i$-th time slot. Let $B_1$ and $B_2$ denote two buffers of sizes $Q_1^{\max}$ and $Q_2^{\max}$ at the relay in which the  information received from user 1 and user 2 is stored, respectively. Moreover, $Q_j(i), \,\, j\in\{1,2\}$, denotes the amount of normalized information in bits/symbol available in buffer $B_j$ in the $i$-th time slot. Using this notation, the transmission modes, their respective rates, and the dynamics of the queues at the buffers are presented in the following:

${\cal M}_1$: User 1 encodes message $W_{12}(i)$ into codeword $X_1(i)$ and transmits it to the relay. User 2 is silent and the relay receives $Y_r(i)$ according to (\ref{Gaussian}c). The relay decodes this information to $\widetilde{W}_{12}(i)$ and stores it in buffer $B_1$. For this mode, the transmission rate in each time slot is limited by the capacity of the user 1-to-relay channel for arbitrary small decoding error, $\Pr\{W_{12}(i)\neq \widetilde{W}_{12}(i)\}\to 0$, and the space available in buffer $B_1$ to store information. Therefore, the transmission rate from user 1 to the relay in the $i$-th time slot must satisfy $R_{1r}(i)\leq \min\{C_{1r}(i),Q_1^{\max}-Q_1(i)\}$, where $C_{1r}(i)=C(P_1(i)S_1(i))$. Moreover, the amount of information in buffer $B_1$ increases to $Q_1(i)=Q_1(i-1)+R_{1r}(i)$.

${\cal M}_2$: User 2 encodes message $W_{21}(i)$ into codeword $X_2(i)$ and transmits it to the relay. User 1 is silent and the relay receives $Y_r(i)$ according to (\ref{Gaussian}c). The relay decodes this information to $\widetilde{W}_{21}(i)$ and stores it in buffer $B_2$. In this mode, the transmission rate from user 2 to the relay in  the $i$-th time slot must satisfy $R_{2r}(i)\leq \min\{C_{2r}(i),Q_2^{\max}-Q_2(i)\}$, where $C_{2r}(i)=C(P_2(i)S_2(i))$.  Moreover, the amount of information in buffer $B_2$ increases to $Q_2(i)=Q_2(i-1)+R_{2r}(i)$.

${\cal M}_3$: Users 1 and 2 encode massages $W_{12}(i)$ and $W_{21}(i)$ to $X_1(i)$ and $X_2(i)$, respectively, and transmit them simultaneously to the relay. The relay receives $Y_r(i)$ according to (\ref{Gaussian}c). For this mode, we assume that multiple-access channel coding is used, see \cite{Cover}. Thereby, the relay decodes the information received from users 1 and users 2 to $\widetilde{W}_{12}(i)$ and $\widetilde{W}_{21}(i)$ and stores it in buffer $B_1$ and $B_2$, respectively.  The transmission rates in the $i$-th time slot must satisfy  $R_{1r}(i)\leq \min\{C_{1r}(i),Q_1^{\max}-Q_1(i)\}$, $R_{2r}(i)\leq \min\{C_{2r}(i),Q_2^{\max}-Q_2(i)\}$, and $R_{1r}(i)+R_{2r}(i)\leq C_{r}(i)$, where $C_{r}(i)=C(P_1(i)S_1(i)+P_2(i)S_2(i))$.
 Since user 1 and user 2 transmit independent messages, the sum rate, $C_r(i)$, can be decomposed into two rates, one from user 1 to the relay and the other one from user 2 to the relay. These two capacity rates can be achieved via successive decoding. To this end, we introduce a binary variable $t(i)\in\{0,1\}$ which determines the order of successive decoding. Specifically, if $t(i)=0$, the relay first decodes the
codeword received from user 1 and considers the codeword from user 2 as noise. Then, the relay subtracts the contribution of user 1's codeword from the received codeword and decodes the
 codeword received from user 2. Similarly, if $t(i)=1$,  the relay first
decodes the codeword received from user 2 and treats the codeword of user 1 as noise, and then the relay decodes the codeword received from user 1\footnote[1]{We note that $0\leq t(i)\leq 1$ is possible via time sharing, i.e., in $t(i)$ fraction of the $i$-th time slot, we first decode the  codeword received from user 2 and in the remaining $1-t(i)$ fraction, we first decode the codeword received from user 1. However, we prove in Appendix \ref{AppBinRelax} that time sharing cannot improve the achievable rate region of the considered bidirectional relay channel. Therefore, for simplicity of implementation, we assume binary values for $t(i)$.}.  Therefore,  we decompose $C_r(i)$ as $C_r(i)=C_{12r}(i)+C_{21r}(i)$ where the transmission rates from users 1 and 2 to the relay in the $i$-th time slot must satisfy $R_{1r}(i) \leq \min\{C_{12r}(i),Q_1^{\max}-Q_1(i)\}$ and $R_{2r}(i)\leq \min\{C_{21r}(i),Q_2^{\max}-Q_2(i)\}$, respectively, and $C_{12r}(i)$ and $C_{21r}(i)$ are given by
\begin{IEEEeqnarray}{Cll}\label{Cr}
  C_{12r}(i)&=t(i) C\left(P_1(i)S_{1}(i)\right) + (1-t(i)) C\left(\frac{P_1(i)S_{1}(i)}{1+P_2(i)S_{2}(i)}\right)  \IEEEeqnarraynumspace\IEEEyesnumber\IEEEyessubnumber \\
   C_{21r}(i)&=(1-t(i)) C\left(P_2(i)S_{2}(i)\right) + t(i) C\left(\frac{P_2(i)S_{2}(i)}{1+P_1(i)S_{1}(i)}\right) \IEEEyessubnumber
\end{IEEEeqnarray}
Moreover, the amounts of information in buffers $B_1$ and $B_2$ increase to $Q_1(i)=Q_1(i-1)+R_{1r}(i)$ and $Q_2(i)=Q_2(i-1)+R_{2r}(i)$, respectively.

${\cal M}_4$: The relay extracts the information from buffer $B_2$, encodes it into codeword $X_r(i)$, and transmits it to user 1. User 2 is silent and user 1 receives $Y_1(i)$ according to (\ref{Gaussian}a) and decodes the information to $\widehat{W}_{21}(i)$. For this mode, the transmission rate from the relay to user 1 in the $i$-th time slot is limited by both the capacity of the relay-to-user 1 channel and the amount of  information  stored in buffer $B_2$. Thus, the transmission  rate from the relay to user 1 must satisfy $R_{r1}(i) \leq \min\{C_{r1}(i),Q_2(i-1)\}$, where $C_{r1}(i)=C(P_r(i)S_1(i))$. The amount of information in buffer $B_2$ decreases to $Q_2(i)\Equal Q_2(i\Minus 1)\Minus R_{r1}(i)$.

${\cal M}_5$: This mode is identical to ${\cal M}_4$ with user 1 and 2 switching roles. The transmission rate from the relay to user 2 must satisfy $R_{r2}(i) \leq \min\{C_{r2}(i),Q_1(i-1)\}$, where $C_{r2}(i)=C(P_r(i)S_2(i))$. The amount of information in buffer $B_1$ decreases to $Q_1(i)\Equal Q_1(i\Minus 1)\Minus R_{r2}(i)$.

${\cal M}_6$: The relay extracts the information intended for user 2 from buffer $B_1$ and  the information intended for user 1 from buffer $B_2$. Then, based on the scheme in \cite{BocheIT}, it constructs superimposed codeword $X_r(i)$ which contains the information of both users and broadcasts it to the users. Therefore, user 1 and user 2 receive $Y_1(i)$ and $Y_2(i)$ according to (\ref{Gaussian}a) and (\ref{Gaussian}b), and using the side information $W_{12}$ and $W_{21}$, decode them to $\widehat{W}_{21}(i)$ and $\widehat{W}_{12}(i)$, respectively. Thus, in the $i$-th time slot, the transmission rates from the relay to users 1 and 2 must satisfy  $R_{r1}(i)\leq \min \{C_{r1}(i), Q_2(i-1)\}$ and $R_{r2}(i)\leq\min\{C_{r2}(i),Q_1(i-1)\}$, respectively. The amounts of information in buffers $B_1$ and $B_2$ decrease to $Q_1(i)\Equal Q_1(i\Minus 1)\Minus R_{r2}(i)$ and $Q_2(i)\Equal Q_2(i\Minus 1)\Minus R_{r1}(i)$, respectively.


\subsection{Degrees of Freedom for Optimization}

Our aim is to characterize the long-term achievable rate region of the considered half-duplex bidirectional relay channel with block fading and to develop a corresponding protocol which achieves all points of the achievable rate region. To this end, we have to optimize all  variables with a degree of freedom in the channel. The variables with a degree of freedom in the considered  channel are  1) the transmission mode selection in each time slot, 2) the transmission rates of the transmitting nodes, and 3) the transmit powers of the transmitting nodes given the adopted power constraint. 

For transmission mode selection, we introduce six binary variables,  $q_k(i) \in\{0,1\}, \,\,k=1,...,6$, where  $q_k(i)$ indicates whether or not transmission mode  $\mathcal{M}_k$ is selected in the $i$-th time slot. In particular, $q_k(i)=1$ if  mode $\mathcal{M}_k$ is selected and $q_k(i)=0$ if it is not selected in the $i$-th time slot. Furthermore, since in each time slot  only one of the six transmission modes can be selected, only one of the mode selection variables is equal to one and the others are zero, i.e., $\mathop \sum_{k = 1}^6 q_k(i)=1$ holds. 

Furthermore, transmitting with less than the maximum possible rate in each time slot cannot improve the performance in terms of the achievable rates. Therefore, without loss of generality, we assume that for any given transmission mode, the transmitting nodes transmit with the maximum possible rates in each time slot. Therefore, the transmission rates of the point-to-point modes and broadcast mode are unique \cite{Cover,BocheIT} and there is no degree of freedom. However, the maximum transmission rates of the multiple-access mode in the $i$-th time slot depend on the successive decoding order variable $t(i)$. Therefore, the value of $t(i)\in\{0,1\}$ has to be optimized as a degree of freedom in the considered bidirectional relay channel if $q_3(i)=1$. 

The transmit powers of the nodes also constitute degrees of freedom in the channel and have to be optimized for the given power constraint. Different types of power constraints have been considered in the literature depending on the particular practical and theoretical aspect of interests. One classification is joint versus individual node power constraints. Under a joint node power constraint, a total power budget is divided between all nodes in the network. In contrast, under individual node power constraints, each node has its own power budget. For a given total power budget, the joint node power constraint for all nodes provides a larger feasible solution set compared to individual node power constraints, and thus, leads to a larger rate region. On the other hand, individual node power constraints are more relevant in practice, where each node typically has its own power supply. Another possible classification is long-term versus short-term power constraints. Specifically, under long-term power constraints, the transmit power can change from time slot to time slot but the average consumed power is limited.  In contrast, under short-term power constraints, the transmit powers have to satisfy the power constraint in each time slot. Obviously, long-term power constraints lead to a larger achievable rate region compared to short-term power constraints assuming that the total power budget, whether it is a total power budget for all nodes or per-node power budgets, is the same in both cases. On the other hand, short-term power constraints are preferable from the  implementation point of view as they lead to less fluctuations in the transmitted signals. Herein, we consider the following two cases: 1) a joint long-term  power constraint for all nodes, and 2) an  individual short-term  power constraint  for each node. Our motivation for choosing these particular power constraints has both theoretical and practical aspects. In particular, the first power constraint leads to the largest rate region for a given total power budget, and thus, provides a performance bound for all other types of power constraints. The second power constraint is more practical, e.g., the complexity of calculating the optimal powers is avoided and the transmitters can be simpler as the transmit signals fluctuate less, facilitating the application of low-cost power amplifiers. However, the second power constraint does not offer a degree of freedom in choosing the powers of the nodes, i.e., the powers of the nodes  are fixed for all time slots. 

We note that the optimal transmission strategy fundamentally depends on the adopted power constraint, see Sections III and IV. This motivated the consideration of both power constraints in this paper.

\subsection{Characterization of the Achievable Rate Region}

In this subsection, we formulate an optimization problem which allows us to characterize the long-term achievable rate region of the considered  bidirectional relay channel with block fading. The long-term achievable rate region is denoted by $\mathcal{R}$, and the average transmission rates from user 1 to user 2 and from user 2 to user 1 are denoted by $\bar{R}_{12}$ and $\bar{R}_{21}$, respectively. We assume that user 1 and user 2 always have enough information to send in all time slots and that the number of time slots, $N$, satisfies $N\to \infty$. Moreover, the user 1-to-relay, user 2-to-relay, relay-to-user 1, and relay-to-user 2 average transmission rates are denoted by $\bar{R}_{1r}$, $\bar{R}_{2r}$, $\bar{R}_{r1}$, and $\bar{R}_{r2}$, respectively, and are given by
\begin{IEEEeqnarray}{lll}\label{RatReg123}
    &&\bar{R}_{1r} = \underset{N\to \infty}{\lim} \frac{1}{N}\mathop \sum \limits_{i = 1}^N \left[ q_1(i) \min\{C_{1r}(i),Q_1^{\max} \Minus Q_1(i)\}\Add q_3(i)\min\{C_{12r}(i),Q_1^{\max} \Minus Q_1(i)\}\right]  \IEEEyesnumber\IEEEyessubnumber \\
		&&\bar{R}_{2r} =  \underset{N\to \infty}{\lim} \frac{1}{N}\mathop \sum \limits_{i = 1}^N \left[ q_2(i) \min\{C_{2r}(i),Q_2^{\max} \Minus Q_2(i)\}\Add q_3(i)\min\{C_{21r}(i),Q_2^{\max} \Minus Q_2(i)\}\right]  \IEEEyessubnumber\\
    &&\bar{R}_{r1}\hspace{-1mm}  =\hspace{-1mm} \underset{N\to \infty}{\lim}\frac{1}{N}\hspace{-1mm} \mathop \sum \limits_{i = 1}^N \left[ q_4(i)\hspace{-0.5mm} +\hspace{-0.5mm} q_6(i)\right]\hspace{-0.5mm} \min\{C_{r1}(i),Q_2(i-1)\} \IEEEeqnarraynumspace\IEEEyessubnumber \\
		&&\bar{R}_{r2}\hspace{-1mm}  =\hspace{-1mm}   \underset{N\to \infty}{\lim}\frac{1}{N}\hspace{-1mm} \mathop \sum \limits_{i = 1}^N \left[ q_5(i)\hspace{-0.5mm} +\hspace{-0.5mm} q_6(i)\right]\hspace{-0.5mm} \min\{C_{r2}(i),Q_1(i-1)\}. \IEEEeqnarraynumspace \IEEEyessubnumber
\end{IEEEeqnarray}
Furthermore, the average rate from user 1 to user 2 is the average rate that user 2 receives from the relay, i.e.,  $\bar{R}_{12}=\bar{R}_{r2}$. Similarly, the average rate from user 2 to user 1 is the average rate that user 1 receives from the relay, i.e.,  $\bar{R}_{21}=\bar{R}_{r1}$.  To obtain the rate region, we first define the boundary surface of the rate region.

\begin{defin}
The boundary surface, $\mathcal{R}^{\mathrm{bound}}$, of rate region $\mathcal{R}$ is the set of all points $(\bar{R}_{12},\bar{R}_{21})\in\mathcal{R}$ such that if one of the rates is fixed and the other rate is increased, the resulting new point is no longer in $\mathcal{R}$, i.e.,
\begin{IEEEeqnarray}{Cll}\label{DefBound}
    \mathcal{R}^{\mathrm{bound}} =\Big\{ (\bar{R}_{12},\bar{R}_{21})\in\mathcal{R} \big|  (\bar{R}_{12}+\epsilon,\bar{R}_{21})\notin\mathcal{R} \wedge (\bar{R}_{12},\bar{R}_{21}+\epsilon)\notin\mathcal{R},\,\,\forall \epsilon>0 \Big\}.
\end{IEEEeqnarray}
\end{defin}

Obtaining the boundary surface of the achievable rate region is sufficient for characterizing the complete achievable rate region. All other points of $\mathcal{R}$ can be achieved by deliberately decreasing the average transmission rates of user 1 and/or user 2. Hence, the boundary surface specifies the set of optimal operating points of the rate region.

In the following, we use the time sharing argument to introduce an alternative representation of the boundary surface of the rate region. In particular, the time sharing argument indicates that if rate pairs  $(\bar{R}_{12},\bar{R}_{21})$ and $(\bar{R}_{12}^{'},\bar{R}_{21}^{'})$ are achievable, rate pair $(\rho\bar{R}_{12}+(1-\rho)\bar{R}_{12}^{'},\,\rho\bar{R}_{21}+(1-\rho)\bar{R}_{21}^{'})$ is also achievable for any $\rho\in [0,\,1]$. One direct result of the time sharing argument is that the rate region is convex \cite{ElGamal}. Using the convexity of the rate region, for each point $(\bar{R}_{12}^{*},\bar{R}_{21}^{*})\in  \mathcal{R}^{\mathrm{bound}}$, there exists one and only one value of $\eta\in(0,\,1)$ for which rate pair $(\bar{R}_{12}^{*},\bar{R}_{21}^{*})$ is obtained by solving the following optimization problem \cite{Tse}, \cite{6ModeMIMO}
\begin{IEEEeqnarray}{rll}\label{OptProbOrigin}
    {\underset{\mathcal{F}}{\mathrm{maximize}}}\,\, & \eta\bar{R}_{12}+(1-\eta)\bar{R}_{21} \nonumber\\
\mathrm{subject \,\, to\,\,}    &\mathrm{C1}:\,\,  \bar{R}_{1r} = \bar{R}_{r2} \nonumber \\
    & \mathrm{C2}:\,\, \bar{R}_{2r}= \bar{R}_{r1} 
\end{IEEEeqnarray}
where $\mathcal{F}$ is the set of variables with a degree of freedom available in the channel. For the problem at hand, in general, $\mathcal{F}$ is comprised of the mode selection variables $q_{k}(i),\,\, \forall i,k$, the successive decoding order variable $t(i),\,\, \forall i$, and the transmit powers  $P_j(i),\,\, \forall i,j$. Moreover, $\bar{R}_{1r} < \bar{R}_{r2}$ and $\bar{R}_{2r} < \bar{R}_{r1} $ is not possible due to the conservation of flow. Furthermore, $\bar{R}_{1r} > \bar{R}_{r2}$ and $\bar{R}_{2r} > \bar{R}_{r1} $ can also not be permitted, since all information bits transmitted by the sources have to arrive at the intended destinations. Therefore, constraints $\mathrm{C1}$ and $\mathrm{C2}$ must hold for any long-term achievable  rate pair. The complete boundary surface of the rate region can be obtained by solving optimization problem (\ref{OptProbOrigin}) for all $\eta\in(0,\,1)$.

\begin{figure}
\centering
\psfrag{R12}[c][c][1]{$\bar{R}_{12}$}
\psfrag{R21}[c][c][1]{$\bar{R}_{21}$}
\psfrag{Eta1}[c][c][1]{$\eta\to 0$}
\psfrag{Eta2}[c][c][1]{$\eta=\frac{1}{2}$}
\psfrag{Eta3}[c][c][1]{$\eta\to 1$}
\includegraphics[width=0.4 \linewidth]{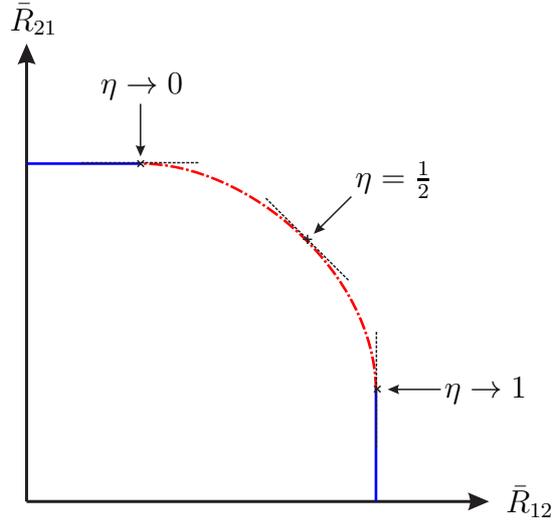}
\caption{The points on the boundary surface of the rate region are denoted by the dash-dotted line for different values of $\eta\in(0,\,1)$.}
\label{FigOpt}
\end{figure}

\begin{remk}
In Fig. \ref{FigOpt}, the concept of using the optimization problem in (\ref{OptProbOrigin}) to obtain different points on the boundary surface of the rate region is illustrated. We observe that the convexity of the rate region is a key point for the adopted alternative representation of the boundary surface of the rate region. In other words, if the rate region was not convex, some points on the boundary surface of the rate region could not be obtained with the optimization problem in (\ref{OptProbOrigin}). We also note that the boundary surface of the rate region in Fig. \ref{FigOpt} is the dash-dotted  line. One can obtain the extreme points on the boundary surface by setting $\eta$ arbitrarily close to $0$ or $1$. However, by setting $\eta=0,1$ in optimization problem (\ref{OptProbOrigin}), we obtain all points on the solid lines which are not on the boundary surface. This is the reason why we exclude $\eta=0,1$ in the alternative representation of the boundary surface of the rate region.
\end{remk}

In the following, we first assume infinite-size buffers at the relay and formulate the problem for delay-unconstrained transmission. For infinite-size buffers, the queues of the buffers have to satisfy a useful condition which significantly simplifies the optimization problem in (\ref{OptProbOrigin}). 

\begin{lem}\label{Queue} 
For infinite-size buffers $B_1$ and $B_2$ at the relay, the boundary surface of the rate region for the considered half-duplex bidirectional relay channel with block fading can be achieved when the queues of the buffers $B_1$ and $B_2$ at the relay are at the edge of non-absorbtion. Furthermore, in this case,  $\bar{R}_{1r},\bar{R}_{r2},\bar{R}_{2r},$ and $\bar{R}_{r1}$ are given by
\begin{IEEEeqnarray}{lll}\label{RatReg-buffer}
    &\bar{R}_{1r} = \underset{N\to \infty}{\lim} \frac{1}{N}\mathop \sum \limits_{i = 1}^N \left[ q_1(i) C_{1r}(i) + q_3(i) C_{12r}(i)\right]  \IEEEyesnumber\IEEEyessubnumber \\
		&\bar{R}_{2r} =  \underset{N\to \infty}{\lim} \frac{1}{N}\mathop \sum \limits_{i = 1}^N \left[ q_2(i) C_{2r}(i) + q_3(i)C_{21r}(i)\right]  \IEEEyessubnumber\\
		&\bar{R}_{r1}= \underset{N\to \infty}{\lim}\frac{1}{N}\mathop \sum \limits_{i = 1}^N \left[ q_4(i)+q_6(i)\right]C_{r1}(i) \IEEEeqnarraynumspace\IEEEyessubnumber \\
&\bar{R}_{r2} = \underset{N\to \infty}{\lim}\frac{1}{N}\mathop \sum \limits_{i = 1}^N \left[ q_5(i)+q_6(i)\right]C_{r2}(i).  \IEEEeqnarraynumspace \IEEEyessubnumber 
\end{IEEEeqnarray}
\end{lem}

\begin{IEEEproof}
Please refer to Appendix \ref{AppQueue}.
\end{IEEEproof}

\begin{remk}
Lemma \ref{Queue} reveals that for the optimal solution, as $N\to\infty$, the effect of the queues at the relay becomes negligible for the optimal solution, i.e., the
relay always has space available in its buffers for information reception and enough information for transmission, and thus, (\ref{RatReg123}a)-(\ref{RatReg123}d) simplify to (\ref{RatReg-buffer}a)-(\ref{RatReg-buffer}d), respectively. Therefore, for infinite-size buffers at the relay, the dynamics of the queues do not affect the long-term achievable rate region.  This property of infinite-size buffers at the relay significantly facilitates finding the boundary surface of the rate region.
\end{remk}

Due to the binary constraints $q_k(i)\in\{0,1\},\,\,\forall i,k$ and $t(i)\in\{0,1\},\,\,\forall i$, the optimization problem in (\ref{OptProbOrigin}) is an integer programming problem which belongs to the category of non-deterministic polynomial-time hard (NP hard) problems. To make the problem tractable, we relax the binary constraints to $0\leq q_k(i)\leq 1$ and $0\leq t(i)\leq 1$ which in general implies that the solution of the relaxed problem might not be obtainable with the original problem, i.e., the relaxed problem has a larger feasible set. However, we prove in Appendix \ref{AppBinRelax} that the solution of the relaxed problem can be achieved with  binary values for $q_k(i)$ and $t(i)$ if the channel gains have  continuous probability density functions.  Therefore, the solution of the relaxed problem solves the original problem as well. For infinite-size buffers, the optimization problem in (\ref{OptProbOrigin}) with the relaxed constraints can be written as
\begin{IEEEeqnarray}{Cll}\label{OptProb}
    {\underset{q_k(i),P_j(i),t(i),\,\, \forall i,j,k}{\mathrm{maximize}}}\,\, &\eta\bar{R}_{1r}+(1-\eta)\bar{R}_{2r} \nonumber \\
    \mathrm{subject\,\, to} \,\, &\mathrm{C1}:\quad \bar{R}_{1r}=\bar{R}_{r2}  \nonumber \\
    &\mathrm{C2}:\quad \bar{R}_{2r}=\bar{R}_{r1} \nonumber \\
		&\mathrm{C3}:\quad \sum\limits_{k = 1}^6 {q_k}\left( i \right) = 1, \,\, \forall i   \nonumber \\
    &\mathrm{C4}:\quad 0\leq q_k(i) \leq 1, \,\, \forall i, k \nonumber \\
    &\mathrm{C5}:\quad 0\leq t(i)\leq 1, \,\, \forall i \nonumber\\
        &\mathrm{C6}:\quad \left(P_1(i),P_2(i),P_r(i)\right)\in \mathcal{P}, \,\, \forall i \IEEEyesnumber 
\end{IEEEeqnarray}
where constraints $\mathrm{C1}$ and $\mathrm{C2}$ are the conditions introduced in Lemma \ref{Queue}. Constraints $\mathrm{C3}$ and $\mathrm{C4}$ specify the relaxed restrictions on the mode selection variables, constraint $\mathrm{C5}$ specifies the relaxed constraint on the decoding order variable $t(i)$, and $\bar{R}_{1r},\bar{R}_{2r},\bar{R}_{r1},$ and $\bar{R}_{r2}$ are given in (\ref{RatReg-buffer}). Furthermore, we maximize  $\eta\bar{R}_{1r}+(1-\eta)\bar{R}_{2r}$ in (\ref{OptProb}) since, according to Lemma \ref{Queue} (and constraints $\mathrm{C1}$ and $\mathrm{C2}$), $\bar{R}_{12}=\bar{R}_{r2}=\bar{R}_{1r}$ and  $\bar{R}_{21}=\bar{R}_{r1}=\bar{R}_{2r}$ hold. Finally, $\mathcal{P}$ is the feasible set specified by the adopted power constraint. In Sections III and IV, for the two power constraints considered in this paper, we develop protocols which solve the optimization problem in (\ref{OptProb}).

The assumption of infinite-size buffers at the relay leads to delay-unconstrained  transmission and can be used as a performance bound for delay-constrained  transmission. Unfortunately, including the delay constraint in the problem formulation complicates the solution significantly as the dynamics of the queues of the buffers at the relay play a role in this case. Thus, in this paper, we introduce a heuristic but efficient modification to the optimal protocol for delay-unconstrained transmission to take into account the effect of  finite-size buffers at the relay, see Sections III and IV. Consequently, by an appropriate choice of $Q_j^{\max},\,\,j\in\{1,2\}$, delay-constrained transmission is guaranteed.  We show in Section V that even for  a small tolerable average delay, the performance of delay-constrained transmission is close to the performance of delay-unconstrained transmission in terms of the achievable rates.

Before providing the achievable rate regions for the considered power constraints and the corresponding protocols, we note that throughout this paper, we assume that all nodes have
full knowledge of the CSI of both links\footnote[1]{We note that in order to achieve the broadcast capacity of  the bidirectional relay channel as proposed in \cite{BocheIT}, both users have to know the  transmission rates from the relay to both users. Therefore, the knowledge of the global CSI at all nodes is  necessary for any conventional/buffer-aided bidirectional relaying protocol using optimal coding and decoding in the broadcast mode.}. Thus, based on the
CSI  and  the  proposed protocols, cf.   Theorem~\ref{AdaptProt} and Theorem~\ref{FixProt},  each  node is  able  to  individually  decide  which  transmission  mode  is
selected and adapt its transmission strategy accordingly. Moreover, the proposed optimal protocols may require coin flips. Thus, we assume that one node (e.g., the relay node) is responsible for performing the coin flipping and conveying the results to the other nodes. Moreover, we assume that the channel states change slow enough such that the signaling overhead caused by channel estimation and feedback  is negligible compared to the amount of transmitted information.

\section{Achievable Rate Region under Joint Long-term Power Constraint}\label{RateRegion}

In this section, we consider a joint long-term power constraint for all nodes which can be mathematically stated as
\begin{IEEEeqnarray}{Cll}\label{TotalPower}
    \mathcal{P} = \Big\{ \left(P_1(i),P_2(i),P_r(i)\right) \big| P_j(i)\geq 0, \,\, \forall i, j\,\, \wedge\,\, \bar{P}_1+\bar{P}_2+\bar{P}_r = P_t \Big\} 
\end{IEEEeqnarray}
 where $P_t$ is the total average power budget of all nodes. The average powers consumed by user 1, user 2, and the relay are given by
\begin{IEEEeqnarray}{lll}\label{PowerIndv}
    \bar{P}_1=\frac{1}{N} \mathop \sum \limits_{i = 1}^N (q_1(i)+q_3(i))P_1 (i) \IEEEyesnumber\IEEEyessubnumber \\
    \bar{P}_2=\frac{1}{N} \mathop \sum \limits_{i = 1}^N (q_2(i)+q_3(i))P_2 (i) \IEEEyessubnumber \\
    \bar{P}_r=\frac{1}{N} \mathop \sum \limits_{i = 1}^N (q_4(i)+q_5(i)+q_6(i))P_r (i), \IEEEyessubnumber
    \end{IEEEeqnarray}
respectively. In the following, we first develop a protocol for delay-unconstrained transmission and then we modify the protocol to take into account delay constraints.

\subsection{Delay-Unconstrained Transmission}

In this subsection, we consider infinite-size buffers at the relay and solve the optimization problem in (\ref{OptProb}).  The corresponding protocol achieves all points on the boundary surface of the rate region of the bidirectional relay channel with block fading under a joint long-term power constraint. However, before formally stating the optimal protocol, cf. Theorem \ref{AdaptProt}, we introduce some variables that we require in the protocol. First, as we will see later, the optimal protocol may involve coin flips. Therefore, we define $\mathcal{C}_n(i)\in \{\mathrm{0,1}\}$ as the outcome of the $n$-th coin flip in the $i$-th time slot. The probabilities of the possible outcomes of the $n$-th coin flip are $\Pr\{\mathcal{C}_n(i)=1\}=p_n$ and $\Pr\{\mathcal{C}_n(i)=0\}=1-p_n$. Second, in the optimal solution, the instantaneous capacities and powers are combined via long-term selection weights $\mu_1$ and $\mu_2$ and power weight $\gamma$, respectively\footnote[1]{Appendix \ref{AppKKTAdapt} reveals that the selection weights $\mu_1$ and $\mu_2$ and the power weight $\gamma$ are in fact Lagrange multipliers corresponding to constraints $\mathrm{C1}$ and $\mathrm{C2}$ in (\ref{OptProb}) and the long-term power constraint in (\ref{TotalPower}), respectively.}. The values of $\mu_1$, $\mu_2$, and $\gamma$ depend on the channel statistics, the total power budget, and the value of $\eta$. Third, we define three mutually exclusive selection regions $\mathcal{S}_0$, $\mathcal{S}_1$, and $\mathcal{S}_2$ based on the values of the selection weights $\mu_1$ and $\mu_2$. As will be shown, each region requires a different optimal selection policy. For given channel statistics, total power budget, and $\eta$,  exactly one  of the selection regions contains the optimal values of $\mu_1$ and $\mu_2$. 

\begin{theo}\label{AdaptProt}
Assuming $N\to \infty$, the boundary surface of the long-term achievable rate region of the considered half-duplex bidirectional relay channel with AWGN and block fading under a joint long-term power constraint is the closure of points $(\bar{R}_{12},\bar{R}_{21})$ obtained for all $\eta\in(0,1)$ with the mode selection, power allocation, and decoding order policies provided in the following. In particular, the optimal mode selection policy is
\begin{IEEEeqnarray}{lll}\label{SelecCrit}
q_{k^*}(i)=
   \begin{cases}
     1, & \textrm{if }k^*= {\underset{k=1,\dots,6}{\arg \, \max}} \{\mathcal{I}_k(i)\Lambda_k(i)\} \\
     0, &\mathrm{otherwise}
\end{cases}
 \IEEEyesnumber
\end{IEEEeqnarray}
where $\Lambda_k(i)$ is referred to as \textit{selection metric} and is given by
\begin{IEEEeqnarray}{lll}\label{SelecMet}
    \Lambda_1(i) = (\eta-\mu_1)C_{1r}(i)  - \gamma  P_1(i) \big |_{P_1(i)=P_1^{\mathcal{M}_1}(i)} \IEEEyesnumber\IEEEyessubnumber  \\
    \Lambda_2(i) =  (1-\eta-\mu_2)C_{2r}(i)  - \gamma  P_2(i)\big |_{P_2(i)=P_2^{\mathcal{M}_2}(i)} \IEEEyessubnumber  \\
    \Lambda_3(i) = (\eta-\mu_1)C_{12r}(i)+(1-\eta-\mu_2)C_{21r}(i) - \gamma ( P_1(i)+P_2(i))\Big |_{P_2(i)=P_2^{\mathcal{M}_3}(i)}^{P_1(i)=P_1^{\mathcal{M}_3}(i)}\quad\,\,\IEEEyessubnumber  \\
\Lambda_4(i) = \mu_2C_{r1}(i)  - \gamma  P_r(i) \big |_{P_r(i)=P_r^{\mathcal{M}_4}(i)}\IEEEyessubnumber  \\
\Lambda_5(i) = \mu_1 C_{r2}(i)  - \gamma  P_r(i) \big |_{P_r(i)=P_r^{\mathcal{M}_5}(i)} \IEEEyessubnumber  \\
  \Lambda_6(i) = \mu_1 C_{r2}(i)+\mu_2 C_{r1}(i)  - \gamma  P_r(i)\big |_{P_r(i)=P_r^{\mathcal{M}_6}(i)}\IEEEyessubnumber
\end{IEEEeqnarray}
and $\mathcal{I}_k(i)\in\{0,1\}$ is a binary indicator variable which determines whether $\mathcal{M}_k$ is a \textit{possible candidate} for selection in the $i$-th time slot, i.e., mode $\mathcal{M}_k$ cannot be selected if $\mathcal{I}_k(i)=0$. The optimal value of $\mathcal{I}_k(i)$ is given by
\begin{IEEEeqnarray}{lll}\label{AdapCoin}
   [\mathcal{I}_1(i),\dots,\mathcal{I}_6(i)] = {\begin{cases} 
    [1,\,\,1,\,\,1,\,\,0,\,\,0,\,\,1], & \text{Selection\,\,Region}\,\,\mathcal{S}_0 \\
[1\Minus \mathcal{C}_1(i),\,\,1,\,\,1,\,\,\mathcal{C}_1(i)[1\Minus \mathcal{C}_2(i)],\,\,0,\,\,\mathcal{C}_1(i)\mathcal{C}_2(i)], & \text{Selection\,\,Region}\,\,\mathcal{S}_1 \\
[1,\,\,1\Minus \mathcal{C}_3(i),\,\,1,\,\,0,\,\,\mathcal{C}_4(i)[1\Minus \mathcal{C}_3(i)],\,\,\mathcal{C}_3(i)\mathcal{C}_4(i)], & \text{Selection\,\,Region}\,\,\mathcal{S}_2
\end{cases}}
\end{IEEEeqnarray}
Furthermore, $P_j^{\mathcal{M}_k}(i)$ denotes the optimal transmit power of node $j$ for  transmission mode $\mathcal{M}_k$ in the $i$-th time slot and is given by
\begin{IEEEeqnarray}{lll}\label{OptPower}
   P_1^{\mathcal{M}_1} (i) = \left[\frac{\eta-\mu_1}{\gamma \mathrm{ln}2}-\frac{1}{S_1(i)}\right]^+
\IEEEyesnumber \IEEEyessubnumber \\
P_2^{\mathcal{M}_2} (i) = \left[\frac{1-\eta-\mu_2}{\gamma \mathrm{ln}2}-\frac{1}{S_2(i)}\right]^+
\IEEEyessubnumber \\
P_1^{\mathcal{M}_3} (i) =
{\begin{cases}
\left[\frac{\eta-\mu_1}{\gamma \mathrm{ln}2}-\frac{1-2\eta+\mu_1-\mu_2}{\gamma \mathrm{ln}2}\frac{1}{\frac{S_1(i)}{S_2(i)}-1}\right]^+, \mathrm{if} \,\, \eta-\mu_1 \leq 1-\eta-\mu_2 \\
\left[\frac{1-2\eta+\mu_1-\mu_2}{\gamma \mathrm{ln}2}\frac{1}{\frac{S_1(i)}{S_2(i)}-1} - \frac{1}{S_1(i)}\right]^+, \mathrm{otherwise}
\end{cases}} \IEEEyessubnumber \\ 
P_2^{\mathcal{M}_3} (i) =
{\begin{cases}
\left[\frac{1-2\eta+\mu_1-\mu_2}{\gamma \mathrm{ln}2}\frac{1}{1-\frac{S_2(i)}{S_1(i)}} - \frac{1}{S_2(i)}\right]^+, \mathrm{if} \,\, \eta-\mu_1 \leq 1-\eta-\mu_2 \\
\left[\frac{1-\eta-\mu_2}{\gamma \mathrm{ln}2}-\frac{1-2\eta+\mu_1-\mu_2}{\gamma \mathrm{ln}2}\frac{1}{1-\frac{S_2(i)}{S_1(i)}}\right]^+, \mathrm{otherwise}
\end{cases}}  \IEEEyessubnumber \\
P_r^{\mathcal{M}_4} (i) = \left[\frac{\mu_2}{\gamma \mathrm{ln}2}-\frac{1}{S_1(i)}\right]^+
\IEEEyessubnumber \\
P_r^{\mathcal{M}_5} (i) = \left[\frac{\mu_1}{\gamma \mathrm{ln}2}-\frac{1}{S_2(i)}\right]^+
 \IEEEyessubnumber \\
P_r^{\mathcal{M}_6} (i) = \left [ \frac{-b+\sqrt{b^2-4ac}}{2a} \right]^+ \IEEEyessubnumber
\end{IEEEeqnarray}
where $[x]^+=\max\{x,0\}$, $a=\gamma \mathrm{ln}2\times S_1(i)S_2(i), b= \gamma \mathrm{ln}2 \times (S_1(i)+S_2(i)) -
(\mu_1+\mu_2)S_1(i)S_2(i)$, and $c=\gamma \mathrm{ln}2 - \mu_1 S_2(i) - \mu_2 S_1(i)$. 
The optimal value of $t(i)$ in the multiple-access mode is given by
\begin{IEEEeqnarray}{lll}\label{OptTime}
   t(i) = {\begin{cases} 0, & \eta-\mu_1 \leq 1-\eta-\mu_2 \\
    1, & \mathrm{otherwise}
\end{cases}}
\end{IEEEeqnarray}
Moreover, $\mu_1,\mu_2,\gamma$, and the coin flip probabilities $p_1,p_2,p_3,$ and $p_4$ are long-term variables and depend only on the channel statistics, the total power budget, and $\eta$. The optimal values of the long-term variables are obtained such that constraints $\mathrm{C1}$ and $\mathrm{C2}$ in (\ref{OptProb}), and the long-term power constraint in (\ref{TotalPower}) hold, cf. Proposition \ref{AdaptProtLong}.
\end{theo}

\begin{IEEEproof}
Please refer to Appendix \ref{AppKKTAdapt}.
\end{IEEEproof}

Theorem \ref{AdaptProt} specifies the optimal transmission strategy, i.e., optimal mode selection policy, optimal power allocation policy, and optimal decoding order policy for the multiple-access mode, in each time slot given the instantaneous CSI of the involved links and some constants, i.e., $\mu_1,\mu_2,\gamma$, and the coin flip probabilities $p_1,p_2,p_3,$ and $p_4$. As stated in Theorem \ref{AdaptProt}, the values of the long-term variables are obtained such that constraints $\mathrm{C1}$ and $\mathrm{C2}$ in (\ref{OptProb}), and the long-term power constraint in (\ref{TotalPower}) hold. Unfortunately, calculating the expectations in constraints $\mathrm{C1}$ and $\mathrm{C2}$ in (\ref{OptProb}), and the long-term power constraint in (\ref{TotalPower}) are very complicated and thus, close-form solutions are not available for the long-term variables as  a function of the channel statistics, the total power budget, and $\eta$. In the following proposition, we propose a three-dimensional search to obtain the optimal values of $\mu_1,\mu_2$, and $\gamma$. Moreover, the search is limited to specific intervals  and for each search step, the optimal values of the coin flip probabilities are given.

\begin{Prop}\label{AdaptProtLong}
The optimal values of the selection weights and the power weight used in Theorem \ref{AdaptProt} are obtain by a three-dimensional search over $\mu_1\in[0,\,\eta)$, $\mu_2\in[0,\,1-\eta)$, and $\gamma>0$ such that constraints $\mathrm{C1}$ and $\mathrm{C2}$ in (\ref{OptProb}), and the long-term power constraint in (\ref{TotalPower}) hold. In order to determine the coin flip probabilities used in Theorem \ref{AdaptProt}, three mutually exclusive selection regions are specified based on the values of $\mu_1$ and $\mu_2$. In particular, in the following,  we provide a procedure to obtain the coin flip probabilities for a given pair of $(\mu_1,\mu_2)$ in each search step. We note that for  given channel statistics, total power budget, and $\eta$, only one of the selection regions contains the optimal values of the selection weights. Moreover, in the following, we provide the conditions for the optimality of $(\mu_1,\mu_2)$ in each search step. The search process can be terminated as soon as a valid point is found. 

\noindent
\textbf{Selection Region $\mathcal{S}_0$:} For this selection region,  $\mu_1\neq 0$ and $\mu_2\neq 0$ have to hold. In this case, the optimal protocol does not involve a coin flip. For any given $\mu_1$, $\mu_2$, and $\gamma$, the optimal values of $q_k(i),P_j(i),$ and $t(i)$ are given by (\ref{SelecCrit}), (\ref{OptPower}), and (\ref{OptTime}), respectively, and we are able to calculate the corresponding expectations in (\ref{RatReg-buffer}) and (\ref{PowerIndv}). A set of selection weights $\mu_1$ and $\mu_2$ and power weight $\gamma$  is optimal in this region if constraints $\mathrm{C1}$ and $\mathrm{C2}$ in (\ref{OptProb}) and the long-term power constraint in (\ref{TotalPower}) hold. 

\noindent
\textbf{Selection Region $\mathcal{S}_1$:}  For  this selection region, $\mu_1 = 0$ has to hold. Moreover, the optimal value of $\mu_2$ in this region can only be from interval $(0,\,\eta]$.  Then, the following two cases are possible:

\noindent
\textit{Case 1:} If $\mu_2\neq \eta$ holds, we obtain $p_1=1$. In this case, for any possible value of $\mu_2$, we obtain $\Lambda_4(i)=\Lambda_6(i),\,\,\forall i$, and the selection between modes $\mathcal{M}_4$ and $\mathcal{M}_6$ is determined by a coin flip as shown in (\ref{AdapCoin}) where $p_2=\frac{E\{q_3C_{12r}\}}{E\{qC_{r2}\}}$ with $q(i)=q_4(i)+q_6(i)$.  Specifically, in order to calculate the value of $p_2$, we first calculate the selection metrics in (\ref{SelecMet}), the optimal powers in (\ref{OptPower}), and the optimal value of $t(i)$ in (\ref{OptTime}) for a given $\gamma>0$. We note that although  the values of $q_4(i)$ and $q_6(i)$ cannot be obtained from (\ref{SelecCrit}) since we do not know $p_2$ yet, the value of $q(i)$ can be directly obtained from (\ref{SelecCrit}). Hence, we are able to calculate the expectations required for evaluating $p_2$. A set of selection weights $\mu_1$ and $\mu_2$ and power weight $\gamma$ is optimal in this case if $0\leq p_2 \leq 1$, constraint $\mathrm{C2}$ in (\ref{OptProb}), and the long-term power constraint in (\ref{TotalPower}) hold.

\noindent
\textit{Case 2:} If $\mu_2=\eta$ holds, we obtain $\Lambda_1(i)=\Lambda_4(i)=\Lambda_6(i),\,\,\forall i$ and the  selection between modes $\mathcal{M}_1,\mathcal{M}_4$, and $\mathcal{M}_6$ is determined by two coin flips as shown in (\ref{AdapCoin}) where $p_1=\frac{E\{q_2C_{2r}+q_3C_{21r}\}} {E\left\{qC_{r1}\right\}}$ and $p_2=\frac{(1-p_1)E\{qC_{1r}\}+E\{q_3C_{12r}\}}{p_1 E\{qC_{r2}\}}$ with $q(i)=q_1(i)+q_4(i)+q_6(i)$. The expectations for $p_1$ and $p_2$ can be calculated with a similar procedure as in Case 1. A set of  selection weights $\mu_1$ and $\mu_2$ and power weight $\gamma$ is optimal in this case if $0\leq p_1 \leq 1$, $0\leq p_2 \leq 1$, and the long-term power constraint in (\ref{TotalPower}) hold. 

\noindent
\textbf{Selection Region $\mathcal{S}_2$:}  For  this selection region, $\mu_2 = 0$ has to hold. Moreover, the optimal value of $\mu_1$ in this region can only be in the interval $(0,\,1-\eta]$.  Then, the following two cases are possible:

\noindent
\textit{Case 1:} If $\mu_1\neq 1-\eta$ holds, we obtain $p_3=1$. In this case, for any given value of $\mu_1$, we obtain $\Lambda_5(i)=\Lambda_6(i),\,\,\forall i$, and the selection between modes $\mathcal{M}_5$ and $\mathcal{M}_6$ is determined by a coin flip as shown in (\ref{AdapCoin}) where $p_4=\frac{E\{q_3C_{21r}\}}{E\{qC_{r1}\}}$ with $q(i)=q_5(i)+q_6(i)$.  The expectations for $p_4$ can be calculated with a similar procedure as in Case 1 of selection region $\mathcal{S}_1$. A set of selection weights $\mu_1$ and $\mu_2$ and power weight $\gamma$ is optimal in this case if constraint $\mathrm{C1}$ in (\ref{OptProb}), $0\leq p_4 \leq 1$, and the long-term power constraint in (\ref{TotalPower}) hold. 

\noindent
\textit{Case 2:} If $\mu_1=1-\eta$ holds, we obtain $\Lambda_2(i)=\Lambda_5(i)=\Lambda_6(i),\,\,\forall i$, and the selection between modes $\mathcal{M}_2,\mathcal{M}_5$, and $\mathcal{M}_6$ is determined by two coin flips as shown in (\ref{AdapCoin}) where $p_3=\frac{E\{q_1C_{1r}+q_3C_{12r}\}} {E\left\{qC_{r2}\right\}}$ and $p_4=\frac{(1-p_3)E\{qC_{2r}\}+E\{q_3C_{21r}\}}{p_3 E\{qC_{r1}\}}$ with $q(i)=q_2(i)+q_5(i)+q_6(i)$. The expectations for $p_3$ and $p_4$ can be calculated with a similar procedure as in Case 1 of selection region $\mathcal{S}_1$. A set of selection weights $\mu_1$ and $\mu_2$ and power weight $\gamma$ is optimal in this case if $0\leq p_3\leq 1$, $0\leq p_4 \leq 1$, and the long-term power constraint in (\ref{TotalPower}) hold. 

\end{Prop}

\begin{IEEEproof}
Please refer to Appendix \ref{AppAdapLong}.
\end{IEEEproof}

\begin{remk}
The protocol specified in Theorem \ref{AdaptProt} may involve coin flips to determine the optimal transmission mode. Moreover, different selection polices are required depending on which modes have to be selected via coin flips, see (\ref{AdapCoin}). We note that given the channel statistics, the total power budget, and $\eta$, \textit{only one} of the proposed selection policies is optimal. In particular, the protocol specified in Theorem \ref{AdaptProt} reveals that for each instantaneous channel condition, the selection metrics for the six possible transmission modes have to be calculated and the mode with the largest selection metric is selected as the optimal transmission mode. However, for some channel statistics and some values of $\eta$, the optimal value of one of the selection weights $\mu_1$ and $\mu_2$ is zero which means that some of the selection metrics are identical $\forall i$. For such cases, the optimal selection policy selects one of the candidate modes with identical selection metric via a coin flip.  In order to deal with different cases, we introduced three mutually exclusive selection regions in Proposition \ref{AdaptProtLong}. In selection region $\mathcal{S}_0$, there is no coin flip in the selection policy.  However, roughly speaking, if $\Omega_1\gg\Omega_2$ or $\eta\to 0$, then the selection region that contains the optimal selection weight values  ultimately becomes  $\mathcal{S}_1$ where one or two coin flips determine the selection between modes $\mathcal{M}_1,\mathcal{M}_4,$ and $\mathcal{M}_6$. A similar statement is true for selection region $\mathcal{S}_2$ but now for $\Omega_1\ll\Omega_2$ or $\eta\to 1$. We note that for  given channel statistics and power budget, the valid selection region which contains the optimal selection weights changes from $\mathcal{S}_1$ to $\mathcal{S}_0$ and then to $\mathcal{S}_2$ as the value of $\eta$ changes from $0$ to $1$. 
\end{remk}

\begin{remk}
The mode selection metric $\Lambda_k(i)$ introduced in (\ref{SelecMet}) has two parts. The first part depends on the instantaneous capacity (capacities) of mode $\mathcal{M}_k$, and the second part depends on the allocated power(s). The capacity and the power terms are combined via selection weights $\mu_1$ and/or $\mu_2$ and power weight $\gamma$. Therefore, the selection policy chooses the optimal mode which achieves the best trade-off between the available capacity and the required power in each time slot. We note that the values of $\mu_1$, $\mu_2$, and $\gamma$ and coin flip probabilities $p_1,p_2,p_3,$ and $p_4$ depend only on the long-term statistics of the channels. Hence, the selections weights and the power weight can be obtained offline and used as long as the channel statistics remain unchanged. 
\end{remk}

\begin{remk}\label{M3Power}
The protocol in Theorem \ref{AdaptProt} contains the following two special cases:

\noindent
\textit{One-way relaying:} In this case, the rate of one of the users is zero. For instance, if we assume $\eta \to 1$, we obtain one-way transmission from user 1 to user 2, i.e., $\bar{R}_{21}\to 0$. For this case, the optimal selection policy chooses only modes $\mathcal{M}_1$ and $\mathcal{M}_5$ and simplifies to the solution presented in \cite{NikolaJSAC}. 

\noindent
\textit{Sum rate:} For sum rate maximization, we set $\eta=\frac{1}{2}$. For this case, modes  $\mathcal{M}_4$ and $\mathcal{M}_5$ are not selected in the optimal selection policy.  If the channels are statistically symmetric, i.e., $\Omega_1=\Omega_2$, we obtain $\mu_1=\mu_2$ which leads to zero powers for user 1 or user 2 in mode $\mathcal{M}_3$. Therefore, the optimal selection policy chooses only $\mathcal{M}_1,\mathcal{M}_2,$ and $\mathcal{M}_6$, i.e., the same modes that are used in the TDBC protocol \cite{TDBC}. A similar observation was also made in \cite{Tse} and \cite{TotalPower} for the multiple-access channel. Specifically, if a long-term power constraint is considered, it is optimal to not use the multiple-access mode for sum rate maximization in the multiple-access channel. We note that the sum rate maximization problem was also considered in \cite{GlobeCom}.

\end{remk}

\subsection{Delay-Constrained Transmission}

The proposed protocol introduced in Theorem \ref{AdaptProt} provides a performance bound for the long-term achievable rate region of the considered bidirectional relay channel. In particular, infinite-size buffers at the relay were assumed which is unrealistic in practice. However, our numerical results in Section IV show that with the simple modifications proposed in this subsection, the protocol for delay-unconstrained transmission can be also employed for delay-constrained transmission at the expense of a small performance degradation due to the delay constraint. Specifically, we modify the protocol in Theorem \ref{AdaptProt} for limited-size buffers at the relay and with an appropriate choice for $Q_1^{\max}$ and $Q_2^{\max}$, the average delay can be limited to any desired value.

Let $T_j(i),\,\,j=1,2$, denote the waiting time that a bit transmitted from user $j$ in the $i$-th time slot  stays in buffer $B_j$ before it is transmitted to the respective user. Then, according to Little's Law \cite{Little}, the average waiting time/delay is given by
\begin{IEEEeqnarray}{lll} \label{LittleLaw}  
		E\{T_j\} = \frac{E\{Q_j\}}{E\{R_{jr}\}}, \quad j=1,2.
\end{IEEEeqnarray}
For infinite-size buffers, the maximum average arrival rates, i.e., $E\{R_{jr}\},\,\,j=1,2$, are fixed and given by the protocol in Theorem \ref{AdaptProt}. According to Lemma \ref{Queue}, the queues of the buffers at the relay are at the edge of non-absorption. However, with this condition for the queues, we cannot guarantee that $E\{Q_j\}$ is bounded, and consequently, we cannot guarantee a delay-constrained transmission.  

On the other  hand, for a buffer of size $Q_j^{\max}$, the average size of the queue is bounded, i.e., $E\{Q_j\}< Q_j^{\max}$. However, the protocol in Theorem \ref{AdaptProt} does not consider the effect of the states of the queues for limited-size buffers which leads to a performance degradation. For instance, if 
$Q_1^{\max}-Q_1(i-1)<C_{1r}(i)$ occurs for mode $\mathcal{M}_1$ or if $Q_1^{\max}-Q_1(i-1)<C_{12r}(i)$ occurs for mode $\mathcal{M}_3$, user 1 cannot transmit to the relay with the capacity rates. On the other hand, if 
$Q_1(i-1)<C_{r2}(i)$ occurs for modes $\mathcal{M}_5$ and $\mathcal{M}_6$, the relay cannot transmit to user 2 with the capacity rate. Similar statements are true for buffer $B_2$. Motivated by these observations, we propose the following modified protocol which guarantees a delay-constrained transmission. 

\noindent
\textit{Delay-Constrained Protocol:} For finite-size buffers at the relay, a delay-constrained protocol for the half-duplex bidirectional relay channel with AWGN and block fading under a joint long-term power constraint is obtained by replacing the capacities and optimal powers in Theorem \ref{AdaptProt} by virtual capacities and modified powers, respectively. In particular, the virtual capacities are given by
\begin{IEEEeqnarray}{lll} \label{VirCh}  
	\widehat{C}_{1r}(i) &=\min\{C_{1r}(i),Q_1^{\max}-Q_1(i-1)\} \IEEEyesnumber \IEEEyessubnumber \\
	\widehat{C}_{2r}(i) &=\min\{C_{2r}(i),Q_2^{\max}-Q_2(i-1)\} \IEEEyessubnumber \\
	\widehat{C}_{12r}(i) &=\min\{C_{12r}(i),Q_1^{\max}-Q_1(i-1)\} \IEEEyessubnumber \\
	\widehat{C}_{21r}(i) &=\min\{C_{21r}(i),Q_2^{\max}-Q_2(i-1)\} \IEEEyessubnumber \\
	\widehat{C}_{r1}(i) &=\min\{C_{r1}(i),Q_2(i-1)\} \IEEEyessubnumber \\
	\widehat{C}_{r2}(i) &=\min\{C_{r2}(i),Q_1(i-1)\}. \IEEEyessubnumber
\end{IEEEeqnarray}
Furthermore, the modified powers are given by
\begin{IEEEeqnarray}{lll} \label{ModPower}  
	\widehat{P}_1^{\mathcal{M}_1}(i) = 
	\frac{1}{S_1(i)}\left( 2^{\widehat{C}_{1r}(i)}-1 \right) \IEEEyesnumber \IEEEyessubnumber \\
	\widehat{P}_2^{\mathcal{M}_2}(i) = \frac{1}{S_2(i)}\left( 2^{\widehat{C}_{2r}(i)}-1 \right) \IEEEyessubnumber \\
		\widehat{P}_1^{\mathcal{M}_3}(i) = \begin{cases}
	 \frac{2^{\widehat{C}_{21r}(i)}}{S_1(i)}\left( 2^{\widehat{C}_{12r}(i)}-1 \right), & \mathrm{if} \,\, t(i)=0 \\
	 \widehat{P}_1^{\mathcal{M}_1}(i),   & \mathrm{if} \,\, t(i)=1
	 \end{cases} \IEEEyessubnumber \\
	 \widehat{P}_2^{\mathcal{M}_3}(i) = \begin{cases}
	 \frac{2^{\widehat{C}_{12r}(i)}}{S_2(i)}\left( 2^{\widehat{C}_{21r}(i)}-1 \right), & \mathrm{if} \,\, t(i)=1 \\
	 \widehat{P}_2^{\mathcal{M}_2}(i),   & \mathrm{if} \,\, t(i)=0
	 \end{cases} \IEEEyessubnumber \\
	\widehat{P}_r^{\mathcal{M}_4}(i) = \frac{1}{S_1(i)}\left( 2^{\widehat{C}_{r1}(i)}-1 \right) \IEEEyessubnumber \\
	\widehat{P}_r^{\mathcal{M}_5}(i) = \frac{1}{S_2(i)}\left( 2^{\widehat{C}_{r2}(i)}-1 \right) \IEEEyessubnumber \\
	\widehat{P}_r^{\mathcal{M}_6}(i) = \max\left\{ \frac{1}{S_1(i)}\left( 2^{\widehat{C}_{r1}(i)}-1 \right), \frac{1}{S_2(i)}\left( 2^{\widehat{C}_{r2}(i)}-1 \right)\right\}. \IEEEyessubnumber 
\end{IEEEeqnarray}

Moreover, the long-term variables used in Theorem \ref{AdaptProt}, i.e., the selection weights $\mu_1$ and $\mu_2$, the power weight $\gamma$, and the coin flip probabilities $p_1,p_2,p_3$, and $p_4$, are obtained from Proposition \ref{AdaptProtLong}.

\begin{remk}
It is worth mentioning that the values of the virtual capacities are the maximum transmission rates given in Section II-B. The virtual capacities take into account both the state of the queues and the channel conditions. For instance, if buffer $B_1$ is empty, we obtain  $\widehat{C}_{r2}(i)=0$ which leads to the exclusion of mode $\mathcal{M}_5$ from the candidates for mode selection. Similarly, if buffer $B_1$ is full, we obtain $\widehat{C}_{1r}(i)=0$ which leads to the exclusion of mode $\mathcal{M}_1$ from the candidates for mode selection. 
\end{remk}

\begin{remk}
As stated before, by appropriately choosing $Q_1^{\max}$ and $Q_2^{\max}$, the average delay of the transmissions from user 1-to-user 2 and user 2-to-user 1 can be limited to any desired value larger than one time slot, respectively. As an example, for $Q_1^{\max} \approx E\{C_{1r}\}\big |_{P_1(i)=P_1^{\mathcal{M}_1}(i)}$ and $Q_2^{\max} \approx  E\{C_{2r}\}\big |_{P_2(i)=P_2^{\mathcal{M}_2}(i)}$, the proposed protocol leads to an average delay of more than one and less than two time slots, i.e., $1< E\{T_j\}< 2,\,\,j=1,2$. 
Roughly speaking, since buffers $B_1$ and $B_2$ are nearly full after one or two consecutive data transmissions from users 1 and 2, respectively, the virtual capacities in (\ref{VirCh}a)-(\ref{VirCh}d) become very small, and thus, the proposed delay-constrained protocol selects the relay to transmit to the users. Hence, the transmitted information from the users stays on average less than two time slots in the relays' buffers. We note that for such a small delay, the performance gain obtained by the proposed protocol  compared to the conventional protocols is very small. However, if larger average delays are permitted, we can allow larger values of $Q_1^{\max}$ and $Q_2^{\max}$, which leads to a higher performance gain compared to the conventional protocols.  The numerical results is Section V reveal that  even for an average delay of five time slots, the proposed protocol achieves a considerable performance gain compared to the conventional protocols.
\end{remk}

\begin{remk}
Although the above modifications are  heuristic, the numerical results in Section \ref{Numerical} confirm their effectiveness. Nevertheless, how to optimally include the delay constraint in the problem formulation is an interesting problem for future work.
\end{remk}

\section{Achievable Rate Region under Fixed Transmit Power Constraint}

In this section, we consider a fixed transmit power constraint for each node. Thus, there is no degree of freedom in the node powers and they are predefined and given, i.e., $ \mathcal{P}=\big\{(P_1(i),P_2(i),P_r(i))|P_j(i)=P_j,\,\,\forall i,j\big\}$. Similar to the previous section, we first develop a protocol for delay-unconstrained transmission. Then, by employing similar modifications as in Section III-B, we adapt the protocol for delay-unconstrained transmission to the case of delay-constrained transmission.

\subsection{Delay-Unconstrained Transmission}

In this subsection, we provide the solution to the optimization problem for delay-unconstrained transmission in (\ref{OptProb}) under a fixed transmit power constraint for each node. To this end, we define again three mutually exclusive selection regions $\mathcal{S}_0$, $\mathcal{S}_1$, and $\mathcal{S}_2$,  and each region  requires  a different optimal selection policy. For  given channel statistics, node powers $P_1$, $P_2$, $P_r$, and $\eta$,  only one of the selection regions contains the optimal values of the selection weights.

\begin{theo}\label{FixProt}
\normalfont Assuming $N\to \infty$, the boundary surface of the long-term achievable rate region of the considered half-duplex bidirectional relay channel with AWGN and block fading under a fixed transmit power constraint for each node is the closure of points $(\bar{R}_{12},\bar{R}_{21})$ obtained for all $\eta\in(0,1)$ with the mode selection and decoding order policies  provided in the following. In particular, the optimal mode selection policy is 
\begin{IEEEeqnarray}{lll}\label{FixSelectTheo}
q_{k^*}(i)=
   \begin{cases}
     1, &k^*= {\underset{k=1,\dots,6}{\arg \, \max}} \{\mathcal{I}_k(i)\Lambda_k(i)\} \\
     0, &\mathrm{otherwise}
\end{cases}
 \IEEEyesnumber
\end{IEEEeqnarray}
where the selection metric $\Lambda_k(i)$ is  given by
\begin{IEEEeqnarray}{lll}\label{SelecMetFix}
    \Lambda_1(i) = (\eta-\mu_1)C_{1r}(i)  \IEEEyesnumber\IEEEyessubnumber  \\
    \Lambda_2(i) =  (1-\eta-\mu_2)C_{2r}(i) \IEEEyessubnumber  \\
    \Lambda_3(i) = (\eta-\mu_1)C_{12r}(i)+(1-\eta-\mu_2)C_{21r}(i)\IEEEyessubnumber \\
    \Lambda_4(i) = \mu_2 C_{r1}(i) \IEEEyessubnumber\\
    \Lambda_5(i) = \mu_1 C_{r2}(i)\IEEEyessubnumber\\
    \Lambda_6(i) = \mu_1 C_{r2}(i)+\mu_2 C_{r1}(i)\IEEEyessubnumber
\end{IEEEeqnarray}
and the binary variable $\mathcal{I}_k(i)$ is given by
\begin{IEEEeqnarray}{lll}\label{FixCoin}
     [\mathcal{I}_1(i),\dots,\mathcal{I}_6(i)]  \nonumber \\ =  \big[[1\Minus \mathcal{C}_1(i)]a(i),\,\,[1\Minus \mathcal{C}_2(i)]a(i),\,\,\mathcal{C}_1(i)\mathcal{C}_2(i)a(i),\,\,[1\Minus \mathcal{C}_3(i)]b(i),\,\,[1\Minus \mathcal{C}_4(i)]b(i),\,\,\mathcal{C}_3(i)\mathcal{C}_4(i)b(i)\big] \quad\,\,
\end{IEEEeqnarray}
where $a(i)=\Equal \mathcal{C}_5(i)$ and $b(i)\Equal 1\Minus \mathcal{C}_5(i)$ if $\{\text{Selection Region}\,\, \mathcal{S}_1, \,\,\text{Case 2 c)} \,\,\wedge\,\,P_2\Equal P_r\,\,\wedge\,\,\eta\Equal \frac{1}{2}\}\,\vee\,\{\text{Selection Region}\,\, \mathcal{S}_2, \,\,\text{Case 2 c)} \,\,\wedge\,\,P_1\Equal P_r\,\,\wedge\,\,\eta\Equal \frac{1}{2}\}$ holds, otherwise, $a(i)=1$ and $b(i)=1,\,\,\forall i$. 

The optimal value of the decoding order is obtained probabilistically as follows 
\begin{IEEEeqnarray}{lll}\label{FixOptTime}
   t(i)=\mathcal{C}_6(i).  
\end{IEEEeqnarray}

Moreover, $\mu_1,\mu_2$, and the coin flip probabilities $p_1,\dots,p_6$ are long-term variables and depend only on the channel statistics, node powers $P_1,P_2$, and $P_r$, and $\eta$. The optimal values of the long-term variables are obtained such that constraints $\mathrm{C1}$ and $\mathrm{C2}$ in (\ref{OptProb}) hold, cf. Proposition \ref{FixProtLong}.

\end{theo}
\begin{IEEEproof}
Please refer to Appendix \ref{AppKKTFix}. 
\end{IEEEproof}

Similar to Theorem \ref{AdaptProt} for the joint long-term power constraint, Theorem \ref{FixProt} specifies the optimal transmission strategy, i.e., the optimal mode selection policy and the optimal decoding order policy for the multiple-access mode, in each time slot given the instantaneous CSI of the involved links and some constants, i.e., $\mu_1,\mu_2$, and the coin flip probabilities $p_1,\dots,p_6$. As stated in Theorem \ref{FixProt}, the values of the long-term variables are obtained such that constraints $\mathrm{C1}$ and $\mathrm{C2}$ in (\ref{OptProb}) hold. However, a close-form solution is not available as a function of the channel statistics, the powers of the nodes, and $\eta$, since  calculating the expectations in constraints $\mathrm{C1}$ and $\mathrm{C2}$ in (\ref{OptProb}) are very complicated. In the following proposition, we propose a two-dimensional search to obtain the optimal values of $\mu_1$ and $\mu_2$. Moreover, the search is limited to specific intervals and for each search step, the optimal values of the coin flip probabilities are given. 

\begin{table}
\label{LongVar}
\caption{The Values of the Coin Flip Probabilities for all Possible Values of the Selection Weights in Theorem \ref{FixProt}.}
\begin{center}
\begin{tabular}{|| c | c | c | c | c | c | c | c | c | c | c ||}
  \hline                  
\multicolumn{3}{||c|}{} &$\mu_1$ & $\mu_2$ & $p_1$ & $p_2$ &$p_3$ &$p_4$ &$p_5$ &$p_6$  \\ \hline  
    \multirow{3}{*}{Selection Region $\mathcal{S}_0$} &\textit{Case 1}& $-$ & $\mu_1\neq 0,\eta$  & $\mu_2\neq 0,1-\eta$ & $1$ &$1$&$1$&$1$&$-$&$p_6$\\  \cline{2-11}
   & \multirow{2}{*}{\textit{Case 2}}&$a$& $0$ & $1-2\eta$ & $1$ & $1$ & $p_3$ & $1$& $-$& $p_6$\\    \cline{3-11}
   &&$b$& $2\eta-1$ & $0$ & $1$ & $1$ & $1$ & $p_4$& $-$& $p_6$ \\ \hline 
     \multirow{6}{*}{Selection Region $\mathcal{S}_1$} 
     &\textit{Case 1}& $-$ & $\mu_1\neq 0,\eta$  & $\mu_2\neq 0,1-\eta$ & $1$ &$1$&$1$&$1$&$-$&$0$ \\ \cline{2-11}
  &\multirow{4}{*}{\textit{Case 2}}&$a$& $\eta$ & $\mu_2$ & $1$ & $p_2$ & $1$ & $1$& $-$& $0$ \\ \cline{3-11}
&&$b$& $\mu_1$ & $0$ & $1$ & $1$ & $1$ & $p_4$& $-$& $0$ \\ \cline{3-11}
&&\multirow{2}{*}{$c$}& \multirow{2}{*}{$\eta$} & \multirow{2}{*}{$0$} & $1$ & $p_2$ & $1$ & $p_4$& $-$& $0$ \\ \cline{6-11}
&&&&& $1$ & $p_2$ & $1$ & $p_4$& $p_5$& $0$ \\ \cline{3-11}
   && $d$ & $0$  & $\mu_2$ & $1$ &$1$&$p_3$&$1$&$-$&$0$    
\\ \hline   
   \multirow{5}{*}{Selection Region $\mathcal{S}_2$} 
   &\textit{Case 1}& $-$ & $\mu_1\neq 0,\eta$  & $\mu_2\neq 0,1-\eta$  & $1$ &$1$&$1$&$1$&$-$&$1$ \\ \cline{2-11}
   &\multirow{4}{*}{\textit{Case 2}} & $a$& $\mu_1$ & $1-\eta$ & $p_1$ & $1$ & $1$ & $1$& $-$& $1$ \\ \cline{3-11}
&& $b$& $0$ & $\mu_2$ & $1$ & $1$ & $p_3$ & $1$& $-$& $1$ \\ \cline{3-11}
&&\multirow{2}{*}{$c$}& \multirow{2}{*}{$0$} & \multirow{2}{*}{$1-\eta$} & $p_1$ & $1$ & $p_3$ & $1$& $-$& $1$ \\ \cline{6-11}
&&&&& $p_1$ & $1$ & $p_3$ & $1$& $p_5$& $1$ \\ \cline{3-11}
   && $d$ & $\mu_1$  & $0$ & $1$ &$1$&$1$&$p_4$&$-$&$1$  \\ \hline

\end{tabular}
\end{center}
\end{table}

\vspace{0.1cm}

\begin{Prop}\label{FixProtLong}
The optimal values of the selection weights used in Theorem \ref{FixProt} are obtained by a two-dimensional search over $\mu_1\in[0,\,\,\eta]$ and $\mu_2\in[0,\,\,1-\eta]$ such that constraints $\mathrm{C1}$ and $\mathrm{C2}$ in (\ref{OptProb}) hold. In order to determine the coin flip probabilities used in Theorem \ref{FixProt}, three mutually exclusive selection regions are specified based on the values of $\mu_1$ and $\mu_2$. Moreover, each selection region is divided into different cases to facilitate the derivation of the coin flip probabilities. In particular, Table I specifies all possible combinations of $\mu_1$ and $\mu_2$ that require different policies to determine the optimal values of the coin flip probabilities. We note that for any given channel statistic, node powers $P_1,P_2$, and $P_r$, and $\eta$, only one of the cases in Table I contains the optimal values of the selection weights. Moreover, in the following, we provide the conditions for the optimality of $(\mu_1,\mu_2)$ and the coin flip probabilities in each search step. Thus, the search process can be terminated as soon as a valid point is found. 
\vspace{0.1cm}

\noindent
\textbf{Selection Region $\mathcal{S}_0$:} For this selection region, $\eta-\mu_1=1-\eta-\mu_2$ has to hold. Then, the following two cases are possible.

\noindent
\textit{Case 1:} If $\mu_1\neq 0,\eta$ and $\mu_2\neq 0,1-\eta$ hold, we set $p_6=\frac{E\{q_3(C_r-C_{2r})\}-E\{q_6C_{r2}\}}{E\left\{q_3 (C_r-C_{1r}-C_{2r})\right\}}$ and check whether sum rate constraint $\bar{R}_{1r}+\bar{R}_{2r}=\bar{R}_{r1}+\bar{R}_{r2}$ holds. We note that since $\Lambda_3(i)=(\eta-\mu_1)C_r(i)$ and $R_{1r}(i)+R_{2r}(i)=C_r(i)$ hold in this region, the values of and $q_3(i)$ and $q_6(i)$ can be obtained from (\ref{FixSelectTheo}) and the sum rate constraint can be checked without any knowledge of $p_6$.  A set of selection weights $\mu_1$ and $\mu_2$  is optimal in this case if the sum rate constraint and $0\leq p_6\leq 1$ hold. 

\noindent
\textit{Case 2:} For this case, $\mu_1= 0,\eta$ or $\mu_2=0,1-\eta$ have to hold. We note that selection pairs $(\mu_1,\mu_2)=(0,0),(\eta,1-\eta)$ cannot be optimal. Therefore, only the following two cases are possible in this case.

\textit{a)}
If $\mu_1=0$ holds, we obtain $\Lambda_4(i)=\Lambda_6(i),\,\,\forall i$, and  the selection between modes $\mathcal{M}_4$ and $\mathcal{M}_6$ is determined by a coin flip as shown in (\ref{FixCoin}) where $p_3=\frac{E\{q_3C_{12r}\}}{E\{qC_{r2}\}}$  with $q(i)=q_4(i)+q_6(i)$. The coin flip probability for the decoding order policy is given by $p_6 = \frac{E\{qC_{r1}\} - E\{q_3 C_{2r}\}}{E\{q_3(C_r-C_{1r}-C_{2r})\}}$. In particular, in order to calculate $p_3$ and $p_6$, we first calculate the selection metrics in (\ref{SelecMetFix}). We note that although the values of $q_4(i)$ and $q_6(i)$ cannot be obtained from (\ref{FixSelectTheo}) yet, the value of $q(i)$ can be directly obtained from (\ref{FixSelectTheo}). Hence, we are able to calculate the expectations for $p_3$ and $p_6$. A set of selection weights $\mu_1$ and $\mu_2$  is optimal in this case if constraint $\mathrm{C2}$ in (\ref{OptProb}), $0\leq p_3\leq 1$, and $0\leq p_6\leq 1$ hold. 

\textit{b)}
If $\mu_2=0$ holds, we obtain $\Lambda_5(i)=\Lambda_6(i),\,\,\forall i$, and the selection between modes $\mathcal{M}_5$ and $\mathcal{M}_6$ is determined by a coin flip as shown in (\ref{FixCoin}) where $p_4=\frac{E\{q_3C_{21r}\}}{E\{qC_{r1}\}}$  with $q(i)=q_5(i)+q_6(i)$. The coin flip probability for the decoding order policy is given by $p_6=\frac{E\{q_3(C_r-C_{2r})\}-E\{qC_{r2}\}}{E\left\{q_3 (C_r-C_{1r}-C_{2r})\right\}}$. The expectations for $p_4$ and $p_6$ can be calculated with a similar procedure as in Case 2 a). A set of selection weights $\mu_1$ and $\mu_2$ is optimal in this case if constraint $\mathrm{C1}$ in (\ref{OptProb}), $0\leq p_4\leq 1$, and $0\leq p_6\leq 1$ hold.

\vspace{0.1cm}

\noindent
\textbf{Selection Region $\mathcal{S}_1$:}
For this selection region, $\eta-\mu_1<1-\eta-\mu_2$ has to hold. Then, the following two cases are possible.

\noindent
\textit{Case 1:}  For this case, $\mu_1\neq 0,\eta$ and $\mu_2\neq 0,1-\eta$ have to hold. A set of selection weights $\mu_1$ and $\mu_2$ is optimal in this case if constraints $\mathrm{C1}$ and $\mathrm{C2}$ in (\ref{OptProb}) hold.

\noindent
\textit{Case 2:}  For this case, $\mu_1 = 0,\eta$ or $\mu_2 = 0,1-\eta$ have to hold. Among the possible combinations of selection weights for this case, only the following four cases can be optimal.

\textit{a)} If $\mu_1=\eta$ and $\mu_2\neq 0,1-\eta$ hold, we obtain $\Lambda_2(i)=\Lambda_3(i),\,\,\forall i$, and the selection between modes $\mathcal{M}_2$ and $\mathcal{M}_3$ is determined by a coin flip as shown in (\ref{FixCoin}) where $p_2=\frac{E\{q_6C_{r2}\}}{E\left\{q[C_r-C_{2r}]\right\}}$ with $q(i)=q_2(i)+q_3(i)$. The expectations for $p_2$  can be calculated with a similar procedure as in selection region $\mathcal{S}_1$, Case 2 a). A  set of selection weights $\mu_1$ and $\mu_2$ is optimal in this case if constraint $\mathrm{C2}$ in (\ref{OptProb}) and $0\leq p_2\leq 1$ hold. 

\textit{b)} If $\mu_1\neq 0,\eta$ and $\mu_2=0$ hold, we obtain $\Lambda_5(i)=\Lambda_6(i),\,\,\forall i$, and the selection between modes $\mathcal{M}_5$ and $\mathcal{M}_6$ is determined by a coin flip as shown in (\ref{FixCoin}) where $p_4=\frac{E\{q_3C_{2r}\}}{E\left\{q C_{r1}\right\}}$ with $q(i)=q_5(i)+q_6(i)$. The expectations for $p_4$  can be calculated with a similar procedure as in selection region $\mathcal{S}_1$, Case 2 a). A set of selection weights $\mu_1$ and $\mu_2$ is optimal in this case  if constraint $\mathrm{C1}$ in (\ref{OptProb}) and $0\leq p_4\leq 1$ hold.

\textit{c)} For this case, $\mu_1 = \eta$ and $\mu_2=0$ have to hold.  Then, if $\eta\neq\frac{1}{2}$ or $P_2\neq P_r$ hold, we obtain  $\Lambda_2(i)=\Lambda_3(i),\,\,\forall i$ and $\Lambda_5(i)=\Lambda_6(i),\,\,\forall i$, and the selection between modes $\mathcal{M}_2$ and $\mathcal{M}_3$ and selection between modes $\mathcal{M}_5$ and $\mathcal{M}_6$ is determined by two coin flips as shown in (\ref{FixCoin}) where $p_2=\frac{E\{(1-q)C_{r2}\}}{E\{q(C_r-C_{2r})\}}$ and $p_4=\frac{E\{qC_{2r}\}}{E\{(1-q)C_{r1}\}}$ with $q(i)=q_2(i)+q_3(i)$.  The expectations for $p_2$ and $p_4$ can be calculated with a similar procedure as in selection region $\mathcal{S}_1$, Case 2 a). On the other hand, if $\eta=\frac{1}{2}$ and $P_2=P_r$ hold, we obtain  $\Lambda_2(i)=\Lambda_3(i)=\Lambda_5(i)=\Lambda_6(i),\,\,\forall i$, and  the selection between modes $\mathcal{M}_2,\mathcal{M}_3,\mathcal{M}_5$, and $\mathcal{M}_6$ is determined by three coin flips as shown in (\ref{FixCoin}) where $p_2=\frac{1-p_5}{p_5} \frac{\omega_l}{1-\omega_l}$ and $p_4=\frac{p_5}{1-p_5} \frac{1-\omega_u}{\omega_u}$, and $\omega_l\leq p_5\leq \omega_u$ with $\omega_l= \frac{E\{C_{r2}\}}{E\{C_{r}\}}$ and $\omega_u=\frac{E\{C_{r1}\}}{E\{C_{r1}+C_{2r}\}}$. The considered selection weights $\mu_1$ and $\mu_2$  are optimal if constraint $0\leq p_2\leq 1$, $0\leq p_4\leq 1$, and/or $0\leq p_5\leq 1$ hold.

\textit{d)} If $\mu_1 = 0$ and $\mu_2\neq 0,1-\eta$ hold, we obtain $\Lambda_4(i)=\Lambda_6(i),\,\,\forall i$, and the selection between modes $\mathcal{M}_4$ and $\mathcal{M}_6$ is determined by a coin flip as shown in (\ref{FixCoin}) where $p_3=\frac{E\{q_3C_{1r}\}}{E\left\{q C_{r2}\right\}}$ with $q(i)=q_4(i)+q_6(i)$. The expectations for $p_3$  can be calculated with a similar procedure as in selection region $\mathcal{S}_1$, Case 2 a). A set of selection weights $\mu_1$ and $\mu_2$ is optimal if constraint $\mathrm{C2}$ in (\ref{OptProb}) and $0\leq p_3\leq 1$ hold.

\vspace{0.1cm}

\noindent
\textbf{Selection Region $\mathcal{S}_2$:}
For this selection region, $\eta-\mu_1>1-\eta-\mu_2$ has to hold. Then, the following two cases are possible.

\noindent
\textit{Case 1:}  For this case, $\mu_1\neq 0,\eta$ and $\mu_2\neq 0,1-\eta$ have to hold. A set of selection weights $\mu_1$ and $\mu_2$ is optimal in this case if constraints $\mathrm{C1}$ and $\mathrm{C2}$ in (\ref{OptProb}) hold.

\noindent
\textit{Case 2:}  For this case, $\mu_1 = 0,\eta$ or $\mu_2 = 0,1-\eta$ have to hold. Among the possible combinations of selection weights for this case, only the following four cases can be optimal.

\textit{a)} If $\mu_1\neq 0,\eta$ and $\mu_2 = 1-\eta$ hold we obtain $\Lambda_1(i)=\Lambda_3(i),\,\,\forall i$, and the result of mode selection between modes $\mathcal{M}_1$ and $\mathcal{M}_3$ is determined by a coin flip as shown in (\ref{FixCoin}) where $p_1=\frac{E\{q_6C_{r1}\}}{E\left\{q [C_r-C_{1r}]\right\}}$ with $q(i)=q_1(i)+q_3(i)$. The expectations for $p_1$  can be calculated with a similar procedure as in selection region $\mathcal{S}_1$, Case 2, a). A set of selection weights $\mu_1$ and $\mu_2$ is optimal in this case if constraint $\mathrm{C1}$ in (\ref{OptProb}) and $0\leq p_1\leq 1$ hold. 

\textit{b)} If $\mu_1=0$ and $\mu_2 \neq 0, 1-\eta$ hold, we obtain $\Lambda_4(i)=\Lambda_6(i),\,\,\forall i$, and the selection between modes $\mathcal{M}_4$ and $\mathcal{M}_6$ is determined by a coin flip as shown in (\ref{FixCoin}) where  $p_3=\frac{E\{q_3C_{1r}\}}{E\left\{qC_{r2}\right\}}$ with $q(i)=q_4(i)+q_6(i)$. The expectations for $p_3$  can be calculated with a similar procedure as in selection region $\mathcal{S}_1$, Case 2 a). A set of selection weights $\mu_1$ and $\mu_2$ is optimal in this case if constraint $\mathrm{C2}$ in (\ref{OptProb}) and $0\leq p_3\leq 1$ hold.

\textit{c)} For this case, $\mu_1 = 0$ and $\mu_2=1-\eta$ have to hold.  Then, if $\eta\neq\frac{1}{2}$ or $P_1\neq P_r$ hold, we obtain  $\Lambda_1(i)=\Lambda_3(i),\,\,\forall i$ and $\Lambda_4(i)=\Lambda_6(i),\,\,\forall i$, and selection between modes $\mathcal{M}_1$ and $\mathcal{M}_3$ and the mode selection between modes $\mathcal{M}_4$ and $\mathcal{M}_6$ is determined by two coin flips as shown in (\ref{FixCoin}) where $p_1=\frac{E\{(1-q)C_{r1}\}}{E\{q(C_r-C_{1r})\}}$ and $p_3=\frac{E\{qC_{1r}\}}{E\{(1-q)C_{r2}\}}$ with $q(i)=q_1(i)+q_3(i)$.  The expectations for $p_1$ and $p_3$ can be calculated with a similar procedure as in selection region $\mathcal{S}_1$, Case 2 a). On the other hand, if $\eta=\frac{1}{2}$ and $P_1=P_r$ hold, we obtain  $\Lambda_1(i)=\Lambda_3(i)=\Lambda_4(i)=\Lambda_6(i),\,\,\forall i$, and the selection between modes $\mathcal{M}_1,\mathcal{M}_3,\mathcal{M}_4$, and $\mathcal{M}_6$ is determined by three coin flips as shown in (\ref{FixCoin}) where $p_1=\frac{1-p_5}{p_5} \frac{\omega_l}{1-\omega_l}$ and $p_3=\frac{p_5}{1-p_5} \frac{1-\omega_u}{\omega_u}$, and $\omega_l\leq p_5\leq \omega_u$ with $\omega_l= \frac{E\{C_{r1}\}}{E\{C_{r}\}}$ and $\omega_u=\frac{E\{C_{r2}\}}{E\{C_{r2}+C_{1r}\}}$. The considered selection weights $\mu_1$ and $\mu_2$ are optimal if constraint $0\leq p_1\leq 1$, $0\leq p_3\leq 1$, and/or $0\leq p_5\leq 1$ hold. 

\textit{d)} If $\mu_1 \neq 0,\eta$ and $\mu_2 = 0$ hold, we obtain $\Lambda_5(i)=\Lambda_6(i),\,\,\forall i$, and the selection between modes $\mathcal{M}_5$ and $\mathcal{M}_6$ is determined by a coin flip as shown in (\ref{FixCoin}) where $p_4=\frac{E\{q_3C_{2r}\}}{E\left\{q C_{r1}\right\}}$ with $q(i)=q_5(i)+q_6(i)$. The expectations for $p_3$  can be calculated with a similar procedure as in selection region $\mathcal{S}_1$, Case 2 a). A set of selection weights $\mu_1$ and $\mu_2$ is optimal in this case if constraint $\mathrm{C1}$ in (\ref{OptProb}) and $0\leq p_4\leq 1$ hold.

\vspace{0.1cm}

\end{Prop}
\begin{IEEEproof}
Please refer to Appendix \ref{AppKKTFix}. 
\end{IEEEproof}

Unlike Theorem \ref{AdaptProt}, the mode selection metrics in Theorem \ref{FixProt} are only comprised of capacity terms and there is no power penalty. How much the capacities of the individual links contribute to the optimal selection policy is specified by long-term selection weights $\mu_1$ and $\mu_2$ where the optimal values of the selection weights are given in Proposition \ref{FixProtLong}. Moreover, since the long-term variables, i.e., the selection weights and coin flip probabilities, depend only on the statistics of the channels, the powers of the nodes, and $\eta$, they can be obtained offline and used as long as the channel statistics remain unchanged.

\begin{remk}
The protocol specified in Theorem \ref{FixProt} may involve coin flips to determine the optimal values of $q_k(i)$ and $t(i)$. Moreover, different selection polices are required depending on whether $t(i)$ has to be selected via a coin flip or which modes have to be selected via coin flips, see (\ref{FixCoin}), (\ref{FixOptTime}), and Table I. We note that given the channel statistics, the powers of the nodes, and $\eta$, \textit{only one} of the proposed selection policies is optimal. In particular, the three different selection regions  in Theorem \ref{FixProt} are defined by whether the value of $t(i)$ is determined by a coin flip in each time slot or is fixed to zero/one for all time slots. Specifically, in selection region $\mathcal{S}_0$, $\eta-\mu_1=1-\eta-\mu_2$ has to hold which leads to a probabilistic choice of $t(i)=X_6(i)\in\{0,1\},\,\,\forall i$. However, in selection region $\mathcal{S}_1$, $\eta-\mu_1<1-\eta-\mu_2$ has to hold which leads to  $t(i)=0,\,\,\forall i$, cf. Table I. Similarly, in selection region $\mathcal{S}_2$, $\eta-\mu_1>1-\eta-\mu_2$ has to hold which leads to $t(i)=1,\,\,\forall i$, cf. Table I. We have further divided each selection region into two cases such that in Case 1, none of the point-to-point modes is selected whereas in Case 2, the optimal mode selection policy uses coin flip(s) to select the point-to-point modes.
\end{remk}

\begin{remk}
We note that in the proposed optimal mode selection policy, regardless of the  channel statistics, the value of the powers of the nodes, and the value of $\eta\in(0,\,\,1)$, the multiple-access and broadcast modes are always selected with non-zero probability. The reason for this is that assuming fixed powers for all nodes, the multiple-access and broadcast modes are more rate-efficient compared to the point-to-point modes. Only in the extreme case, when the average SNR of one of the links is much larger than the average SNR of the other link or when $\eta$ gives significantly more priority to one direction of information flow, some of the point-to-point modes are also selected in the optimal policy to satisfy  constraints $\mathrm{C1}$ and $\mathrm{C2}$ in  (\ref{OptProb}).  However, for the joint long-term power constraint for all nodes assumed in Theorem \ref{AdaptProt}, the point-to-point modes from the users to the relay are more rate/power-efficient compared to the multiple-access mode for most of the channel realizations. While the broadcast mode is still more power/rate-efficient compared to the point-to-point modes from the relay to the users. As a conclusion, excluding the broadcast mode leads to a considerable loss in the achievable rates for both adopted power constraints, while excluding the multiple-access mode leads to a considerable loss in the achievable rates only for fixed per-user power constraint, and does not severely reduce the achievable rates for a joint long-term power constraint for all nodes.
\end{remk}

\begin{remk}
The protocol in Theorem \ref{FixProt} contains the following two special cases:

\noindent
\textit{One-way relaying:} Unlike for the joint power constraint case, with the fixed power assumption, we can achieve the maximum rate of one-way relaying in one direction, and provide a non-zero rate in the other direction. For instance, if we assume $\eta \to 1$, the long-term variables are set based on Case 2 of selection region $\mathcal{S}_2$. Therefore, $\bar{R}_{12}$ is identical to the maximum rate obtained in \cite{NikolaJSAC}, but still $\bar{R}_{21}>0$ holds. The main reason for this difference is that with a joint power constraint, we have to reduce the power used in one direction to increase the rate in the other direction while with individual fixed power for each node, the rate in one direction can still be positive although the rate in the other direction assumes the maximum value.
\end{remk}

\noindent
\textit{Sum rate:} For sum rate maximization, we have to set $\eta = \frac{1}{2}$. For identical powers and statistically symmetric channels, i.e., $P_1=P_2=P_r$ and $\Omega_1=\Omega_2$, the optimal selection region is Case 1 of selection region $\mathcal{S}_0$. Thereby, only modes $\mathcal{M}_3$ and $\mathcal{M}_6$ are selected. This solution was also obtained in \cite{EUSIPCO}. In contrast, for a joint long-term power constraint for  sum rate maximization and statistically symmetric channels, only modes $\mathcal{M}_1,\mathcal{M}_2,$ and $\mathcal{M}_6$ are selected. 

We  conclude  that the two considered power constraints lead to two fundamentally different protocols for the bidirectional relay channel. As discussed before, this fact is the main motivation for investigating optimal protocols for both power constraints in this paper.

\subsection{Delay-Constrained Transmission}

The protocol for delay-constrained transmission is obtained similarly as in Section III-B. In particular, we employ the  virtual channel capacities  in  Theorem \ref{FixProt} instead of actual the capacities in (\ref{VirCh}). Thus, both the instantaneous channel capacities and the state of the queues affects the mode selection. We note that by choosing the size of the  buffers appropriately, the delay can be limited to any desired value larger than one time slot.  The numerical results in Section V show that even for a small delay, the performance loss caused by the delay constraint, is comparatively small, cf. Section V.

\section{Numerical Results}\label{Numerical}

In this section, we evaluate the achievable rate region obtained via the proposed protocols for the  bidirectional
relay channel with block fading. We assume Rayleigh fading, i.e.,  the channel gains $S_1(i)$
and $S_2(i)$ follow exponential distributions with means $\Omega_1$  and
$\Omega_2$, respectively. In the following, we first introduce benchmark schemes which are used to evaluate the performance of the proposed protocols. Subsequently, we provide numerical results for both the proposed protocols and the benchmark schemes.

\begin{table}
\label{Benchmarks}
\caption{Transmission Modes Utilized by Existing Bidirectional Protocols.}
\begin{center}
\begin{tabular}{|| c | c | c | c | c | c ||}
  \hline                  
\text{Protocol} &\text{6-Phase}& \text{HBC} & \text{TDBC}  & \text{MABC}  &\text{Traditional}  \\ \hline
 \text{Utilized Modes} &$\mathcal{M}_1,\cdots,\mathcal{M}_6$& $\mathcal{M}_1,\mathcal{M}_2,\mathcal{M}_3,\mathcal{M}_6$ & $\mathcal{M}_1,\mathcal{M}_2,\mathcal{M}_6$  & $\mathcal{M}_3,\mathcal{M}_6$  & $\mathcal{M}_1,\mathcal{M}_2,\mathcal{M}_4,\mathcal{M}_5$
 \\ \hline 
   
\end{tabular}
\end{center}
\end{table}

\subsection{Benchmark Schemes}

As benchmark schemes, we consider protocols which have a fixed schedule for using a subset of the possible transmission modes, also referred to as conventional protocols in this section, in contrast to the proposed protocols, cf. Theorems \ref{AdaptProt} and \ref{FixProt}, which optimally employ adaptive mode selection (AMS) in each time slot. As discussed in Section I, several bidirectional                                                                                                                                                                                                                                                                                                                                                                                                                                                                                                                                   relaying protocols have been introduced in the literature including: 1) traditional two-way relaying, 2) the TDBC protocol \cite{TDBC}, 3) the MABC protocol \cite{MABC}, 4) the HBC protocol \cite{Tarokh}, and 5) the 6-phase protocol \cite{6ModeMIMO}. Table II summarizes the transmission modes that are utilized in each protocol.  In the following, we consider two types of conventional protocols for comparison with the proposed protocols: 1) conventional protocols with delay-constrained transmission, and 2) conventional protocols with delay-unconstrained transmission. In particular, in the protocols considered in \cite{MABC,TDBC,Tarokh,BocheIT,BochePIMRC,6ModeMIMO,PopovskiICC}, the users transmit to the relay for a fraction of each time slot and the relay forwards the information to the respective users in the remaining fraction of the time slot. These protocols lead to delay-constrained transmissions since the information  stays in the buffers at the relay for less than one time slot.  In order to make the comparison fair with respect to  the delay, we also consider benchmark schemes that exploit the buffering capability. In this case, although  the benchmark protocols have a fixed and predetermined schedule of using the transmission modes, the users are allowed to transmit to the relay in multiple time slots as long as the buffers at the relay are not full and then the relay forwards the information to the respective users using again multiple time slots. As a performance bound for this scheme, we assume $Q_j^{\max}\to\infty,\,\,j=1,2$, i.e., infinite-size buffers at the relay. These protocols are extensions of the protocol proposed in \cite{Poor} for one-way relaying. In the following, we present the achievable rate regions for the benchmark schemes.

 In the conventional protocols with delay-constrained transmission, the transmission is accomplished in one time slot. In particular, based on the CSI of the involved links, a fraction of the time slot is allocated to each transmission mode in each time slot. Let $\Delta_k(i)$ be the fraction of the $i$-th                                                                                                                             time slot that is allocated  to mode $\mathcal{M}_k$. Thereby, the boundary surface of the achievable rate region of the considered conventional protocols with delay-constrained transmission is obtained from the following optimization problem	
\begin{IEEEeqnarray}{rll}\label{ConvDC} 
\underset{\mathcal{R}(i),\mathcal{D},\mathcal{P}}{\mathrm{maximize}} \,\,& \eta R_{1r}(i) + (1-\eta) R_{2r}(i) \nonumber \\
		\mathrm{subject \,\,to}\,\,&\mathrm{C1}:\,\, R_{1r}(i)=R_{r2}(i), \nonumber\\
		&\mathrm{C2}:\,\, R_{2r}(i)=R_{r1}(i), \nonumber\\
		&\mathrm{C3}:\,\, \left( R_{1r}(i),R_{2r}(i),R_{r1}(i),R_{r2}(i)\right ) \in\mathcal{R}(i),\nonumber\\
		&\mathrm{C4}:\,\, \left( \Delta_1(i),\dots,\Delta_6(i)\right ) \in\mathcal{D},\nonumber\\
		&\mathrm{C5}:\,\, \left( P_1(i),P_2(i),P_r(i)\right ) \in\mathcal{P},
\end{IEEEeqnarray}
where $\mathcal{R}(i)$ is the set of the achievable rates in the $i$-th time slot for a given $\left( \Delta_1(i),\dots,\Delta_6(i)\right )$ and is given by
\begin{IEEEeqnarray}{rll}
\mathcal{R}(i) = \bigg \{ \left( R_{1r}(i),R_{2r}(i),R_{r1}(i),R_{r2}(i)\right ) \big|
        &R_{1r}(i)\geq 0,R_{2r}(i)\geq 0,R_{r1}(i)\geq 0,R_{r2}(i)\geq 0   \nonumber \\
		&\wedge\,\,R_{1r}(i)\leq (\Delta_1(i)+\Delta_3(i))C_{1r}(i)  \nonumber\\
		&\wedge\,\,R_{2r}(i)\leq (\Delta_2(i)+\Delta_3(i))C_{2r}(i)    \nonumber\\
		&\wedge\,\,R_{1r}(i)+R_{2r}(i)\leq \Delta_1(i)C_{1r}(i)+\Delta_2(i)C_{2r}(i)+\Delta_3(i)C_r(i)   \nonumber\\
		&\wedge\,\,R_{r1}(i)\leq (\Delta_4(i)+\Delta_6(i))C_{r1}(i)   \nonumber\\
		&\wedge\,\,R_{r2}(i)\leq (\Delta_5(i)+\Delta_6(i))C_{r2}(i)    \bigg \}
\end{IEEEeqnarray}
and $\mathcal{D}$ is the set of possible  $\Delta_k(i)$ and is given by
\begin{IEEEeqnarray}{rll}
\mathcal{D} = \bigg\{\left( \Delta_1(i),\dots,\Delta_6(i)\right ) \big| 
        \Delta_k(i)\geq 0, \,\,\forall k,i\,\,\wedge \,\,
         \sum\limits_{k=1}^{6}\Delta_k(i) = 1, \,\,\forall i\,\,\wedge \,\,
          \Delta_k(i)= 0, \,\,k \notin \mathcal{K} \bigg \}
\end{IEEEeqnarray}
where set $\mathcal{K}$ contains the indices of the utilized modes, e.g., $\mathcal{K}=\{1,2,6\}$ for the TDBC protocol. 
On the other hand, in the conventional protocols with delay-unconstrained transmission, first the buffers at the relay are filled by using modes $\mathcal{M}_1$, $\mathcal{M}_2$, and $\mathcal{M}_3$ and then the relay forwards the information to the users using modes $\mathcal{M}_4$, $\mathcal{M}_5$, and $\mathcal{M}_6$. We define $\Delta_k=\underset{N\to\infty}{\lim}\frac{N_k}{N}$ where $N_k$ is the number of time slots allocated to mode $\mathcal{M}_k$. Assuming $Q_j^{\max}\to\infty,\,\,j=1,2$, for these protocols,  the value of $\Delta_k$ depends only on the long-term statistics of the channel and the value of $\eta$. Thereby, the boundary surface of the achievable rate region of the conventional protocols with delay-unconstrained transmission is obtained from the following optimization problem
\begin{IEEEeqnarray}{rll}  \label{ConvDU} 
\underset{\bar{\mathcal{R}},\bar{\mathcal{D}},\mathcal{P}}{\mathrm{maximize}} \,\,& \eta \bar{R}_{1r} + (1-\eta) \bar{R}_{2r} \nonumber \\
		\mathrm{subject \,\,to}\,\,&\mathrm{C1}:\,\, \bar{R}_{1r}=\bar{R}_{r2}, \nonumber\\
		&\mathrm{C2}:\,\, \bar{R}_{2r}=\bar{R}_{r1}, \nonumber\\
		&\mathrm{C3}:\,\, \left( \bar{R}_{1r}, \bar{R}_{2r}, \bar{R}_{r1},\bar{R}_{r2}\right ) \in\bar{\mathcal{R}}, \nonumber\\
		&\mathrm{C4}:\,\, \left( \Delta_1,\dots,\Delta_6\right ) \in \bar{\mathcal{D}},\nonumber\\
		&\mathrm{C5}:\,\, \left( P_1(i),P_2(i),P_r(i)\right ) \in\mathcal{P},
\end{IEEEeqnarray}
where $\bar{\mathcal{R}}$ is the set of the long-term achievable rates for a given $\left( \Delta_1,\dots,\Delta_6\right )$ and is given by
\begin{IEEEeqnarray}{rll}
\bar{\mathcal{R}} = \bigg \{ \left( \bar{R}_{1r}, \bar{R}_{2r}, \bar{R}_{r1},\bar{R}_{r2}\right ) \big|
        &\bar{R}_{1r} \geq 0, \bar{R}_{2r}\geq 0, \bar{R}_{r1}\geq 0, \bar{R}_{r2}\geq 0   \nonumber \\
		&\wedge\,\, \bar{R}_{1r}\leq (\Delta_1+\Delta_3)\bar{C}_{1r}  \nonumber\\
		&\wedge\,\, \bar{R}_{2r} \leq (\Delta_2+\Delta_3)\bar{C}_{2r}   \nonumber\\
		&\wedge\,\, \bar{R}_{1r}+\bar{R}_{2r}\leq \Delta_1 \bar{C}_{1r} +\Delta_2 \bar{C}_{2r}+\Delta_3 \bar{C}_{r}   \nonumber\\
		&\wedge\,\,\bar{R}_{r1}\leq (\Delta_4+\Delta_6)\bar{C}_{r1} \nonumber\\
		&\wedge\,\,\bar{R}_{r2}\leq (\Delta_5+\Delta_6)\bar{C}_{r2}    \bigg \}
\end{IEEEeqnarray}
where $\bar{C}_{jj'}=\underset{N\to\infty}{\lim} \frac{1}{N} \sum_{i=1}^{N} C_{jj'}(i),\,\,j,j'\in\{1,2,r\}$, and $\bar{C}_{r}=\underset{N\to\infty}{\lim} \frac{1}{N} \sum_{i=1}^{N} C_{r}(i)$. Moreover, $\bar{\mathcal{D}}$ is the set of possible  $\Delta_k$ and is given by
\begin{IEEEeqnarray}{rll}
\bar{\mathcal{D}} = \bigg\{\left( \Delta_1,\dots,\Delta_6\right ) \big| 
        \Delta_k\geq 0, \,\,\forall k\,\,\wedge \,\,
         \sum\limits_{k=1}^{6}\Delta_k = 1 \,\,\wedge \,\,
          \Delta_k= 0, \,\,k \notin \mathcal{K} \bigg \}.
\end{IEEEeqnarray}

\begin{remk}\label{RemkScheme}
Recall that the direct link is not present in the considered system model. Therefore,  the point-to-point modes are special cases of the multiple-access or broadcast modes, i.e., the rate is zero in one direction. Hence, one can consider only modes $\mathcal{M}_3$ and $\mathcal{M}_6$ to achieve the maximum possible performance in terms of the achievable rate region. However, in this case, not all nodes may be able to transmit with the capacity rates in all time slots in order to satisfy constraints $\mathrm{C1}$ and $\mathrm{C2}$ in (\ref{ConvDC}) and (\ref{ConvDU}). Thus, the transmission rates in each time constitute degrees of freedom and have to be optimized as is done in (\ref{ConvDC}) or (\ref{ConvDU}). However, this complicates the final protocol. Alternatively, if all the point-to-point modes are also considered, as is done in the proposed protocols in Theorems \ref{AdaptProt} and \ref{FixProt}, the maximum performance in terms of achievable rate region can be achieved by transmitting with the capacity rates in all time slots which significantly simplifies the resulting protocols. However, if the rates are optimized, the conventional 6-phase, HBC, and MABC protocols have the same performance if a direct link is not present between user 1 and user 2. We note that the main idea of proposing the  HBC protocol in \cite{Tarokh} and the 6-phase protocol in \cite{6ModeMIMO} was to utilize the direct link more efficiently. However, since the direct link is not available in the considered system model, we only consider the conventional protocols with $\mathcal{K}=\{1,\dots,6\}$, $\mathcal{K}=\{1,2,6\}$, and $\mathcal{K}=\{1,2,4,5\}$ for the benchmark schemes. 
\end{remk}

\begin{remk}\label{RemkPower}
For the fixed transmit power constraint, the optimization problems in (\ref{ConvDC}) and (\ref{ConvDU}) are linear programming problems, which can be efficiently solved  using available solvers such as CVX \cite{CVX}. However, for the joint long-term power constraint, the optimization problems in (\ref{ConvDC}) and (\ref{ConvDU}) are non-convex. Unfortunately, to the best of our knowledge, there is no work available in the literature that considers  optimization problems in (\ref{ConvDC}) and (\ref{ConvDU}) to obtain  all points of the rate region under a joint long-term power constraint. Therefore, as a benchmark scheme for the proposed protocol with joint power constraint in Theorem \ref{AdaptProt}, we assume that $P_1=P_2=P_r=P$ and choose the value of $P$ such that the total consumed power of nodes meets the total power budget. 
\end{remk}

\begin{remk}
The  protocols proposed in Theorems \ref{AdaptProt} and \ref{FixProt} consider all six modes in the selection policy. However, one can exclude certain modes and derive the corresponding optimal strategy according to the framework that we have developed in this paper. Thereby, in the resulting protocol, not all nodes may transmit with the capacity rates in all time slots to satisfy constraints $\mathrm{C1}$ and $\mathrm{C2}$ in (\ref{OptProb}). In the following subsection, we  also present simulation results for  protocols with AMS for a subset of the available modes for comparison with the  protocols proposed in Theorems \ref{AdaptProt} and \ref{FixProt}. We skip the derivation and the statement of the optimal AMS protocols for a subset of the available modes due to space constraints.
\end{remk}

\subsection{Evaluation of the Proposed Protocols and the Benchmark Schemes}

\begin{figure}
\centering
\psfrag{A}[ll][cc][0.5]{$\text{AMS, proposed protocol in Theorem \ref{FixProt},} \,\,\mathcal{K}=\{1,\dots,6\}$}
\psfrag{B}[ll][cc][0.5]{$\text{AMS,}\,\,\mathcal{K}=\{1,2,6\}$}
\psfrag{C}[ll][cc][0.5]{$\text{AMS,}\,\,\mathcal{K}=\{1,2,4,5\}$}
\psfrag{D}[ll][cc][0.5]{$\text{Conventional protocol,} \,\,\mathcal{K}=\{1,\dots,6\}$}
\psfrag{E}[ll][cc][0.5]{$\text{Conventional protocol,}\,\,\mathcal{K}=\{1,2,6\}$}
\psfrag{F}[ll][cc][0.5]{$\text{Conventional protocol,}\,\,\mathcal{K}=\{1,2,4,5\}$}
\psfrag{AMS}[cc][cc][0.5]{$\quad\qquad\text{AMS Protocols}$}
\psfrag{Conv}[ll][cc][0.5]{$\text{Conventional}$}
\psfrag{DU}[ll][cc][0.5]{$\,\,\text{Protocols}$}
\psfrag{R12}[cc][cc][0.8]{$\bar{R}_{12}$}
\psfrag{R21}[cc][cc][0.8]{$\bar{R}_{21}$}
\includegraphics[width=0.7 \linewidth]{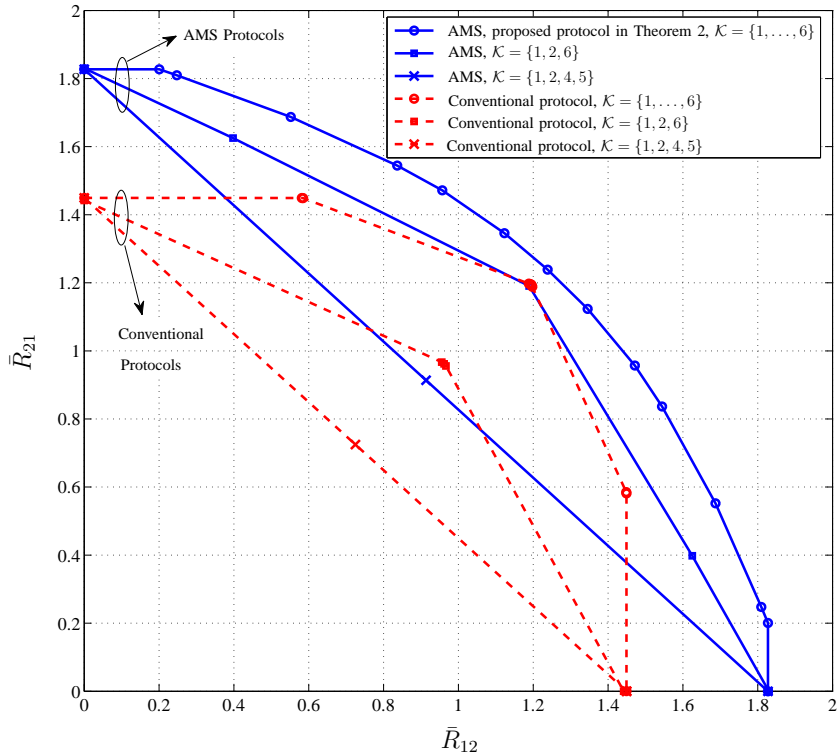}
\caption{Rate region for delay-unconstrained transmission, symmetric channels, i.e.,  $\Omega_1=\Omega_2=1$, and node powers $P_1 = P_2 = P_r = 10$ dB.}
\label{FigFixComparison}
\end{figure}

We first consider the protocols assuming a fixed transmit power for each node.  As performance benchmarks for delay-unconstrained transmission in Theorem \ref{FixProt}, we consider the conventional protocols with delay-unconstrained transmission where the boundary surface of the rate region is obtained from (\ref{ConvDU}). Moreover, we consider different subsets of the available modes for both the AMS and the conventional protocols. Fig. \ref{FigFixComparison} shows the rate regions of the AMS and conventional protocols for delay-unconstrained transmission. We assume fixed powers $P_1=P_2=P_r=10\,\,\text{dB}$ and symmetric channels, i.e., $\Omega_1=\Omega_2=1$.  We observe that the rate regions of the conventional protocols are in the interior of the rate regions of the AMS protocols for any utilized subset of the available modes. Moreover, a considerable gain is obtained by adaptive mode selection compared to transmission with a fixed schedule for using the modes. We can also observe that for a fixed per-node power constraint, excluding the multiple-access or broadcast modes leads to a huge performance loss. We note that the sum rate point for the AMS protocol with $\mathcal{K}=\{1,2,6\}$, was also derived in \cite{PopovskiLetter}.  

\begin{figure}
\centering
\psfrag{A}[ll][cc][0.5]{$\text{AMS, proposed delay-unconstrained protocol in Theorem \ref{FixProt}}$}
\psfrag{B}[ll][cc][0.5]{$\text{AMS,} \,\,E\{T_1\}=E\{T_2\}=10 \,\,\text{time slots}$}
\psfrag{C}[ll][cc][0.5]{$\text{AMS,} \,\,E\{T_1\}=E\{T_2\}=5 \,\,\text{time slots}$}
\psfrag{D}[ll][cc][0.5]{$\text{Conventional protocol with delay-constrained transmission}$}
\psfrag{R1}[ll][cc][0.5]{$\text{Region}\,\,\mathcal{S}_1$}
\psfrag{C1}[ll][cc][0.5]{$\text{Case 2, d}$}
\psfrag{R2}[ll][cc][0.5]{$\text{Region}\,\,\mathcal{S}_0$}
\psfrag{C2}[ll][cc][0.5]{$\text{Case 2, a}$}
\psfrag{R3}[ll][cc][0.5]{$\text{Region}\,\,\mathcal{S}_0$}
\psfrag{C3}[ll][cc][0.5]{$\text{Case 1}$}
\psfrag{R4}[ll][cc][0.5]{$\text{Region}\,\,\mathcal{S}_0$}
\psfrag{C4}[ll][cc][0.5]{$\text{Case 2, b}$}
\psfrag{R5}[ll][cc][0.5]{$\text{Region}\,\,\mathcal{S}_2$}
\psfrag{C5}[ll][cc][0.5]{$\text{Case 2, d}$}
\psfrag{R12}[cc][cc][0.8]{$\bar{R}_{12}$}
\psfrag{R21}[cc][cc][0.8]{$\bar{R}_{21}$}
\includegraphics[width=0.7 \linewidth]{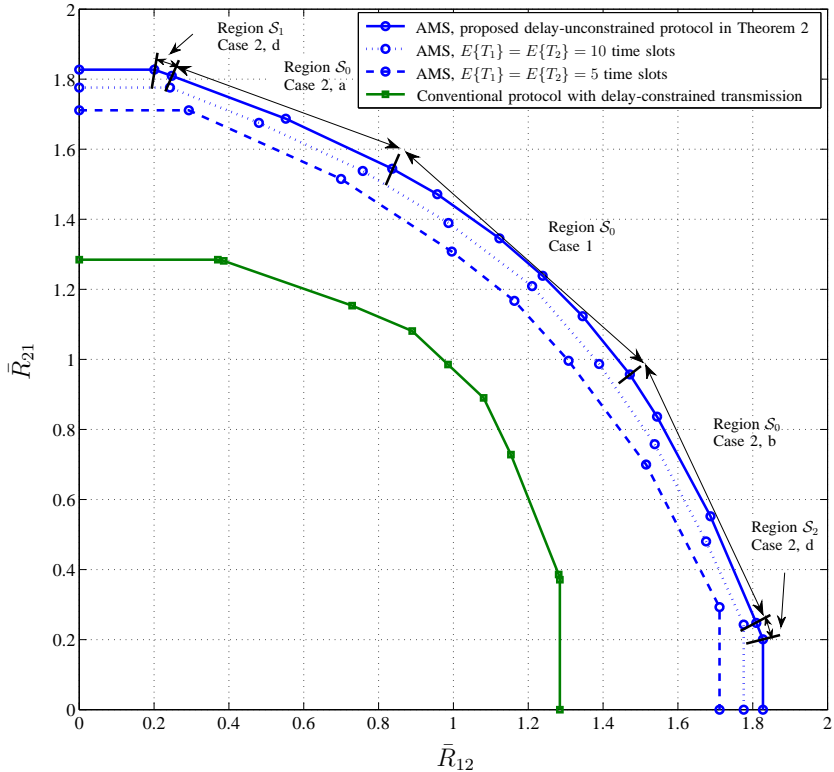}
\caption{Rate region for delay-constrained transmission, symmetric channels, i.e.,  $\Omega_1=\Omega_2=1$, node powers $P_1 = P_2 = P_r = 10$ dB, and $\mathcal{K}=\{1,\dots,6\}$.}
\label{FigFixSym}
\end{figure}
\begin{figure}
\centering
\psfrag{A}[ll][cc][0.5]{$\text{AMS, proposed delay-unconstrained protocol in Theorem \ref{FixProt}}$}
\psfrag{B}[ll][cc][0.5]{$\text{AMS,} \,\,E\{T_1\}=E\{T_2\}=10 \,\,\text{time slots}$}
\psfrag{C}[ll][cc][0.5]{$\text{AMS,} \,\,E\{T_1\}=E\{T_2\}=5 \,\,\text{time slots}$}
\psfrag{D}[ll][cc][0.5]{$\text{Conventional protocol with delay-constrained transmission}$}
\psfrag{R1}[ll][cc][0.5]{$\text{Region}\,\,\mathcal{S}_1$}
\psfrag{C1}[ll][cc][0.5]{$\text{Case 2, a}$}
\psfrag{R2}[ll][cc][0.5]{$\text{Region}\,\,\mathcal{S}_1$}
\psfrag{C2}[ll][cc][0.5]{$\text{Case 2, c}$}
\psfrag{R3}[ll][cc][0.5]{$\text{Region}\,\,\mathcal{S}_1$}
\psfrag{C3}[ll][cc][0.5]{$\text{Case 2, b}$}
\psfrag{R4}[ll][cc][0.5]{$\text{Region}\,\,\mathcal{S}_0$}
\psfrag{C4}[ll][cc][0.5]{$\text{Case 2, b}$}
\psfrag{R5}[ll][cc][0.5]{$\text{Region}\,\,\mathcal{S}_2$}
\psfrag{C5}[ll][cc][0.5]{$\text{Case 2, d}$}
\psfrag{R12}[cc][cc][0.8]{$\bar{R}_{12}$}
\psfrag{R21}[cc][cc][0.8]{$\bar{R}_{21}$}
\includegraphics[width=0.7 \linewidth]{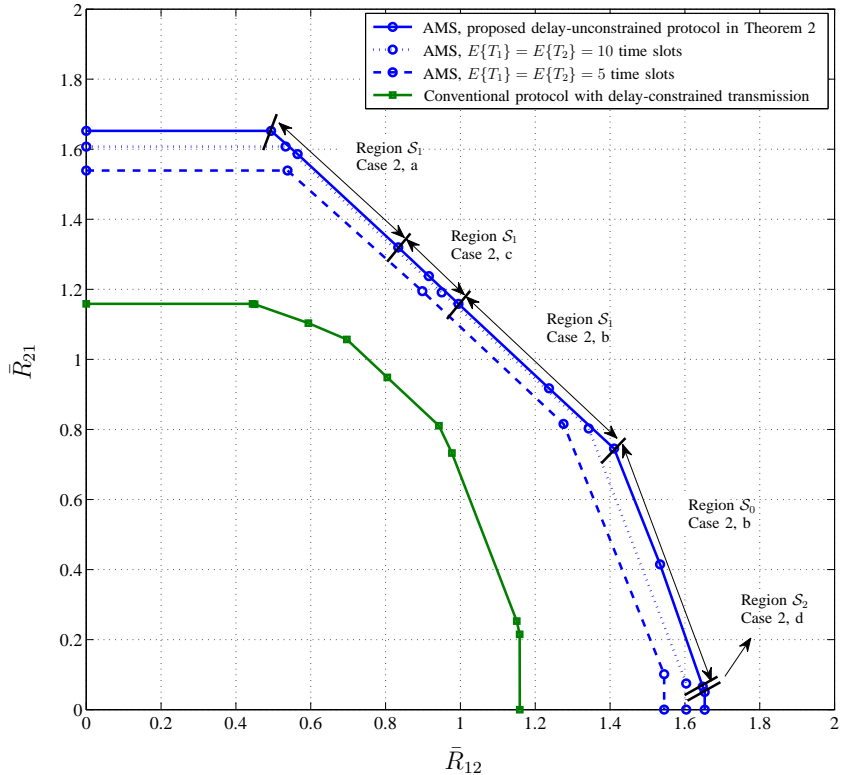}
\caption{Rate region for delay-constrained transmission, asymmetric channels, i.e.,  $\Omega_1=\frac{3}{2},\Omega_2=\frac{1}{2}$, node powers $P_1 = P_2 = P_r = 10$ dB, and $\mathcal{K}=\{1,\dots,6\}$.}
\label{FigFixAsym}
\end{figure}

In Figs. \ref{FigFixSym} and \ref{FigFixAsym}, we evaluate the proposed protocol for delay-constrained transmission assuming a fixed transmit power for each node. We assume $P_1=P_2=P_r=10\,\,\text{dB}$.  As a performance benchmark, we consider the rate region of the conventional protocol with delay-constrained transmission obtained from (\ref{ConvDC}) for $\mathcal{K}=\{1,\dots,6\}$. We note that this scheme introduces a maximum delay of one time slot. We assume symmetric channels in Fig. \ref{FigFixSym}, i.e., $\Omega_1=\Omega_2=1$.  We observe that with  average delays of five and ten time slots around $93\%$ and $97\%$ of the rate with infinite delay is achieved, respectively. Therefore, the proposed heuristic protocol for delay-constrained transmission is indeed efficient. We also conclude that even for a small average delay, a considerable gain is obtained by the proposed protocol compared to the conventional protocol which causes a delay of one time slot. 

As stated in Proposition \ref{FixProtLong}, the optimal values of the selection weights $\mu_1$ and $\mu_2$ used in the proposed delay-unconstrained protocol in Theorem \ref{FixProt}, depend on the channel statistics, the power of the nodes, and the value of $\eta$. In Fig. \ref{FigFixSym}, the channel statistics and the powers of the nodes are fixed and only $\eta$ affects the optimal values of the selection weights. Moreover, we specified different selection regions in Proposition \ref{FixProtLong} and Table I. In Fig \ref{FigFixSym}, we also indicate the relevant selection region for each part of the boundary surface of the rate region. As stated before, the valid selection region which contains the optimal selection weights changes from $\mathcal{S}_1$ to $\mathcal{S}_0$ and then to $\mathcal{S}_2$ as the value of $\eta$ changes from $0$ to $1$. In particular, for the part corresponding to selection region $\mathcal{S}_0$, Case 1, we obtain  from Proposition \ref{FixProtLong} that $\mu_1^*\neq 0,\eta$ and $\mu_2^*\neq 0,1-\eta$ have to hold which leads the selection of modes $\mathcal{M}_3$ and $\mathcal{M}_6$ only in the optimal selection policy.  The maximum sum rate point is a special point of this part corresponding to $\eta=\frac{1}{2}$ which leads to $\bar{R}_{12}=\bar{R}_{21}$ and the coin flip probability for the decoding order variable as $p_6=\frac{1}{2}$. As the value of $\eta$ decreases,  the optimal value of $\mu_1$ decreases and the optimal value of $\mu_2$ increases which ultimately causes the transition from Case 1 to Case 2 a) of selection region $\mathcal{S}_0$. Then, in the part corresponding to selection region $\mathcal{S}_0$, Case 2 a), mode $\mathcal{M}_4$ is also selected in addition to  modes $\mathcal{M}_3$ and $\mathcal{M}_6$ and the optimal values of the selection weights are obtained as $\mu_1^*=0$ and $\mu_2^*=1-2\eta$. Moreover, as $\eta$ decreases further, the optimal value of $p_6$ decreases to ultimately become zero. Then, the optimal values of selection weights are taken from selection region $\mathcal{S}_1$, Case 2 d). Similar considerations hold for $\eta\in[\frac{1}{2},\,1)$.

Fig. \ref{FigFixAsym} shows the rate region for asymmetric channels, i.e., $\Omega_1=\frac{3}{2}$ and $\Omega_2=\frac{1}{2}$. Similar to the case of symmetric channels, the performance of delay-constrained transmission is close to the performance of delay-unconstrained transmission. We also observe that the rate region is asymmetric in favor of rate $\bar{R}_{12}$. For example, for delay-unconstrained transmission, when the rate of the user 2-to-user 1 transmission achieves its maximum value, $\bar{R}_{21}=1.652$, the maximum rate of the user 1-to-user 2 transmission is still $\bar{R}_{12}=0.4942$. However, when the rate of the user 1-to-user 2 transmission achieves its maximum value, $\bar{R}_{12}=1.652$, the maximum the rate of the user 2-to-user 1 transmission is only $\bar{R}_{21}=0.0504$. We also note that the maximum achievable rate for one-way relaying in one direction is the same as that in the other direction in both symmetric and asymmetric channels. In Fig. \ref{FigFixAsym}, we also specify the relevant selection regions for each part of the boundary surface of the rate region of the delay-unconstrained protocol. One can follow how the valid selection region changes and which transmission modes are selected based on Proposition \ref{FixProtLong} for asymmetric channel statistics in Fig. \ref{FigFixAsym}, similar to the case for symmetric channel statistics  in Fig. \ref{FigFixSym}.

\begin{figure}
\centering
\psfrag{A}[ll][cc][0.5]{$\text{AMS, proposed protocol in Theorem \ref{AdaptProt},} \,\,\mathcal{K}=\{1,\dots,6\}$}
\psfrag{B}[ll][cc][0.5]{$\text{AMS,}\,\,\mathcal{K}=\{1,2,6\}$}
\psfrag{C}[ll][cc][0.5]{$\text{AMS,}\,\,\mathcal{K}=\{1,2,4,5\}$}
\psfrag{D}[ll][cc][0.5]{$\text{Conventional protocol,}\,\,\mathcal{K}=\{1,2,6\}$}
\psfrag{E}[ll][cc][0.5]{$\text{Conventional protocol,}\,\,\mathcal{K}=\{1,2,4,5\}$}
\psfrag{AMS}[cc][cc][0.5]{$\quad\qquad\text{AMS Protocols}$}
\psfrag{Conv}[ll][cc][0.5]{$\text{Conventional}$}
\psfrag{DU}[ll][cc][0.5]{$\,\,\text{Protocols}$}
\psfrag{R12}[cc][cc][0.8]{$\bar{R}_{12}$}
\psfrag{R21}[cc][cc][0.8]{$\bar{R}_{21}$}
\includegraphics[width=0.7 \linewidth]{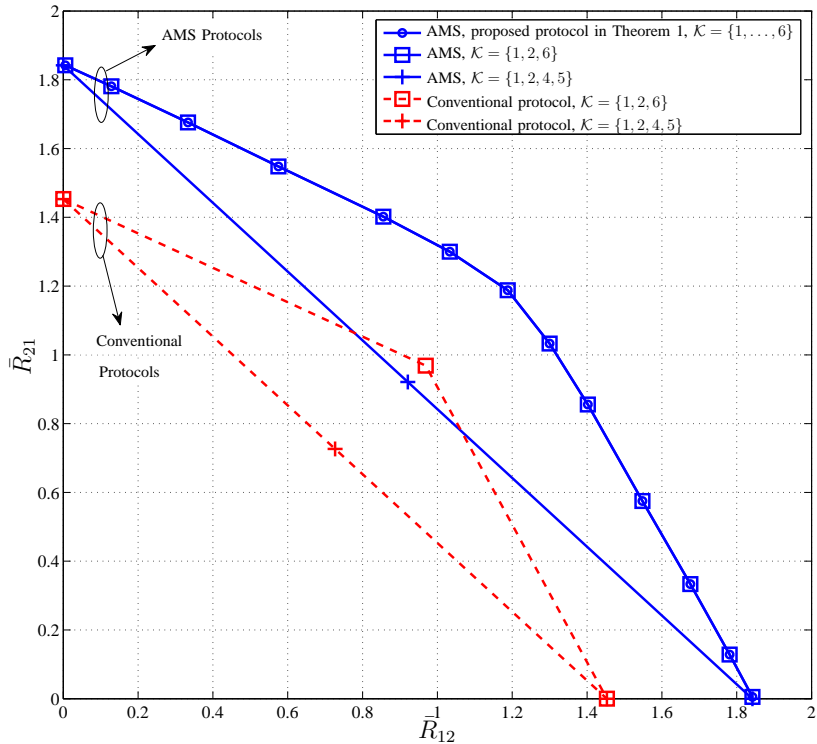}
\caption{Rate region for delay-unconstrained transmission, symmetric channels, i.e.,  $\Omega_1=\Omega_2=1$, and total power constraint $P_t = 10$ dB.}
\label{FigAdapComparison}
\end{figure}
\begin{figure}
\centering
\psfrag{A}[ll][cc][0.5]{$\text{AMS, proposed delay-unconstrained protocol in Theorem \ref{AdaptProt}}$}
\psfrag{B}[ll][cc][0.5]{$\text{AMS,} \,\,E\{T_1\}=E\{T_2\}=10 \,\,\text{time slots}$}
\psfrag{C}[ll][cc][0.5]{$\text{AMS,} \,\,E\{T_1\}=E\{T_2\}=5 \,\,\text{time slots}$}
\psfrag{D}[ll][cc][0.5]{$\text{Conventional protocol with delay-constrained transmission}$}
\psfrag{R1}[ll][cc][0.5]{$\text{Region}\,\,\mathcal{S}_1$}
\psfrag{C1}[ll][cc][0.5]{$\text{Case 2}$}
\psfrag{R2}[ll][cc][0.5]{$\text{Region}\,\,\mathcal{S}_0$}
\psfrag{R3}[ll][cc][0.5]{$\text{Region}\,\,\mathcal{S}_2$}
\psfrag{C3}[ll][cc][0.5]{$\text{Case 2}$}
\psfrag{R12}[cc][cc][0.8]{$\bar{R}_{12}$}
\psfrag{R21}[cc][cc][0.8]{$\bar{R}_{21}$}
\includegraphics[width=0.7 \linewidth]{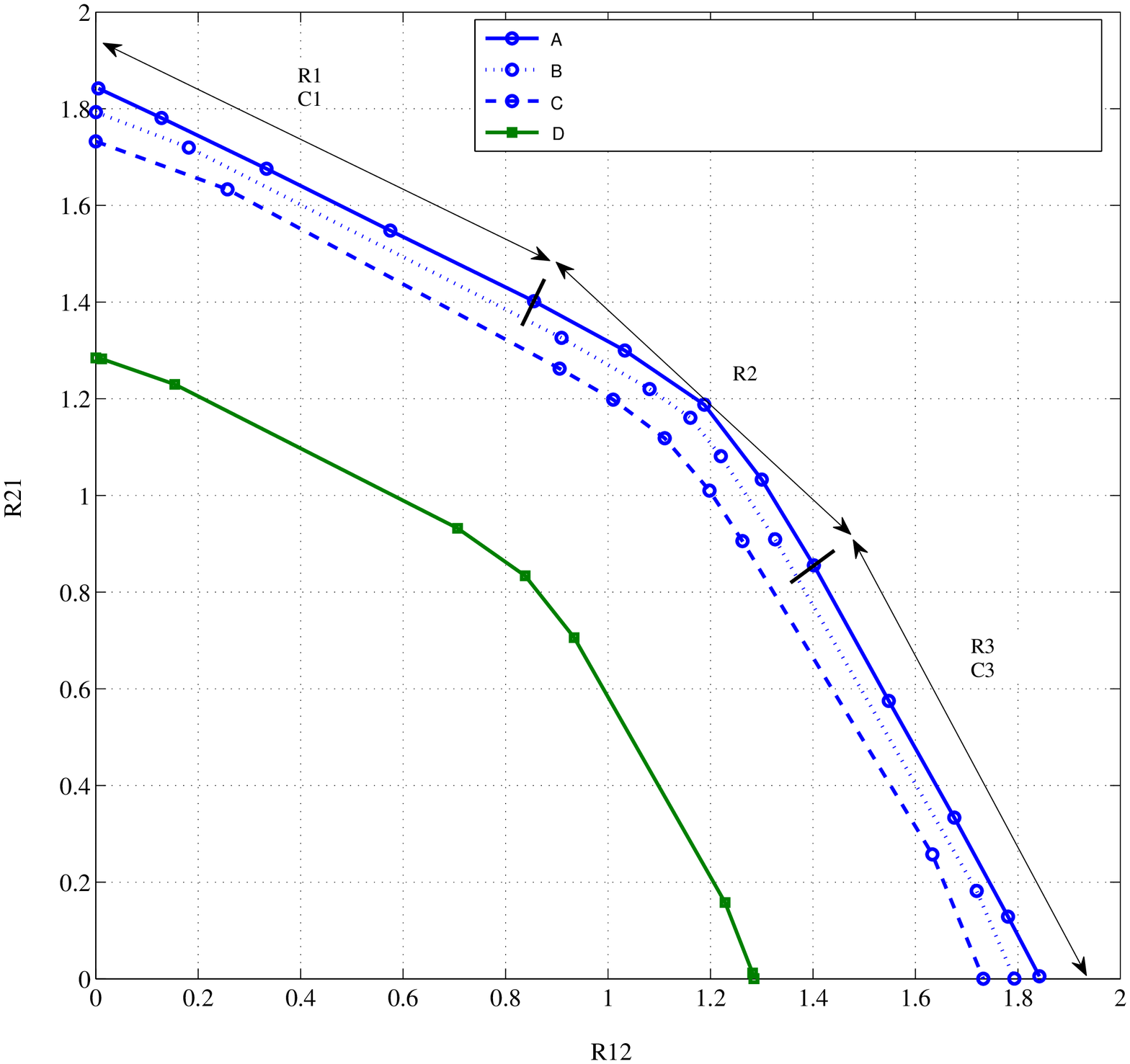}
\caption{Rate region for delay-constrained transmission, symmetric channels, i.e.,  $\Omega_1=\Omega_2=1$, and total power constraint $P_t = 10$ dB, $\mathcal{K}=\{1,\dots,6\}$ for AMS protocols, and $\mathcal{K}=\{1,2,6\}$ for conventional protocol.}
\label{FigAdapSym}
\end{figure}
\begin{figure}
\centering
\psfrag{A}[ll][cc][0.5]{$\text{AMS, proposed delay-unconstrained protocol in Theorem \ref{AdaptProt}}$}
\psfrag{B}[ll][cc][0.5]{$\text{AMS,} \,\,E\{T_1\}=E\{T_2\}=10 \,\,\text{time slots}$}
\psfrag{C}[ll][cc][0.5]{$\text{AMS,} \,\,E\{T_1\}=E\{T_2\}=5 \,\,\text{time slots}$}
\psfrag{D}[ll][cc][0.5]{$\text{Conventional protocol with delay-constrained transmission}$}
\psfrag{R1}[ll][cc][0.5]{$\text{Region}\,\,\mathcal{S}_1$}
\psfrag{C1}[ll][cc][0.5]{$\text{Case 1}$}
\psfrag{R2}[ll][cc][0.5]{$\text{Region}\,\,\mathcal{S}_0$}
\psfrag{R3}[ll][cc][0.5]{$\text{Region}\,\,\mathcal{S}_2$}
\psfrag{C3}[ll][cc][0.5]{$\text{Case 2}$}
\psfrag{R4}[ll][cc][0.5]{$\text{Region}\,\,\mathcal{S}_2$}
\psfrag{C4}[ll][cc][0.5]{$\text{Case 1}$}
\psfrag{R12}[cc][cc][0.8]{$\bar{R}_{12}$}
\psfrag{R21}[cc][cc][0.8]{$\bar{R}_{21}$}
\includegraphics[width=0.7 \linewidth]{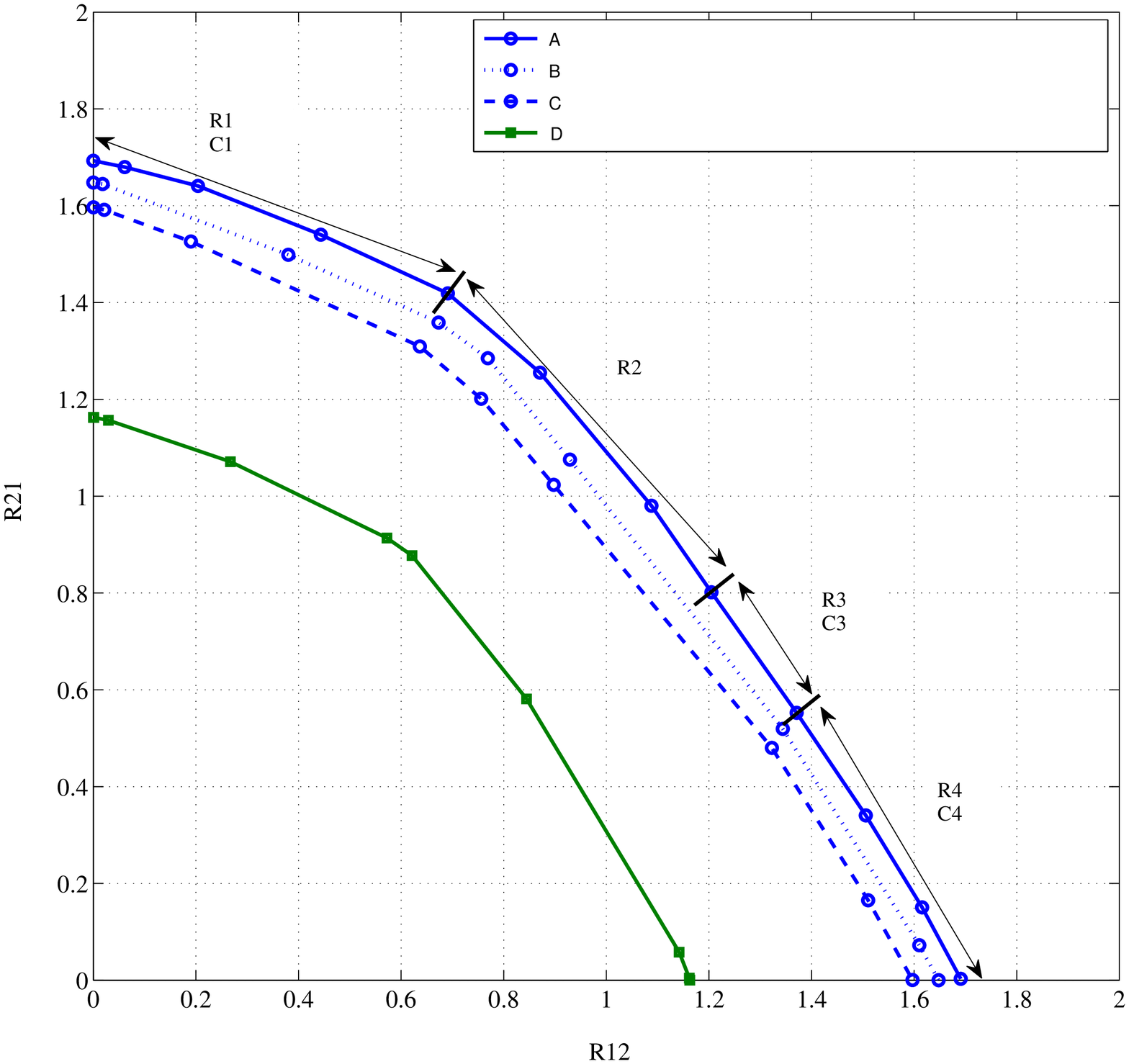}
\caption{Rate region for delay-constrained transmission, asymmetric channels, i.e.,  $\Omega_1=\frac{3}{2},\Omega_2=\frac{1}{2}$, and total power constraint $P_t = 10$ dB, $\mathcal{K}=\{1,\dots,6\}$ for AMS protocols, and $\mathcal{K}=\{1,2,6\}$ for conventional protocol.}
\label{FigAdapAsym}
\end{figure}

Fig. \ref{FigAdapComparison} presents the performance of the proposed protocol in Theorem \ref{AdaptProt} for the case of a joint long-term power constraint for all nodes. As benchmark schemes, we adopt the conventional protocols with delay-unconstrained transmission where the boundary surface of the rate region is obtained from in (\ref{ConvDU}) for $\mathcal{K}=\{1,2,6\}$ and $\mathcal{K}=\{1,2,4,5\}$.  We assume a total power budget $P_t=10$ dB and symmetric channels, i.e., $\Omega_1=\Omega_2=1$.  For the power allocation scheme used in the conventional protocols, we refer to Remark \ref{RemkPower}. In contrast to the rate region of the protocol with fixed per-node power constraint in Fig. \ref{FigFixComparison}, we observe that the shape of the boundary surface of the rate region approximately consists of two straight lines for the considered joint long-term power constraint. Moreover, as stated in Remark \ref{M3Power}, the maximum sum rate point is achieved without using the multiple-access mode. Moreover, the multiple-access  mode is also not used for the one-way points. Since all the points on the line between the maximum sum rate point and the one-way relaying points are achievable by time sharing, we can conclude that the multiple-access mode does not improve the achievable rate region noticeably. Therefore, we can also conclude that the problems of finding the largest achievable rate region under the fixed per-node transmit power constraint and the joint total power constraint for all nodes are fundamentally different. Similar to Fig. \ref{FigFixComparison}, we observe from Fig. \ref{FigAdapComparison} that the rate region of conventional protocols lay in the interior of the rate region of the AMS protocols for all considered subsets of  transmission modes. In particular, a considerable performance gain is obtained by adaptive mode selection compared to transmission based on a predefined schedule of using the transmission modes.

In Figs. \ref{FigAdapSym} and  \ref{FigAdapAsym}, we  evaluate the performance of the proposed protocol with delay-constrained transmission reported in Section III-B.  As a performance benchmark, we consider the rate region of the conventional protocol with delay-constrained transmission obtained from (\ref{ConvDC}) for $\mathcal{K}=\{1,2,6\}$. Fig. \ref{FigAdapSym} is depicted for symmetric channels, i.e., $\Omega_1=\Omega_2=1$, while Fig. \ref{FigAdapAsym} is depicted for asymmetric channels, i.e., $\Omega_1=\frac{3}{2},\Omega_2=\frac{1}{2}$. Similar to the results for the proposed protocol under fixed per-node transmit power constraint, we observe that the performance degradation due to the delay constraint compared to the performance bound for delay-unconstrained transmission is relatively small. This confirms that the proposed heuristic approach to limit the delay is efficient. In other words, even a protocol which can optimally limit the delay, cannot achieve a large performance gain compared to the  heuristic approach proposed in this paper. Moreover, we see that for even a small tolerable average delay, we obtain a considerable performance gain compared to the conventional protocol. 

Similar to Figs. \ref{FigFixSym} and \ref{FigFixAsym}, in Figs. \ref{FigAdapSym} and \ref{FigAdapAsym}, we also specify the relevant selection region for each part of the boundary surface of the rate region of the delay-unconstrained protocol. Again, the valid selection region which contains the optimal selection weights changes from $\mathcal{S}_1$ to $\mathcal{S}_0$ and then to $\mathcal{S}_2$ as the value of $\eta$ increases from $0$ to $1$. In particular, for the part corresponding to selection region $\mathcal{S}_0$, from Proposition \ref{AdaptProtLong}, we obtain  $\mu_1^*\neq 0$,  $\mu_2^*\neq 0$,  and the optimal selection policy does not involve a coin flip. As the value of $\eta$ decreases,  the optimal value of $\mu_1$ decreases and the optimal value of $\mu_2$ increases, which ultimately leads to the transition from selection region $\mathcal{S}_0$ to Case 2 of selection region $\mathcal{S}_1$. Then, as stated in Proposition \ref{AdaptProtLong}, for this selection region, two coin flips determine the optimal selection between modes $\mathcal{M}_1$, $\mathcal{M}_4$, and $\mathcal{M}_6$. Moreover, as $\eta \to 0$, the coin flip probabilities for the selection of modes  $\mathcal{M}_1$ and $\mathcal{M}_6$ become zero, i.e., $p_1 \to 1$ and $p_2 \to 0$, which leads to a one-way relaying point of the rate region. Similar considerations apply to the rate region of asymmetric channels in Fig. \ref{FigAdapAsym}.

\section{Conclusion}
In this paper, we have derived the long-term achievable rate region of the half-duplex bidirectional buffer-aided relay channel with
block fading. We proposed protocols which are not restricted to adhere to a predefined schedule for using a subset of the available transmission modes. In particular, the proposed protocols specify the optimal transmission strategy, i.e., the optimal transmission mode, transmission rates, and/or transmit powers of the nodes, in each time slot based on the 
instantaneous CSI of the involved links and
their long-term statistics. To this end, the relay has to be equipped with two buffers for storage of the information received from both users which leads to an increase in the end-to-end delay. Therefore, we developed protocols for both delay-unconstrained and delay-constrained transmissions. Moreover, we considered the case of a fixed transmit power for each node as well as the case of power allocation under joint long-term power constraint for all nodes.  Simulation results confirmed that the
proposed protocols outperform protocols with a fixed schedule of transmission for all points of the rate region even for small average delays.

\appendices


\section{Proof of Lemma \ref{Queue} }
\label{AppQueue}

For infinite-size buffers at the relay, i.e., $Q_j^{\max}\to\infty,\,\,j=1,2$, the state of the queues does not influence the average rates from the users to the relay. In particular, based on (\ref{RatReg123}a) and (\ref{RatReg123}b), rates $\bar{R}_{1r}$ and $\bar{R}_{2r}$ are given by
\begin{IEEEeqnarray}{lll}
		\bar{R}_{1r}= \underset{N\to \infty}{\lim}\frac{1}{N}\mathop \sum \limits_{i = 1}^N q_1(i)C_{1r}(i)+q_3(i)C_{12r}(i) \IEEEyesnumber \IEEEyessubnumber\\
\bar{R}_{2r}= \underset{N\to \infty}{\lim}\frac{1}{N}\mathop \sum \limits_{i = 1}^N q_2(i)C_{2r}(i)+q_3(i)C_{21r}(i). \IEEEyessubnumber
\end{IEEEeqnarray}
Moreover, let $\bar{C}_{r1}$ and $\bar{C}_{r2}$ denote the average rates achieved by the relay assuming that buffers $B_1$ and $B_2$ always have enough information to supply. Then, $\bar{C}_{r1}$ and $\bar{C}_{r2}$ are given by
\begin{IEEEeqnarray}{lll} \label{Capacity}  
		\bar{C}_{r1}= \underset{N\to \infty}{\lim}\frac{1}{N}\mathop \sum \limits_{i = 1}^N \left[ q_4(i)+q_6(i)\right]C_{r1}(i) \IEEEyesnumber \IEEEyessubnumber\\
\bar{C}_{r2}= \underset{N\to \infty}{\lim}\frac{1}{N}\mathop \sum \limits_{i = 1}^N \left[ q_5(i)+q_6(i)\right]C_{r2}(i). \IEEEyessubnumber
\end{IEEEeqnarray}
 Moreover, the average transmission rates from the relay to the users, i.e., $\bar{R}_{r1}$ and $\bar{R}_{r2}$, can be written as
\begin{IEEEeqnarray}{lll}\label{MinRate}
   \bar{R}_{r1} = \begin{cases}
\bar{R}_{2r}, \,\, &\mathrm{if}\,\, \bar{R}_{2r}<\bar{C}_{r1} \\
\bar{R}_{2r}=\bar{C}_{r1}, &\mathrm{if}\,\, \bar{R}_{2r}=\bar{C}_{r1} \\
\bar{C}_{r1}, \,\, &\mathrm{if}\,\, \bar{R}_{2r}>\bar{C}_{r1} 
\end{cases} \IEEEyesnumber\IEEEyessubnumber \\
\bar{R}_{r2} =\begin{cases}
\bar{R}_{1r}, \,\, &\mathrm{if}\,\, \bar{R}_{1r}<\bar{C}_{r2} \\
\bar{R}_{1r}=\bar{C}_{r2}, & \mathrm{if}\,\, \bar{R}_{1r}=\bar{C}_{r2}\\
\bar{C}_{r2}, \,\, &\mathrm{if}\,\, \bar{R}_{1r}>\bar{C}_{r2}
\end{cases} \IEEEyessubnumber
\end{IEEEeqnarray}
In particular, if $\bar{R}_{2r}<\bar{C}_{r1}$ holds, i.e., the average information flowing into the buffer is less than the average capacity of the relay-to-user 2 channel, then by the law of conservation of flow, we obtain $\bar{R}_{r1}=\bar{R}_{2r}$. On the other hand, if  $\bar{R}_{2r}>\bar{C}_{r1}$ holds, i.e., the average rate flowing into the buffer is larger than  the average capacity of the relay-to-user 2 channel, then the buffer always has enough information to supply and $\bar{R}_{r1}$ in (\ref{RatReg123}c) becomes $\bar{R}_{r1}=\bar{C}_{r1}$, where $\bar{C}_{r1}$ is given in (\ref{Capacity}a). However, $\bar{R}_{2r}>\bar{R}_{r1}$ leads to information loss since user 2 transmits information bits that will not be received at user 1. If  $\bar{R}_{2r}=\bar{C}_{r1}$ holds, we obtain $\bar{R}_{r1}=\bar{R}_{2r}=\bar{C}_{r1}$, for a detailed proof of this property, we refer to the proof of Theorem 1 in \cite[Appendix A]{NikolaJSAC}.  Similar  results hold for $\bar{R}_{r2}$, $\bar{R}_{1r}$, and $\bar{C}_{r2}$.  For the long-term achievable rate pairs, we have to exclude the cases $\bar{R}_{2r}>\bar{C}_{r1} $ and $\bar{R}_{1r}>\bar{C}_{r2}$ in (\ref{MinRate}) to avoid information loss. Thus, we obtain
\begin{IEEEeqnarray}{lll}\label{EqualRate}
  \bar{R}_{1r} = \bar{R}_{r2} \overset{(a)}{\leq} \bar{C}_{r2}  \IEEEyesnumber\IEEEyessubnumber \\
 \bar{R}_{2r} = \bar{R}_{r1} \overset{(b)}{\leq} \bar{C}_{r1}  \IEEEyessubnumber
\end{IEEEeqnarray}
In the following, we will show that, for the rate pairs on the boundary surface of the rate region,   inequalities $(a)$ and $(b)$ hold with equality.  
 To show this, we denote the set of time slots $i$ in which $q_k(i)=1$ holds by $I_k$, i.e., the elements in $I_k$ represent the time slots in which transmission mode $\mathcal{M}_k$ is selected for the optimal solution. Note that $\sum_{k=1}^6 |I_k|=N$, where $|\cdot|$ denotes the cardinality of a set. For the two inequalities in (\ref{MinRate}), we have to consider the following four cases:\\*
\textit{Case 1:} If $\bar{R}_{1r}= \bar{R}_{r2}<\bar{C}_{r2}$ and $\bar{R}_{2r}=\bar{R}_{r1}<\bar{C}_{r1}$, then we can move some indices $i$ from  $I_6$ to $I_3$ and thereby increase both $\bar{R}_{1r}$ and $\bar{R}_{2r}$. Moreover, since $\bar{R}_{1r}<\bar{C}_{r2}$ and $\bar{R}_{2r}<\bar{C}_{r1}$ still hold, from (\ref{MinRate}), we obtain that the equalities are maintained $\bar{R}_{1r}= \bar{R}_{r2}$ and $\bar{R}_{2r}=\bar{R}_{r1}$. Thus, constraints $\mathrm{C1}$ and $\mathrm{C2}$ in (\ref{OptProbOrigin}) hold and the cost function in (\ref{OptProbOrigin}) increases. Hence, $\bar{R}_{1r}= \bar{R}_{r2}<\bar{C}_{r2}$ and $\bar{R}_{2r}=\bar{R}_{r1}<\bar{C}_{r1}$ cannot hold for the optimal solution since the cost function in (\ref{OptProbOrigin}) can be further increased.\\*
\textit{Case 2:} If $\bar{R}_{1r}= \bar{R}_{r2}=\bar{C}_{r2}$ and $\bar{R}_{2r}=\bar{R}_{r1}<\bar{C}_{r1}$, then if $I_4 \neq \varoslash$,  we can move some indices from $I_4$ to $I_2$ which maintains $\bar{R}_{1r}= \bar{R}_{r2}=\bar{C}_{r2}$ and increases $\bar{R}_{2r}$. Moreover, since $\bar{R}_{2r}<\bar{C}_{r1}$ still holds, from (\ref{MinRate}a), we obtain that the equality $\bar{R}_{2r}=\bar{R}_{r1}$ holds. Consequently, constraints $\mathrm{C1}$ and $\mathrm{C2}$ in (\ref{OptProbOrigin}) hold and the cost function in (\ref{OptProbOrigin}) increases which contradicts optimality. On the other hand, if $I_4 = \varoslash$,  we must have $I_6 \neq \varoslash$, otherwise, we obtain $\bar{C}_{r1}=0$ which leads to a contradiction with the earlier assumption $0\leq\bar{R}_{r1}\hspace{-1mm}<\hspace{-1mm}\bar{C}_{r1}$. However, if $I_6 \neq \varoslash$, we can move some of the indices in $I_6$ to $I_5$, which maintains $\bar{R}_{1r}= \bar{R}_{r2}=\bar{C}_{r2}$ and decreases $\bar{C}_{r1}$, until we obtain $\bar{R}_{2r}=\bar{R}_{r1}=\bar{C}_{r1}$ without reducing the cost function in (\ref{OptProbOrigin}).\\*
\textit{Case 3:} For this case, $\bar{R}_{1r}= \bar{R}_{r2}<\bar{C}_{r2}$ and $\bar{R}_{2r}=\bar{R}_{r1}=\bar{C}_{r1}$ hold. The statements for Case 3 are similar to the ones provided for Case 2.

Therefore, the final case remaining for the optimal solution is $\bar{R}_{1r}= \bar{R}_{r2}=\bar{C}_{r2}$ and $\bar{R}_{2r}=\bar{R}_{r1}=\bar{C}_{r1}$. This completes the proof.

\section{Proof of Theorem \ref{AdaptProt}}
\label{AppKKTAdapt}

In this appendix, we solve the optimization problem given in (\ref{OptProb}). In the following, we investigate the Karush-Kuhn-Tucker (KKT) necessary conditions \cite{Boyd} for the optimization problem and show that the necessary conditions result in a unique value for the weighted sum rate, i.e., $\eta\bar{R}_{1r}+(1-\eta)\bar{R}_{2r}$, which, for any given $\eta$, corresponds to a point or a line segment on the boundary surface of the achievable rate region.

To simplify the usage of the KKT conditions, we consider a standard minimization problem equivalent to the maximization
problem in (\ref{OptProb}), i.e., we adopt $-(\eta\bar{R}_{1r}+(1-\eta)\bar{R}_{2r})$ as the cost function to be minimized, and we rewrite all inequality and equality constraints in the form of $f(x)\leq 0$ and $g(x)=0$, respectively.
The corresponding Lagrangian function for the equivalent minimization problem is obtained as
\begin{IEEEeqnarray}{l}\label{KKTFunction}
   \underset{\mathrm{for}\,\, \forall i,k,j,l}{\mathcal{L}(q_k(i),P_j(i),t(i),\mu_l,\gamma,\lambda(i),\alpha_k(i),\beta_k(i),\phi_l(i))}  = \nonumber \\ 
\qquad \,\, -(\eta\bar{R}_{1r}+(1-\eta)\bar{R}_{2r}) + \mu_1(\bar{R}_{1r}-\bar{R}_{r2}) + \mu_2(\bar{R}_{2r}-\bar{R}_{r1}) + \gamma \left(\bar{P}_1+\bar{P}_2+\bar{P}_r - P_t\right) \nonumber \\
   \qquad \,\, + \mathop \sum \limits_{i = 1}^N \lambda \left( i \right)\left( {\mathop \sum \limits_{k = 1}^6 {q_k}\left( i \right) - 1} \right)
+ \mathop \sum \limits_{i = 1}^N \mathop \sum \limits_{k = 1}^6 {\alpha _k}\left( i \right)\left( {{q_k}\left( i \right) - 1} \right) - \mathop \sum \limits_{i = 1}^N \mathop \sum \limits_{k = 1}^6 {\beta _k}\left( i \right){q_k}\left( i \right) \nonumber \\ \qquad \,\,
-\mathop \sum \limits_{i = 1}^N \left[ \nu_1(i)P_1(i) + \nu_2(i)P_2(i) + \nu_r(i)P_r(i) \right] + \mathop \sum \limits_{i = 1}^N \phi_1(i) (t(i)-1) - \mathop \sum \limits_{i = 1}^N \phi_0(i) t(i) \IEEEeqnarraynumspace \IEEEyesnumber
\end{IEEEeqnarray}
where $\mu_1,\mu_2,\lambda(i),\alpha_k(i),\beta_k(i),
\phi_0(i)$, and $\phi_1(i)$ are the Lagrange multipliers corresponding to constraints $\mathrm{C1,C2,C3}$, the upper limit in $\mathrm{C4}$, the lower limit in $\mathrm{C4}$,  the upper limit in $\mathrm{C5}$, and the lower limit in $\mathrm{C5}$, respectively. Moreover, $\gamma$ is the Lagrange multiplier corresponding to the long-term total power constraint in (\ref{TotalPower}) and $\nu_j(i)$ is the Lagrange multiplier for the power non-negativity constraint in (\ref{TotalPower}) for node $j$ in the $i$-th time slot. The KKT conditions include the following:

\noindent
\textbf{1) Stationary condition:}  The differentiation of the Lagrangian function with respect to the
primal variables, $q_k(i),P_j(i)$, and $t(i),\,\,\forall i,j,k$, is zero for the optimal solution, i.e.,
\begin{IEEEeqnarray}{CCCl}\label{Stationary Condition}
    \frac{\partial\mathcal{L}}{\partial q_k(i)} &=& 0, \quad &\forall i,k \IEEEyesnumber\IEEEyessubnumber\\
\frac{\partial\mathcal{L}}{\partial P_j(i)} &=& 0, \quad &\forall i,j \IEEEyessubnumber \\
\frac{\partial\mathcal{L}}{\partial t(i)} &=& 0, \quad &\forall i.\IEEEyessubnumber
\end{IEEEeqnarray}

\noindent
\textbf{2) Primal feasibility condition:}  The optimal solution has to satisfy the constraints of the primal problem, i.e., constraints $\mathrm{C1},\dots,\mathrm{C6}$ in (\ref{OptProb}).

\noindent
\textbf{3) Dual feasibility condition:}  The Lagrange multipliers for the inequality constraints have to be non-negative, i.e.,
\begin{IEEEeqnarray}{lll}\label{Dual Feasibility Condition}
            \alpha_k(i)\geq 0, \quad &\forall i,k \IEEEyesnumber\IEEEyessubnumber\\
         \beta_k(i)\geq 0,\quad &\forall i,k \IEEEyessubnumber\\
           \nu_j(i) \geq0,   & \forall i,j \IEEEyessubnumber \\
             \phi_l(i) \geq0,   & \forall i,l. \IEEEyessubnumber
\end{IEEEeqnarray}

\noindent
\textbf{4) Complementary slackness:}  If an inequality is inactive, i.e., the optimal solution is in the interior of the corresponding set, the corresponding Lagrange multiplier is zero. Thus, we obtain
\begin{IEEEeqnarray}{lll}\label{Complementary Slackness}
    {\alpha _k}\left( i \right)\left( {{q_k}\left( i \right) - 1} \right)=0,\quad  &\forall i,k \IEEEyesnumber\IEEEyessubnumber\\
    {\beta _k}\left( i \right){q_k}\left( i \right)=0,  &\forall i,k \IEEEyessubnumber\\
    \nu_j(i) P_j(i) = 0, &\forall i,j \IEEEyessubnumber\\
    \phi_1(i) (t(i)-1) = 0, &\forall i \IEEEyessubnumber\\
		\phi_0(i) t(i) = 0, &\forall i.\IEEEyessubnumber
\end{IEEEeqnarray}
A common approach to find a set of primal variables, i.e., $q_k(i), P_j(i),t(i),\,\,\forall i,j,k$ and Lagrange multipliers, i.e., $\mu_1,\mu_2,\gamma,\lambda(i),\alpha_k(i),\beta_k(i),\nu_j(i),\phi_l(i),\,\,
\forall i,k,l$, which satisfy the KKT conditions is to start with the complementary slackness conditions and see if the inequalities are active or not. Combining these results with the primal feasibility and dual feasibility conditions, we obtain various possibilities. Then, from these possibilities, we obtain one or more candidate solutions from the stationary conditions and the optimal solution is surely one of these candidates. In the following subsections, with this approach, we find the optimal values  $q_k^*(i),P_j^*(i),$ and $t^*(i),\,\,\forall i,j,k$.

\subsection{Optimal $q_k^*(i)$}

In order to determine the optimal selection policy, $q_k^*(i)$, we must calculate the derivatives
in (\ref{Stationary Condition}a). This leads to
\begin{IEEEeqnarray}{lll}\label{Stationary Mode}
    \frac{\partial\mathcal{L}}{\partial q_1(i)} = - \frac{1}{N}(\eta\Minus\mu_1)C_{1r}(i)\Add \lambda(i)\Add \alpha_1(i)\Minus \beta_1(i) \nonumber  + \frac{1}{N}\gamma P_1(i)\Equal 0\, \IEEEeqnarraynumspace \IEEEyesnumber \IEEEyessubnumber \\
    \frac{\partial\mathcal{L}}{\partial q_2(i)} = -\frac{1}{N}(1\Minus\eta\Minus\mu_2)C_{2r}(i)\Add\lambda(i)\Add\alpha_2(i)\Minus\beta_2(i) \nonumber  +\frac{1}{N}\gamma P_2(i)\Equal 0 \IEEEyessubnumber   \\
    \frac{\partial\mathcal{L}}{\partial q_3(i)} = -\frac{1}{N}[(\eta\Minus\mu_1)C_{12r}(i)\Add(1\Minus\eta\Minus\mu_2)C_{21r}(i)]\Add\lambda(i) +\alpha_3(i)\Minus\beta_3(i)\Add\frac{1}{N}\gamma (P_1(i)\Add P_2(i))\Equal 0 \quad  \IEEEyessubnumber \\
    \frac{\partial\mathcal{L}}{\partial q_4(i)} = - \frac{1}{N}\mu_2 C_{r1}(i)\Add\lambda(i)\Add\alpha_4(i)\Minus \beta_4(i)\Add\frac{1}{N}\gamma P_r(i)\Equal 0 \qquad \, \IEEEyessubnumber \\
    \frac{\partial\mathcal{L}}{\partial q_5(i)} = -\frac{1}{N}\mu_1 C_{r2}(i)\Add\lambda(i)\Add\alpha_5(i)\Minus\beta_5(i)\Add \frac{1}{N}\gamma P_r(i)\Equal 0 \qquad \, \IEEEyessubnumber \\
    \frac{\partial\mathcal{L}}{\partial q_6(i)} = -\frac{1}{N}[\mu_1 C_{r2}(i)\Add\mu_2 C_{r1}(i)]\Add\lambda(i)\nonumber +\alpha_6(i)\Minus\beta_6(i)\Add\frac{1}{N}\gamma P_r(i)\Equal 0. \IEEEyessubnumber
\end{IEEEeqnarray}
Without loss of generality, we first obtain the necessary condition for $q_1^*(i)=1$ and then generalize the result
to $q_k^*(i)=1,\,\,k=2,\dots,6$. If $q_k^*(i)=1$, from constraint $\mathrm{C4}$ in (\ref{OptProb}), the other
selection variables are zero, i.e., $q_k^*(i)=0,\,\,k=2,...,6$. Furthermore, from (\ref{Complementary Slackness}),
we obtain $\alpha_k(i)=0,\,\,k = 2,...,6$ and $ \beta_1(i)=0$. By substituting these values into (\ref{Stationary Mode}), we obtain
\begin{IEEEeqnarray}{lll}\label{MetAdaptApp}
   N[\lambda(i)+\alpha_1(i)] = (\eta-\mu_1)C_{1r}(i) -\gamma P_1(i) \triangleq \Lambda_1(i) \IEEEyesnumber\IEEEyessubnumber  \\
    N[\lambda(i)-\beta_2(i)] =  (1-\eta-\mu_2)C_{2r}(i) -\gamma P_2(i) \triangleq \Lambda_2(i)\IEEEyessubnumber  \\
   N[\lambda(i)-\beta_3(i)] =  (\eta\Minus \mu_1)C_{12r}(i)\Add (1\Minus\eta\Minus \mu_2)C_{21r}(i) -\gamma (P_1(i)\Add P_2(i)) \triangleq  \Lambda_3(i) \quad\,\,\,\, \IEEEyessubnumber \\
   N[\lambda(i)-\beta_4(i)]  = \mu_2 C_{r1}(i) -\gamma P_r(i) \triangleq \Lambda_4(i) \IEEEyessubnumber\\
    N[\lambda(i)-\beta_5(i)] = \mu_1 C_{r2}(i) -\gamma P_r(i) \triangleq \Lambda_5(i) \IEEEyessubnumber\\
    N[\lambda(i)-\beta_6(i)] = \mu_1 C_{r2}(i)+\mu_2 C_{r1}(i) -\gamma P_r(i) \triangleq \Lambda_6(i),\qquad\IEEEyessubnumber
\end{IEEEeqnarray}
where $\Lambda_k(i)$ is referred to as selection metric. By subtracting (\ref{MetAdaptApp}b)-(\ref{MetAdaptApp}f) from (\ref{MetAdaptApp}a), we obtain
\begin{IEEEeqnarray}{rCl}\label{eq_2_1}
    \Lambda_1(i) - \Lambda_k(i) = N[\alpha_1(i)+\beta_k(i)], \quad k=2,3,4,5,6. \IEEEyesnumber
\end{IEEEeqnarray}
From the dual feasibility conditions given in (\ref{Dual Feasibility Condition}a) and (\ref{Dual Feasibility Condition}b), we have $\alpha_k(i),\beta_k(i)\geq 0$. By inserting $\alpha_k(i),\beta_k(i)\geq 0$ in (\ref{eq_2_1}), we obtain the necessary condition for $q_1^*(i)=1$ as
\begin{IEEEeqnarray}{lll}
    \Lambda_1(i) \geq \max \left \{ \Lambda_2(i), \Lambda_3(i), \Lambda_4(i), \Lambda_5(i), \Lambda_6(i) \right \}. \IEEEyesnumber
\end{IEEEeqnarray}
Repeating the same procedure for $q_k^*(i)=1,\,\,k=2,\dots,6$, we obtain a necessary condition for selecting transmission mode $\mathcal{M}_{k^*}$ in the $i$-th time slot as 
\begin{IEEEeqnarray}{lll}\label{OptMetAdapt}
   \Lambda_{k^*}(i) \geq {\underset{k\in\{1,\cdots,6\}}{\max}}\{\Lambda_{k}(i)\}, \IEEEyesnumber
\end{IEEEeqnarray}
where the Lagrange multipliers $\mu_1,\mu_2$, and $\gamma$ are chosen such that constraint $\mathrm{C1}$ and $\mathrm{C2}$ in (\ref{OptProb}) and the joint long-term power constraint in (\ref{TotalPower}) hold. We refer  to $\mu_1$ and $\mu_2$ also as selection weights and to $\gamma$ as power weight.
We note that if the selection metrics are not equal in the $i$-th time slot, only one of
the modes satisfies (\ref{OptMetAdapt}). Therefore, in this case, the necessary condition  for    mode  selection  in  (\ref{OptMetAdapt})  is  sufficient.  Moreover,  in
Appendix \ref{AppBinRelax}, we prove that the probability that two selection
metrics are equal is non-zero for channel gains with continuous probability density function only if $\mu_1=0$ or $\mu_2=0$.  Therefore, unless $\mu_1=0$ or $\mu_2=0$, the necessary condition for selecting transmission mode $\mathcal{M}_k$ in (\ref{OptMetAdapt}) is in fact sufficient and is the optimal mode selection policy. However, if $\mu_1 = 0$, then we obtain $\Lambda_4(i)=\Lambda_6(i)\overset{(a)}{=}\Lambda_1(i),\,\,\forall i$, where equality $(a)$ holds if $\mu_2=\eta$. Similarly, if $\mu_2 = 0$, then we obtain $\Lambda_5(i)=\Lambda_6(i)\overset{(b)}{=}\Lambda_2(i),\,\,\forall i$, where equality $(b)$ holds if $\mu_1=1-\eta$. For these cases, we prove in Appendix \ref{AppAdapLong} that we have the freedom to choose between the modes which yield the same value for the selection metric in (\ref{MetAdaptApp}) as long as the long-term constraints $\mathrm{C1}$ and $\mathrm{C2}$ in (\ref{OptProb}) hold. One way of selecting between the modes for which the selection metrics $\Lambda_{k}(i)$ are identical is to employ a probabilistic approach using coin flips.  In order to include the coin flips in (\ref{OptMetAdapt}), we write the condition for the selection of mode $\mathcal{M}_{k}$ in the $i$-th time slot as
\begin{IEEEeqnarray}{lll}\label{OptMetSufAdapt}
 \Lambda_{k}(i) > {\underset{k\in\{1,\dots,6\}}{\max}}\{\mathcal{I}_{k}(i)\Lambda_{k}(i)\} \IEEEyesnumber
\end{IEEEeqnarray}
where $\mathcal{I}_{k}(i)\in\{0,1\}$ is a binary indicator variable which leads to the selection of only one of the transmission modes with identical values of the selection metrics in each time slot.  In particular, the value of $\mathcal{I}_{k}(i)$ is a function of the outcomes of the coin flips used to select the modes with identical values of the selection metrics. In order to be able to find $\mathcal{I}_{k}(i)$ mathematically, we define $\mathcal{C}_n(i)\in \{\mathrm{0,1}\}$ as the outcome of the $n$-th coin flip in the $i$-th time slot with the probabilities of the possible outcomes defined as $\Pr\{\mathcal{C}_n(i)=1\}=p_n$ and $\Pr\{\mathcal{C}_n(i)=0\}=1-p_n$. Using this notation, we will prove in Appendix \ref{AppAdapLong} that $\mathcal{I}_{k}(i),\,\,k=1,\dots,6$ is obtained as follows 
\begin{IEEEeqnarray}{lll}
   [\mathcal{I}_1(i),\dots,\mathcal{I}_6(i)] = {\begin{cases} 
    [1,\,\,1,\,\,1,\,\,0,\,\,0,\,\,1], & \mathrm{if}\,\,\mu_1\neq 0,\,\mu_2\neq 0 \\
[1\Minus \mathcal{C}_1(i),\,\,1,\,\,1,\,\,\mathcal{C}_1(i)[1\Minus \mathcal{C}_2(i)],\,\,0,\,\,\mathcal{C}_1(i)\mathcal{C}_2(i)], & \mathrm{if}\,\,\mu_1 = 0,\,\mu_2\neq 0 \\
[1,\,\,1\Minus \mathcal{C}_3(i),\,\,1,\,\,0,\,\,\mathcal{C}_4(i)[1\Minus \mathcal{C}_3(i)],\,\,\mathcal{C}_3(i)\mathcal{C}_4(i)], & \mathrm{if}\,\,\mu_1\neq 0,\,\mu_2 = 0
\end{cases}}
\end{IEEEeqnarray}
where the coin flip probabilities $p_1,\dots,p_4$ are long-term variables, which depend only on the channel statistics and the value of $\eta$, and are given in Appendix \ref{AppAdapLong}.

\subsection{Optimal $P_j^*(i)$}

In order to determine the optimal $P_j(i)$, we have to calculate the derivatives in (\ref{Stationary Condition}b). This leads to
\begin{subequations}\label{Stationary Power}
\begin{align*}
    \frac{\partial\mathcal{L}}{\partial P_1(i)} =        &-\frac{1}{N\mathrm{ln}2} \Big[ \big \{(\eta-\mu_1)q_1(i)-t(i)(1-2\eta+\mu_1-\mu_2)q_3(i) \big \}\nonumber  \\
&\times \frac{S_1(i)}{1\Add P_1(i)S_1(i)}\Add \big\{t(i)(1-2\eta+\mu_1\Minus \mu_2)\Add \eta \Minus \mu_1 \big\}q_3(i) \nonumber \\
&\times \frac{S_1(i)}{1+P_1(i)S_1(i)+P_2(i)S_2(i)} \Big ] + \gamma \frac{1}{N} (q_1(i)+q_3(i))  - \nu_1(i) =0 \quad \tag{\stepcounter{equation}\theequation}\\
    \frac{\partial\mathcal{L}}{\partial P_2(i)} \Equal        &-\frac{1}{N\mathrm{ln}2} \Big[ \big \{ (1\Minus\eta\Minus\mu_2)q_2(i) \Add(1\Minus t(i))(1-2\eta+\mu_1\Minus \mu_2)q_3(i) \big\} \nonumber \\
&\times\frac{S_2(i)}{1\Add P_2(i)S_2(i)} \Add \big\{t(i)(1-2\eta+\mu_1\Minus\mu_2)\Add \eta\Minus \mu_1 \big\}q_3(i) \nonumber \\
&\times\frac{S_2(i)}{1+P_1(i)S_1(i)+P_2(i)S_2(i)} \Big]+\gamma \frac{1}{N} (q_2(i)+q_3(i))  - \nu_2(i)=0 \quad\tag{\stepcounter{equation}\theequation}\\
    \frac{\partial\mathcal{L}}{\partial P_r(i)} =        &-\frac{1}{N\mathrm{ln}2} \Big[ \mu_2\left(q_4(i)+q_6(i)\right)\frac{S_1(i)}{1+P_r(i)S_1(i)}\nonumber \\ &+\mu_1\left(q_5(i)+q_6(i)\right)\frac{S_2(i)}{1+P_r(i)S_2(i)} \Big] +\gamma \frac{1}{N} (q_4(i)+q_5(i)+q_6(i))  - \nu_r(i)=0 \tag{\stepcounter{equation}\theequation}
\end{align*}
\end{subequations}
The above conditions allow the derivation of the optimal powers for each transmission mode in each time slot.
For instance, in order to determine the transmit power of user 1 in  transmission mode $\mathcal{M}_1$, we
set $q_1^*(i)=1$. From constraint $\mathrm{C3}$ in (\ref{OptProb}), we obtain that the other
selection variables are zero and therefore $q_3^*(i)=0$. Moreover, if $\mathcal{M}_1$ is selected,
then $P_1^*(i)\neq 0$ and thus from (\ref{Complementary Slackness}c), we obtain $\nu_1^*(i)=0$.
Substituting these results into (\ref{Stationary Power}a), we obtain
\begin{IEEEeqnarray}{lll}	\label{eq_11}	
		  P_1^{\mathcal{M}_1} (i) = \left[\frac{\eta-\mu_1}{\gamma \mathrm{ln}2}-\frac{1}{S_1(i)}\right]^+.
\end{IEEEeqnarray}
In a similar manner, we obtain the optimal powers for user 2 in mode $\mathcal{M}_2$, and the
optimal powers of the relay in modes $\mathcal{M}_4$ and $\mathcal{M}_5$ as follows
\begin{IEEEeqnarray}{lll}	
		  P_2^{\mathcal{M}_2} (i) = \left[\frac{1-\eta-\mu_2}{\gamma \mathrm{ln}2}-\frac{1}{S_2(i)}\right]^+ \label{P2}	\\
P_r^{\mathcal{M}_4} (i) = \left[\frac{\mu_2}{\gamma \mathrm{ln}2}-\frac{1}{S_1(i)}\right]^+ \label{P4}	 \\
P_r^{\mathcal{M}_5} (i) = \left[\frac{\mu_1}{\gamma \mathrm{ln}2}-\frac{1}{S_2(i)}\right]^+ \label{P5}	
\end{IEEEeqnarray}
In order to obtain the optimal powers of user 1 and user 2 in mode $\mathcal{M}_3$, we set $q_3^*(i)=1$.
From $\mathrm{C3}$ in (\ref{OptProb}), we obtain that the other selection variables   are zero, and
therefore $q_1^*(i)=0$ and $q_2^*(i)=0$. We  note that if one of the powers of user 2 and user 1 is zero,
mode $\mathcal{M}_3$ is identical to modes $\mathcal{M}_1$ and $\mathcal{M}_2$, respectively, and for that
case the optimal powers are already given by (\ref{eq_11}) and (\ref{P2}), respectively. For the case
when $P_1^*(i)\neq 0$ and $P_2^*(i)\neq 0$, we obtain $\nu_1^*(i)=0$ and $\nu_2^*(i)=0$  from (\ref{Complementary Slackness}c).
Furthermore, for $q_3^*(i)=1$, we will show in Appendix \ref{AppBinRelax} that   $t(i)$ can only take the boundary values, i.e.,
zero or one, and cannot be in between. Hence, if we assume $t(i)=0$, from (\ref{Stationary Power}a)
and (\ref{Stationary Power}b), we obtain
\begin{IEEEeqnarray}{lC}\label{PowerM3}
&-\frac{\eta-\mu_1}{\mathrm{ln}2} \frac{S_1(i)}{1+P_1(i)S_1(i)+P_2(i)S_2(i)} + \gamma  =0 \quad \IEEEyesnumber\IEEEyessubnumber \\
&-\frac{1}{\mathrm{ln}2} \Big[ (1-2\eta+\mu_1-\mu_2) \frac{S_2(i)}{1+P_2(i)S_2(i)} + (\eta-\mu_1) \frac{S_2(i)}{1+P_1(i)S_1(i)+P_2(i)S_2(i)} \Big]+ \gamma =0 \quad \IEEEyessubnumber
\end{IEEEeqnarray}
By substituting (\ref{PowerM3}a) in (\ref{PowerM3}b), we obtain $P_2^{\mathcal{M}_3} (i)$ and then we can derive $P_1^{\mathcal{M}_3} (i)$ from (\ref{PowerM3}a). This leads to
\begin{IEEEeqnarray}{lll}  \label{PM3-t0}
P_1^{\mathcal{M}_3} (i) =
\begin{cases}
P_1^{\mathcal{M}_1} (i),  &\mathrm{if}\,\, S_2(i) \leq \frac{S_1(i)}{\frac{1-2\eta+\mu_1-\mu_2}{\gamma \mathrm{ln}2}S_1(i)+1} \\
\left[\frac{\eta-\mu_1}{\gamma \mathrm{ln}2}-\frac{1-2\eta+\mu_1-\mu_2}{\gamma \mathrm{ln}2}\frac{1}{\frac{S_1(i)}{S_2(i)}-1}\right]^+,   \,\, &\mathrm{otherwise}
\end{cases}\IEEEyesnumber\IEEEyessubnumber \\
P_2^{\mathcal{M}_3} (i) =
\begin{cases}
P_2^{\mathcal{M}_2} (i),  &\mathrm{if}\,\, S_2(i) \geq \frac{\eta-\mu_1}{1-\eta-\mu_2}S_1(i)\\
\left[\frac{1-2\eta+\mu_1-\mu_2}{\gamma \mathrm{ln}2}\frac{1}{1-\frac{S_2(i)}{S_1(i)}} - \frac{1}{S_2(i)}\right]^+,  \,\, &\mathrm{otherwise}
\IEEEyessubnumber
\end{cases}\end{IEEEeqnarray}
Similarly, if we assume $t(i)=1$, we obtain
\begin{IEEEeqnarray}{lll} \label{PM3-t1}
P_1^{\mathcal{M}_3} (i) =
\begin{cases}
P_1^{\mathcal{M}_1} (i),  & \mathrm{if}\,\, S_2(i) \leq \frac{\eta-\mu_1}{1-\eta-\mu_2}S_1(i)\\
\left[\frac{1-2\eta+\mu_1-\mu_2}{\gamma \mathrm{ln}2}\frac{1}{\frac{S_1(i)}{S_2(i)}-1} - \frac{1}{S_1(i)}\right]^+,  \,\, &\mathrm{otherwise}
\end{cases}\IEEEyesnumber\IEEEyessubnumber \\
P_2^{\mathcal{M}_3} (i) =
\begin{cases}
P_2^{\mathcal{M}_2} (i),  &\mathrm{if}\,\, S_2(i) \geq \frac{S_1(i)}{\frac{1-2\eta+\mu_1-\mu_2}{\gamma \mathrm{ln}2}S_1(i)+1} \\
\left[\frac{1-\eta-\mu_2}{\gamma \mathrm{ln}2}-\frac{1-2\eta+\mu_1-\mu_2}{\gamma \mathrm{ln}2}\frac{1}{1-\frac{S_2(i)}{S_1(i)}}\right]^+,\,\, &\mathrm{otherwise}
\end{cases} \IEEEyessubnumber
\end{IEEEeqnarray}
We note that when $P_1^{\mathcal{M}_3} (i)=P_1^{\mathcal{M}_1} (i)$, we obtain $P_2^{\mathcal{M}_3} (i)=0$ which
means that mode $\mathcal{M}_3$ is identical to mode $\mathcal{M}_1$. Thus, there is no difference between both
modes and we select $\mathcal{M}_1$ to clarify that user 2 is silent.

For mode $\mathcal{M}_6$, we assume $q_6^*(i)=1$. From constraint $\mathrm{C3}$ in (\ref{OptProb}), we obtain
that the other selection variables are zero and therefore $q_4^*(i)=0$ and $q_5^*(i)=0$. Moreover,
if $q_6^*(i)=1$ then $P_r^*(i)\neq 0$ and thus from (\ref{Complementary Slackness}c), we obtain $\nu_r^*(i)=0$.
Using these results in (\ref{Stationary Power}c), we obtain
\begin{IEEEeqnarray}{lll}\label{PowerM6}
\mu_2\frac{S_1(i)}{1+P_r(i)S_1(i)} + \mu_1\frac{S_2(i)}{1+P_r(i)S_2(i)} = \gamma \mathrm{ln}2
\end{IEEEeqnarray}
The above equation is a quadratic equation and has two solutions for $P_r(i)$ in general. However, since $P_r(i)\geq 0$ has to hold, we can conclude that the left hand side of (\ref{PowerM6}) is monotonically decreasing in $P_r(i)$. Thus, the minimum value of the left hand side of (\ref{PowerM6}) occurs when $P_r(i)=0$ which leads to the necessary condition  $\mu_2S_1(i)+\mu_1S_2(i)>\gamma \mathrm{ln}2$, to obtain a unique positive solution for $P_r(i)$.  In particular, if  $\mu_2S_1(i)+\mu_1S_2(i)<\gamma \mathrm{ln}2$ holds, both of the two roots of (\ref{PowerM6}) are negative, and if  $\mu_2S_1(i)+\mu_1S_2(i)>\gamma \mathrm{ln}2$ holds, one of the roots of (\ref{PowerM6}) is negative and one is positive. Since the relay power has to be positive, we select the maximum of the roots of (\ref{PowerM6}) if it is positive. Thus, we obtain
\begin{IEEEeqnarray}{lll}\label{PM6}
P_r^{\mathcal{M}_6} (i) = \left [ \frac{-b+\sqrt{b^2-4ac}}{2a} \right]^+, 
\end{IEEEeqnarray}
where $a=\gamma \mathrm{ln}2 S_1(i)S_2(i), b= \gamma \mathrm{ln}2 (S_1(i)+S_2(i)) - (\mu_1+\mu_2)S_1(i)S_2(i)$, and $c=\gamma \mathrm{ln}2 - \mu_1 S_2(i) - \mu_2 S_1(i)$.

\subsection{Optimal $t^*(i)$}
To find the optimal $t(i)$, we assume $q^*_3(i)=1$ and calculate the stationary condition in (\ref{Stationary Condition}c). This leads to
\begin{IEEEeqnarray}{lll} \label{Stationary t}
    \frac{\partial\mathcal{L}}{\partial t(i)} =-\frac{1}{N}(1-2\eta+\mu_1-\mu_2) \left[ C_r(i) - C_{1r}(i) - C_{2r}(i) \right] +\phi_1(i)-\phi_0(i)=0.
\end{IEEEeqnarray}
In Appendix \ref{AppBinRelax}, we prove that $t(i)$ only takes binary values. Now, we investigate the following possible cases for $t^*(i)$:

\noindent

\noindent
\textit{Case 1:} If $t^*(i)=0$, then from (\ref{Complementary Slackness}d), we obtain $\phi_1(i)=0$ and from (\ref{Dual Feasibility Condition}d), we obtain $\phi_0(i)\geq 0$. Combining these results into (\ref{Stationary t}), the necessary condition for  $t(i)=0$ is obtained as $\eta-\mu_1\leq 1-\eta-\mu_2$.

\noindent
\textit{Case 2:} If $t^*(i)=1$, then from (\ref{Complementary Slackness}e), we obtain $\phi_0(i)=0$ and from  (\ref{Dual Feasibility Condition}d), we obtain $\phi_1(i)\geq 0$. Combining these results into (\ref{Stationary t}), the necessary condition for $t(i)=1$ is obtained as $\eta-\mu_1\geq 1-\eta-\mu_2$.

We note that if $\eta-\mu_1= 1-\eta-\mu_2$, we obtain either
$P_1^{\mathcal{M}_3}(i)=0$ or $P_2^{\mathcal{M}_3}(i)=0,\,\,\forall i$. Therefore, mode
$\mathcal{M}_3$ is not selected and the value of $t(i)$ does not affect the optimality of the solution.
Therefore, the optimal value of $t(i)$ is given by
\begin{IEEEeqnarray}{lll}
   t^*(i) = {\begin{cases} 0, & \eta-\mu_1\leq 1-\eta-\mu_2 \\
    1, & \mathrm{otherwise}
\end{cases}}
\end{IEEEeqnarray}

We note that given the Lagrange multipliers, i.e., selection weights and power weight, and the coin flip probabilities, the optimal protocol given in Theorem \ref{AdaptProt} was derived in this appendix. The optimal values of Lagrange multipliers and coin flip probabilities are obtained in Appendix \ref{AppAdapLong}. This completes the proof.


\section{Proof of Proposition \ref{AdaptProtLong}}\label{AppAdapLong}

In this appendix, we develop a framework to obtain the optimal values of Lagrange multipliers, $\mu_1,\mu_2,$ and $\gamma$, for  given channel statistics, total power budget, and $\eta$.  First, we note that, in Appendix \ref{AppMURegion}, the acceptable intervals for the optimal values of the Lagrange multipliers are derived as $\mu_1\in[0,\,\,\eta),\mu_2\in[0,\,\,1-\eta)$, and $\gamma>0$. Moreover, we have to choose $\mu_1,\mu_2,$ and $\gamma$ such that constraints $\mathrm{C1}$, $\mathrm{C2}$, and the joint long-term power constraint in (\ref{TotalPower}) hold. Therefore, we can perform a three-dimensional search over $\mu_1\in[0,\,\,\eta),\mu_2\in[0,\,\,1-\eta)$, and $\gamma>0$, and  calculate all the expectations required to check whether  constraints $\mathrm{C1}$, $\mathrm{C2}$, and the power constraint in (\ref{TotalPower}) hold or not. As will be seen later, calculating the expectations required to check  constraints $\mathrm{C1}$, $\mathrm{C2}$, and  the power constraint in (\ref{TotalPower}) depends on the values of $\mu_1$ and $\mu_2$ in each search step. In the following, we define three mutually exclusive selection regions $\mathcal{S}_0$, $\mathcal{S}_1$, and $\mathcal{S}_2$ based on the values of $\mu_1$ and $\mu_2$. Then, we present the conditions to check the optimality of a given pair of selection weights $(\mu_1,\mu_2)$ and power weight $\gamma$ in each search step. The search process can be terminated as soon as a valid point is found. 

\noindent
\textbf{Selection Region $\mathcal{S}_0$:} For this selection region, $\mu_1\neq 0$ and $\mu_2\neq 0$ must hold. Therefore, 
$\Lambda_6(i)\geq \Lambda_4(i)$ and $\Lambda_6(i)\geq \Lambda_5(i)$ hold and the inequalities hold with equality with  probability zero. To prove $\Lambda_6(i)\geq \Lambda_4(i)$, from (\ref{MetAdaptApp}), we obtain
\begin{IEEEeqnarray}{rCl}
   \Lambda_6(i)  &=& \mu_1C_{r2}(i) + \mu_2C_{r1}(i) -\gamma P_r(i) \big |_{P_r(i)=P_r^{\mathcal{M}_6}(i)} \nonumber \\
 &\overset{(a)} {\geq}& \mu_1C_{r2}(i) + \mu_2C_{r1}(i) -\gamma P_r(i)\big |_{P_r(i)=P_r^{\mathcal{M}_4}(i)} \nonumber \\
 &\overset{(b)} {\geq}& \mu_2C_{r1}(i) -\gamma P_r(i) \big |_{P_r(i)=P_r^{\mathcal{M}_4}(i)} = \Lambda_4(i),
\end{IEEEeqnarray}
where   $(a)$ follows from the fact that $P_r^{\mathcal{M}_6}(i)$ maximizes $\Lambda_6(i)$ and   $(b)$ follows from $\mu_1C_{r2}(i)\geq 0$. Both inequalities hold with equality if $S_2(i)=0$ which occurs with probability zero. A similar statement is true for $\Lambda_6(i)\geq \Lambda_5(i)$. Therefore, the necessary condition in  (\ref{OptMetAdapt}) is indeed sufficient. In this selection region, the value of $\mathcal{I}_{k}(i)$ introduced in (\ref{OptMetSufAdapt}) is deterministic as $\mathcal{I}_{k}(i)=1,\,\,k=1,2,3,6$ and $\mathcal{I}_{k}(i)=0,\,\,k=4,5$ since modes $\mathcal{M}_4$ and $\mathcal{M}_5$ cannot be selected. A set of $\mu_1,\mu_2$, and $\gamma$  is optimal in this case if  constraints $\mathrm{C1}$ and $\mathrm{C2}$ in (\ref{OptProb}), and the  power constraint in (\ref{TotalPower}) hold.

\noindent
\textbf{Selection Region $\mathcal{S}_1$:} For this selection region, $\mu_1=0$ must hold. Thus, we obtain $\Lambda_5(i)=0$ and $\Lambda_6(i)=\Lambda_4(i),\,\,\forall i$. Hence, we have
\begin{IEEEeqnarray}{rCl}
   \Lambda_1(i) &=& \eta C_{1r}(i) -\gamma P_1(i) \big |_{P_1(i)=P_1^{\mathcal{M}_1}(i)} \nonumber \\
 &\overset{(a)} {\geq}& \eta C_{1r}(i) -\gamma P_1(i)\big |_{P_1(i)=P_r^{\mathcal{M}_4}(i)} \nonumber \\
 &\overset{(b)} {\geq}& \mu_2 C_{1r}(i) -\gamma P_1(i) \big |_{P_1(i)=P_r^{\mathcal{M}_4}(i)} = \Lambda_4(i) = \Lambda_6(i),
\end{IEEEeqnarray}
where $(a)$ follows from the fact that $P_1^{\mathcal{M}_1}(i)$ maximizes $\Lambda_1(i)$ and $(b)$ holds if $\mu_2\leq \eta$ and both inequalities hold with equality with non-zero probability only if $\mu_2=\eta$. Therefore, we can conclude that if $\mu_2<\eta$, none of the transmission modes from the relay to the users is selected which  contradicts    constraints $\mathrm{C1}$ and $\mathrm{C2}$ in (\ref{OptProb}). Therefore, if $\mu_1=0$, we obtain $\mu_2\geq \eta$. Then, we consider the following cases:

\noindent
\textit{Case 1: } If $\mu_2=\eta$ holds, we obtain  $\Lambda_1(i)=\Lambda_4(i)=\Lambda_6(i),\,\,\forall i$. Then, the the necessary condition in  (\ref{OptMetAdapt}) is sufficient for selecting modes $\mathcal{M}_2,\mathcal{M}_3$, and $\mathcal{M}_5$, which leads to the deterministic values $\mathcal{I}_{1}(i)=\mathcal{I}_{3}(i)=1,$ and $\mathcal{I}_{5}(i)=0,\,\,\forall i$. However, for the case $\Lambda_1(i)=\Lambda_4(i)=\Lambda_6(i)\geq\max\{\Lambda_2(i),\Lambda_3(i)\}$, we prove in the following that a probabilistic selection of the modes $\mathcal{M}_1,\mathcal{M}_4$, and $\mathcal{M}_6$, while satisfying  constraints $\mathrm{C1}$ and $\mathrm{C2}$ in (\ref{OptProb}), is optimal.

We first note that $\bar{R}_{21}=\bar{R}_{2r}$ is fixed and does not depend on how we select between modes $\mathcal{M}_1,\mathcal{M}_4$, and $\mathcal{M}_6$. Therefore, we  first have to select between modes $\mathcal{M}_1$ and $\{\mathcal{M}_4,\mathcal{M}_6\}$ such that constraint  $\mathrm{C2}$ in (\ref{OptProb}) holds. However, since $C_{1r}(i)=C_{r1}(i),\,\,\forall i$, it does not matter in which time slots we select $\mathcal{M}_1$ and in which time slots we select $\{\mathcal{M}_4,\mathcal{M}_6\}$. Thus, we consider a simple probabilistic approach and  select modes $\{\mathcal{M}_4,\mathcal{M}_6\}$ with probability $p_1$ and mode $\mathcal{M}_1$ with probability $1-p_1$. Then, since $\bar{R}_{12}=\bar{R}_{1r}$ is fixed and does not depend on how we select between modes $\mathcal{M}_4$ and $\mathcal{M}_6$, we again select   mode $\mathcal{M}_6$ probabilistically with probability $p_2$ and mode $\mathcal{M}_4$ with probability $1-p_2$. The probabilities $p_1$ and $p_2$ are chosen such that  $\mathrm{C2}$ and $\mathrm{C1}$ in (\ref{OptProb}) hold, respectively. To implement the probabilistic strategy, we set $\mathcal{I}_1(i)=1-\mathcal{C}_1(i), \mathcal{I}_4(i)=\mathcal{C}_1(i)[1-\mathcal{C}_2(i)],$ and $\mathcal{I}_6(i)=\mathcal{C}_1(i)\mathcal{C}_2(i)$, where $\mathcal{C}_1(i)$ and $\mathcal{C}_2(i)$ are the outcomes of coin flips with probability $p_1=\frac{E\{q_2C_{2r}+q_3C_{21r}\}} {E\left\{qC_{r1}\right\}}$ where $q(i)=q_1(i)+q_4(i)+q_6(i)$ and $p_2=\frac{(1-p_1)E\{qC_{1r}\}+E\{q_3C_{12r}\}}{p_1E\{ qC_{r2}\}}$. We note that the value of $q(i)$ can be obtained from (\ref{OptMetSufAdapt}) without knowing $\mathcal{I}_k(i),\,\,k=1,4,6$. In other words, when $\Lambda_1(i)=\Lambda_4(i)=\Lambda_6(i) > \max\{\Lambda_2(i),\Lambda_3(i),\Lambda_5(i)\}$ holds for the $i$-th time slot, regardless of how we choose between modes $\mathcal{M}_1,\mathcal{M}_4$, and $\mathcal{M}_6$, the value of $q(i)$ is one. Otherwise, if $\Lambda_1(i)=\Lambda_4(i)=\Lambda_6(i) < \max\{\Lambda_2(i),\Lambda_3(i),\Lambda_5(i)\}$ holds for the $i$-th time slot, the value of $q(i)$ is zero. A set of $\mu_1,\mu_2$, and $\gamma$  is optimal in this case if  $0\leq p_1\leq 1$, $0\leq p_2\leq 1$, and the power constraint in (\ref{TotalPower}) holds.

\noindent
\textit{Case 2: } If $\mu_2>\eta$, then we obtain $\Lambda_1(i)\leq\Lambda_4(i)=\Lambda_6(i),\,\,\forall i$. Since the inequality holds with equality with probability zero, mode $\mathcal{M}_1$ is not selected. We note that constraint $\bar{R}_{2r}=\bar{R}_{r1}$ does not depend on how we select between modes $\mathcal{M}_4$ and $\mathcal{M}_6$, and only depends on $\mu_2$. Therefore, for a given $\mu_2$ in the search, we check whether constraint $\mathrm{C2}$ in (\ref{OptProb}) holds or not. If constraint $\mathrm{C2}$ in (\ref{OptProb}) holds, the value of  $\bar{R}_{1r}$ is fixed and consequently the cost function is fixed. Therefore, we should select between modes $\mathcal{M}_4$ and $\mathcal{M}_6$ such that constraint $\mathrm{C1}$ in (\ref{OptProb}) holds. Similar to Case 1, we consider a probabilistic approach which leads to $\mathcal{I}_4(i)=1-\mathcal{C}_2(i),$ and $\mathcal{I}_6(i)=\mathcal{C}_2(i)$, where $p_2=\frac{E\{q_3C_{12r}\}}{E\{ qC_{r2}\}}$ with $q(i)=q_4(i)+q_6(i)$. A set of $\mu_1,\mu_2$, and $\gamma$   is optimal in this case if  constraint $\mathrm{C2}$ in (\ref{OptProb}), $0\leq p_2\leq 1$, and the power constraint in (\ref{TotalPower}) hold.

\noindent
\textbf{Selection Region $\mathcal{S}_2$:} For this selection region, $\mu_2 = 0$ must hold.
Following a similar procedure as for selection region $\mathcal{S}_1$, in this selection region, we obtain that $\mu_1\geq 1-\eta$. Then, we have to consider the two cases $\mu_1= 1-\eta$ and $\mu_1>1-\eta$ which leads to the optimal mode selection policy   given in Theorem \ref{AdaptProt}. We skip the  derivation since it is quite similar to the one for selection region $\mathcal{S}_1$.

To summarize, the optimal values of the selection weights and the power weight are obtained by a three-dimensional search over $\mu_1\in[0,\,\,\eta),\mu_2\in[0,\,\,1-\eta)$, and $\gamma>0$, respectively.  Moreover, for a set of $\mu_1,\mu_2$, and $\gamma$ in each search step, we obtain the optimal values of the coin flip probabilities and the conditions to check whether the considered point in each search step is optimal or not.  The search process is terminated as soon as a valid point is found. This completes the proof.

\section{Proof of Theorem \ref{FixProt}}
\label{AppKKTFix}

The proof of the protocol in Theorem \ref{FixProt} is similar to the proof of Theorem \ref{AdaptProt} given in Appendix \ref{AppKKTAdapt}.  Herein, due to space constraints, we only highlight the differences and avoid repeating the entire procedure. 
Similar to Appendix \ref{AppKKTAdapt}, we first consider the Lagrangian function corresponding to the standard minimization problem equivalent to (\ref{OptProb}) under the fixed transmit power constraint. This leads to a  function similar to the one given in (\ref{KKTFunction}) excluding the terms which correspond to the  power constraints. We can now investigate the KKT conditions to obtain the optimal solution. Moreover, in Appendix \ref{AppMURegion}, we find the intervals for the optimal values of the selection weights as $\mu_1\in [0,\,\,\eta]$ and $\mu_2 \in [0\,\,1-\eta]$. We note that unlike the optimization problem under the joint long-term power constraint, the optimal values of the selection weights can also take the values $\mu_1=\eta$ and $\mu_2=1-\eta$ under the fixed per-user power constraint.  In the following subsections, we find the optimal values of $q_k^*(i),$ and $t^*(i),\,\,\forall i,k$ for  given values of selection weights $\mu_1$ and $\mu_2$. The optimal values of the selection weights are obtained in Appendix \ref{AppFixLong}.

\subsection{Optimal $q_k^*(i)$}

In order to determine the optimal selection policy, we differentiate the Lagrangian function with respect to   $q_k(i)$, for $k=1,...,6$, equate the result to zero, and obtain the corresponding $q_k(i)$. Using the same procedure as in Appendix \ref{AppKKTAdapt}, we obtain the necessary condition for selecting transmission mode $\mathcal{M}_{k^*}$ in the $i$-th time slot as 
\begin{IEEEeqnarray}{lll}\label{OptMetFix}
  \Lambda_{k}(i) > {\underset{k\in\{1,\cdots,6\}}{\max}}\{\mathcal{I}_{k}(i)\Lambda_{k}(i)\} 
\end{IEEEeqnarray}
where 
\begin{IEEEeqnarray}{lll}\label{MetFix}
    \Lambda_1(i) = (\eta-\mu_1)C_{1r}(i)  \IEEEyesnumber\IEEEyessubnumber  \\
    \Lambda_2(i) =  (1-\eta-\mu_2)C_{2r}(i) \IEEEyessubnumber  \\
    \Lambda_3(i) = (\eta-\mu_1)C_{12r}(i)+(1-\eta-\mu_2)C_{21r}(i)\IEEEyessubnumber \\
    \Lambda_4(i) = \mu_2 C_{r1}(i) \IEEEyessubnumber\\
    \Lambda_5(i) = \mu_1 C_{r2}(i)\IEEEyessubnumber\\
    \Lambda_6(i) = \mu_1 C_{r2}(i)+\mu_2 C_{r1}(i)\IEEEyessubnumber
\end{IEEEeqnarray}
and $\mathcal{I}_{k}(i)\in\{0,1\}$ is a binary indicator variable which selects between the modes with identical values of selection metrics.  In Appendix \ref{AppFixLong}, we  prove that  $\mathcal{I}_{k}(i),\,\,k=1,\dots,6$ are obtained as follows
\begin{IEEEeqnarray}{lll}
     [\mathcal{I}_1(i),\dots,\mathcal{I}_6(i)]  \nonumber \\ =  \big[[1\Minus \mathcal{C}_1(i)]a(i),\,\,[1\Minus \mathcal{C}_2(i)]a(i),\,\,\mathcal{C}_1(i)\mathcal{C}_2(i)a(i),\,\,[1\Minus \mathcal{C}_3(i)]b(i),\,\,[1\Minus \mathcal{C}_4(i)]b(i),\,\,\mathcal{C}_3(i)\mathcal{C}_4(i)b(i)\big] \quad\,\,
\end{IEEEeqnarray}
where $a(i)=\mathcal{C}_5(i)$ and $b(i)=1-\mathcal{C}_5(i)$ if $\{\mu_1=\eta \,\,\wedge\,\,\mu_2=0 \,\,\wedge\,\,P_2=P_r\,\,\wedge\,\,\eta=\frac{1}{2}\}\,\vee\,\{\mu_1=0 \,\,\wedge\,\,\mu_2=1-\eta \,\,\wedge\,\,P_1=P_r\,\,\wedge\,\,\eta=\frac{1}{2}\}$ holds, otherwise, $a(i)=1$ and $b(i)=1,\,\,\forall i$. The coin flip probabilities $p_1,\dots,p_5$ are long-term variables and depend only on the channel statistics, powers of the nodes, and the value of $\eta$ and are given in Appendix \ref{AppFixLong}.

\subsection{Optimal $t^*(i)$}
To obtain the optimal $t^*(i)$ for the multiple-access mode, we assume $q^*_3(i)=1$ and calculate the derivative in (\ref{Stationary Condition}c) which leads to (\ref{Stationary t}). Now, we investigate the following possible cases for $t^*(i)$:

\noindent
\textit{Case 1:} If $t^*(i)=0$, then from (\ref{Complementary Slackness}d), we obtain $\phi_1(i)=0$ and from (\ref{Dual Feasibility Condition}d), we obtain $\phi_0(i)\geq 0$. Combining these results with (\ref{Stationary t}), we obtain $\eta-\mu_1\leq 1-\eta-\mu_2$.

\noindent
\textit{Case 2:} If $t^*(i)=1$, then considering (\ref{Complementary Slackness}e),  we obtain $\phi_0(i)=0$ and from (\ref{Dual Feasibility Condition}d), we obtain $\phi_1(i)\geq 0$. Combining these results with (\ref{Stationary t}), we obtain  $\eta-\mu_1\geq 1-\eta-\mu_2$.

Moreover, in Appendix \ref{AppFixLong}, we prove that if $\eta-\mu_1= 1-\eta-\mu_2$ holds, the decoding order variable can be obtained by a coin flip. Thus, we can write the decoding order variable in a compact form as $t^*(i)=\mathcal{C}_6(i)$ where we obtain $p_6=0$ if $\eta-\mu_1 < 1-\eta-\mu_2$, $p_6=1$ if $\eta-\mu_1 > 1-\eta-\mu_2$, and $0<p_6<1$ if $\eta-\mu_1 = 1-\eta-\mu_2$. We note that given the selection weights and the coin flip probabilities, the optimal protocol given in Theorem \ref{FixProt} is derived in this appendix. The optimal values of the Lagrange multipliers and coin flip probabilities are obtained in Appendix \ref{AppFixLong}. This completes the proof.


\section{Proof of Proposition \ref{FixProtLong}}\label{AppFixLong}

In this appendix, we develop a similar framework as in Appendix \ref{AppAdapLong} to obtain the optimal values of the selection weights for  given channel statistics, powers of the nodes, and $\eta$.  First, we note that, in Appendix \ref{AppMURegion}, the feasible intervals for the optimal values of the Lagrange multipliers are derived as $\mu_1\in[0,\,\,\eta]$ and $\mu_2\in[0,\,\,1-\eta]$. Moreover, we have to choose $\mu_1$ and $\mu_2$ such that constraints $\mathrm{C1}$ and $\mathrm{C2}$ in (\ref{OptProb}) hold. Therefore, we can perform a two-dimensional search over $\mu_1\in[0,\,\,\eta]$ and $\mu_2\in[0,\,\,1-\eta)$ and  calculate all the expectations required to check whether  constraints $\mathrm{C1}$, $\mathrm{C2}$ in (\ref{OptProb}) hold or not. As will be seen later, the calculation of the expectations required to check  constraints $\mathrm{C1}$ and $\mathrm{C2}$ depends on the values of $\mu_1$ and $\mu_2$ in each search step. In the following, we define three mutually exclusive selection regions $\mathcal{S}_0$, $\mathcal{S}_1$, and $\mathcal{S}_2$ based on the values of $\mu_1$ and $\mu_2$. Then, we present the conditions to check the optimality of a given pair of  selection weights $(\mu_1,\mu_2)$ in each search step. The search process can be terminated as soon as a valid point is found. 

\noindent
\textbf{Selection Region $\mathcal{S}_0$:} For this selection region $\eta-\mu_1=1-\eta-\mu_2$ has to hold. We note that in Appendix \ref{AppBinRelax}, we prove that if $\eta-\mu_1\neq 1-\eta-\mu_2$ holds, the value of $t(i)$ has to be binary. However, assuming $\eta-\mu_1 = 1-\eta-\mu_2$, we show in the following that  binary values of $t(i)$ achieve the same maximum value of the cost function in (\ref{OptProb}) as non-binary values of $t(i)$ do. To this end, assume $0<t^*(i)<1$, then from (\ref{Complementary Slackness}d) and (\ref{Complementary Slackness}e), we obtain $\phi_l(i)=0,$ for $l=0,1$. Considering that $\eta-\mu_1=1-\eta-\mu_2$ holds and by combining these results into (\ref{MetFix}), we obtain
\begin{IEEEeqnarray}{lll} \label{MU12}
    \Lambda_3(i) = (\eta-\mu_1)C_r(i)= (1-\eta-\mu_2)C_r(i)  \geq \max \{\Lambda_1(i),\Lambda_2(i)\} \IEEEyesnumber\IEEEyessubnumber \\
\Lambda_6(i) = \mu_2 C_{r1}(i) + \mu_1 C_{r2}(i)  \geq \max \{\Lambda_4(i),\Lambda_5(i)\}. \IEEEeqnarraynumspace \IEEEyessubnumber
\end{IEEEeqnarray}
The inequality in (\ref{MU12}a) holds with equality with non-zero probability only if  $\{\mu_1=\eta\,\,\wedge\,\,\mu_2=1-\eta\}$ which leads to a contradiction as shown in Appendix \ref{AppMURegion}. Therefore, from (\ref{OptMetAdapt}), modes $\mathcal{M}_1$ and $\mathcal{M}_2$ cannot be selected for the optimal strategy. Hence, we set $\mathcal{I}_k(i)=0, \,\, k=1,2, \forall i$. However, the inequality in (\ref{MU12}b) holds with equality with non-zero probability if  $\{\mu_1=0\,\,\vee\,\,\mu_2=0\}$ which is possible as long as both $\mu_1$ and $\mu_2$ are not zero simultaneously.  In the following, we consider the possible cases based on whether $\mu_1$ and $\mu_2$ are zero or not.

\noindent
\textit{Case 1:}
If $\mu_1\neq 0$ and $\mu_2\neq 0$, then only modes $\mathcal{M}_3$ and $\mathcal{M}_6$ can  be selected for the optimal strategy, i.e., $\mathcal{I}_k(i)=0, \,\, k=4,5,$ and $\mathcal{I}_k(i)=1, \,\, k=3,6, \forall i$.  Note that when  $\mathrm{C1}$ and $\mathrm{C2}$ in (\ref{OptProb}) hold, the following also holds
\begin{IEEEeqnarray}{rCl}\label{EquCons}
    \bar{R}_{1r}+\bar{R}_{2r}&=&\bar{R}_{r1}+\bar{R}_{r2}. \IEEEyesnumber
\end{IEEEeqnarray}
We note that since only modes $\mathcal{M}_3$ and $\mathcal{M}_6$ are selected and $\eta-\mu_1=1-\eta-\mu_2$ holds, we obtain $\Lambda_3(i)=(\eta-\mu_1)C_r(i)$ and $R_{1r}(i)+R_{2r}(i)=C_r(i)$ which are independent of the value of $t(i)$. Therefore, we can check whether the sum rate constraint in (\ref{EquCons}) holds independent of knowing the optimal value of $t(i)$. 
Moreover, for given $\mu_1$ and $\mu_2$, one can determine the values of $\bar{R}_{r1}$ and $\bar{R}_{r2}$ regardless of $t(i)$.   Furthermore, due to constraints $\mathrm{C1}$ and $\mathrm{C2}$ in (\ref{OptProb}), the values of  $\bar{R}_{1r}$ and $\bar{R}_{2r}$ have to be identical to $\bar{R}_{r2}$  and $\bar{R}_{r1}$, respectively. Consequently,  cost function $\eta\bar{R}_{1r}+(1-\eta)\bar{R}_{2r}$ is given and it is independent of $t(i)$, and any choice of $t(i)$, $\forall i$, that satisfies $\mathrm{C1}$ and $\mathrm{C2}$ in (\ref{OptProb}) is optimal. Moreover, assuming (\ref{EquCons})  holds, if one of the constraints $\mathrm{C1}$ and $\mathrm{C2}$ in (\ref{OptProb}) holds, the other one will hold as well. Therefore, as a simple solution, we can assume a fixed non-binary $t(i)$ for $\forall i$ such that $\mathrm{C1}$ is satisfied. However, the maximum weighted sum rate  can be also achieved with binary values of $t(i)$ by probabilistically selecting $t(i)=0$ and $t(i)=1$ such that $\mathrm{C1}$ is satisfied. Therefore, we set $t(i)=\mathcal{C}_6(i)$ where $p_6=\frac{E\{q_3(C_r-C_{2r})\}-E\{q_6C_{r2}\}}{E\left\{q_3 (C_r-C_{1r}-C_{2r})\right\}}$ is chosen to satisfy $\mathrm{C1}$ in (\ref{OptProb}). A set of $\mu_1$ and $\mu_2$   is optimal in this case if  the sum rate constraint in (\ref{EquCons}) and $0\leq p_6\leq 1$ hold.

\noindent
\textit{Case 2:}
For this case, $\mu_1 = 0$ or $\mu_2 = 0$ have to hold. We prove in Appendix \ref{AppMURegion} that $\mu_1$ and $\mu_2$ cannot be simultaneously zero. Thus, the following two cases are possible.

\noindent
\textbf{a)} If $\mu_1=0$ holds, we obtain $\mu_2=1-2\eta$ because of the constraint for selection region $\mathcal{S}_0$, i.e., $\eta-\mu_1=1-\eta-\mu_2$. Then, only modes $\mathcal{M}_3,\mathcal{M}_4,$ and $\mathcal{M}_6$ can  be selected in the optimal strategy. Similar to Case 1 a), the selection between mode $\mathcal{M}_3$ and $\{\mathcal{M}_4,\mathcal{M}_6\}$ does not depend on $t(i)$. Moreover, regardless of how we choose between $\mathcal{M}_4$ and $\mathcal{M}_6$,  rate $\bar{R}_{r1}$ is fixed since both modes have the same relay-to-user 1 channel capacity, i.e., $C_{r1}(i)$. Therefore, the optimal $t(i)$ has to satisfy constraint $\mathrm{C2}$ in (\ref{OptProb}), i.e., $\bar{R}_{2r}=\bar{R}_{r1}$. However, since the weighted sum rate $\eta\bar{R}_{1r}+\eta\bar{R}_{2r}=\eta E\{q_3C_r(i)\}$ does not depend on the values of $t(i)$, any choice of $t(i)$ that satisfies constraint $\mathrm{C2}$ in (\ref{OptProb}) is optimal.  Similar to Case 1 a), we adopt a probabilistic approach and choose $t(i)=\mathcal{C}_6(i)$,  $\mathcal{I}_6(i)=\mathcal{C}_3(i)$, and $\mathcal{I}_4(i)=1-\mathcal{C}_3(i)$, where $p_6 = \frac{E\{qC_{r1}\} - E\{q_3 C_{2r}\}}{E\{q_3(C_r-C_{1r}-C_{2r})\}}$ and $p_3=\frac{E\{q_3C_{12r}\}}{E\{qC_{r2}\}}$ with $q(i)=q_4(i)+q_6(i)$. The considered values of $\mu_1$ and $\mu_2$  are optimal in this case if  constraint $\mathrm{C2}$ in (\ref{OptProb}) and $0\leq p_6\leq 1$ hold.

\noindent
\textbf{b)} If $\mu_2=0$ holds, we obtain $\mu_1=2\eta-1$ because of the constraint for selection region $\mathcal{S}_0$, i.e., $\eta-\mu_1=1-\eta-\mu_2$. Then, only modes $\mathcal{M}_3,\mathcal{M}_5,$ and $\mathcal{M}_6$ can  be selected for the optimal strategy. With a similar reasoning as in Case 1 b), we obtain $t(i)=\mathcal{C}_6(i)$, $\mathcal{I}_6(i)=\mathcal{C}_4(i)$, and $\mathcal{I}_5(i)=1-\mathcal{C}_4(i)$, where $p_6 = \frac{E\{q_3(C_r-C_{2r})\} - E\{q C_{r2}\}}{E\{q_3(C_r-C_{1r}-C_{2r})\}}$, $p_4=\frac{E\{q_3C_{21r}\}}{E\{qC_{r1}\}}$, and $q(i)=q_5(i)+q_6(i)$. The considered values of $\mu_1$ and $\mu_2$  are optimal in this case if  constraint $\mathrm{C1}$ in (\ref{OptProb}) and $0\leq p_6\leq 1$ hold.

\noindent
\textbf{Selection Region $\mathcal{S}_1$:} For this selection region, $\eta-\mu_1 < 1-\eta-\mu_2$ has to hold which leads to $t(i)=0,\,\,\forall i$. Then, from (\ref{MetFix}), we obtain
\begin{IEEEeqnarray}{rll}\label{M25}
    \Lambda_3(i) &=& (\eta-\mu_1)[C_{r}(i)-C_{2r}(i)]+(1-\eta-\mu_2)C_{2r}(i) \nonumber \\
     &=&(\eta-\mu_1)C_{r}(i)+ (1-2\eta+\mu_1-\mu_2)C_{2r}(i) \nonumber \\ &\geq& \max \{\Lambda_1(i),\Lambda_2(i)\}  \IEEEyesnumber \IEEEyessubnumber \\
\Lambda_6(i) &=& \mu_2 C_{r1}(i) + \mu_1 C_{r2}(i)\geq \max \{\Lambda_4(i),\Lambda_5(i)\}. \IEEEeqnarraynumspace \IEEEyessubnumber
\end{IEEEeqnarray}
The expressions in (\ref{M25}a) and  (\ref{M25}b) hold with equality with non-zero probability if and only if $\mu_1=0,\eta$ and $\mu_2=0$, respectively, otherwise the inequality holds. In the following, we consider the possible cases based on whether $\mu_1$ and $\mu_2$ are at the boundaries of the feasible intervals or not.

\noindent
\textit{Case 1:} If $\mu_1 \neq 0,\eta$ and $ \mu_2 \neq 0, 1-\eta$ hold, then  (\ref{M25}) holds with inequality and therefore, only $\mathcal{M}_3$ and $\mathcal{M}_6$ are selected. Hence, for this case, we can set $\mathcal{I}_k(i)=0,\,\, k=1,2,4,5$ and $\mathcal{I}_k(i)=1,\,\,k=3,6$ for $\forall i$. A set of $\mu_1$ and $\mu_2$   is optimal in this case if  constraints $\mathrm{C1}$ and $\mathrm{C2}$ in (\ref{OptProb}) hold.

\noindent
\textit{Case 2:} For this case, $\mu_1 = 0,\eta$ or $ \mu_2 = 0, 1-\eta$ have to hold. Since for selection region $\mathcal{S}_1$, $\eta-\mu_1\leq 1-\eta-\mu_2$ has to hold, the following four combinations of $\mu_1$ and $\mu_2$ are possible.

\noindent
\textbf{a)} If $\mu_1 = \eta$ and $ \mu_2\neq 0,1-\eta$ hold, then from (\ref{MetFix}), we obtain 
\begin{IEEEeqnarray}{rll}
    \Lambda_2(i) &=& \Lambda_3(i) =(1-\eta-\mu_2)C_{2r}(i)  \geq \Lambda_1(i) \IEEEyesnumber\IEEEyessubnumber \\
\Lambda_6(i) &=& \mu_2 C_{r1}(i)+\eta C_{r2}(i)\geq \max \{\Lambda_4(i),\Lambda_5(i)\}. \IEEEyessubnumber
\end{IEEEeqnarray}
The probability that the above expressions hold with equality is zero. Therefore, we can set  $\mathcal{I}_k(i)=0,\,\,k=1,4,5$ and $\mathcal{I}_6(i)=1$, $\forall i$. Moreover, we observe that constraint $\mathrm{C2}$ in (\ref{OptProb}), i.e., $\bar{R}_{2r}=\bar{R}_{r1}$, only depends on $\mu_2$. Therefore, without any knowledge about how we select between modes $\mathcal{M}_2$ and $\mathcal{M}_3$, we are able to check whether constraint $\mathrm{C2}$ in (\ref{OptProb}) holds or not. In the following, we show that the maximum weighted sum rate is constant and does not depend on how we select between $\mathcal{M}_2$ and $\mathcal{M}_3$. In particular, for given selection weights, one can determine the values of $\bar{R}_{r1}$ and $\bar{R}_{r2}$ regardless of how the choice between modes $\mathcal{M}_2$ and $\mathcal{M}_3$ is made if $\Lambda_2(i)=\Lambda_3(i)>\Lambda_6(i)$.
Furthermore, due to constraints $\mathrm{C1}$ and $\mathrm{C2}$ in (\ref{OptProb}), the optimal values of  $\bar{R}_{1r}$ and $\bar{R}_{2r}$ are  determined  when the optimal values of $\bar{R}_{r1}$ and $\bar{R}_{r2}$ are known. Consequently, the optimal value of the cost function $\eta\bar{R}_{1r}+(1-\eta)\bar{R}_{2r}$ is determined, and thus any method of choosing between modes $\mathcal{M}_2$ and $\mathcal{M}_3$ that satisfies $\mathrm{C1}$ in (\ref{OptProb}) is optimal. Thus, we adopt again a simple probabilistic approach and choose ${\cal M}_3$ with probability $p_1$ and ${\cal M}_2$ with probability $1-p_1$. To implement the probabilistic strategy, we set $\mathcal{I}_3(i)=1-\mathcal{I}_2(i)=\mathcal{C}_2(i)$, where $\mathcal{C}_2(i)$ is the outcome of a coin flip with probability $p_2=\frac{E\{q_6C_{r2}\}}{E\left\{q[C_r-C_{2r}]\right\}}$ and $q(i)=q_2(i)+q_3(i)$. A set of the values of $\mu_1$ and $\mu_2$   is optimal in this case if  constraint  $\mathrm{C2}$ in (\ref{OptProb}) and $0\leq p_1 \leq 1$ hold.

\noindent
\textbf{b)} If $ \mu_1 \neq 0,\eta$ and $\mu_2=0$ hold, then from (\ref{MetFix}), we obtain 
\begin{IEEEeqnarray}{rll}
    \Lambda_3(i) &=& (\eta-\mu_1)[C_r(i)-C_{2r}(i)]+(1-\eta)C_{2r}(i) \nonumber \\ 
&\geq& \max \{\Lambda_1(i),\Lambda_2(i)\} \IEEEyesnumber\IEEEyessubnumber \\
\Lambda_5(i) &=& \Lambda_6(i) = \mu_1 C_{r2}(i) \geq \Lambda_4(i).  \IEEEyessubnumber
\end{IEEEeqnarray}
Similar to Case 2 b), the probability that the above expressions hold with equality is zero. Therefore, we set $\mathcal{I}_k(i)=0,\,\,k=1,2,4,\mathcal{I}_3(i)=1$, and $\mathcal{I}_6(i)=1-\mathcal{I}_5(i)=\mathcal{C}_4(i)$, where $p_4=\frac{E\{q_3C_{2r}\}}{E\left\{qC_{r1}\right\}}$ with $q(i)=q_5(i)+q_6(i)$. A set of $\mu_1$ and $\mu_2$   is optimal in this case if  constraint  $\mathrm{C1}$ in (\ref{OptProb}) and $0\leq p_4 \leq 1$ hold.

\noindent
\textbf{c)} If $\mu_1 = \eta$ and $\mu_2=0$ hold, then from (\ref{MetFix}), we obtain 
\begin{IEEEeqnarray}{rll}
    \Lambda_2(i) &=& \Lambda_3(i) = (1-\eta)C_{2r}(i)  \geq \Lambda_1(i)  \IEEEyesnumber\IEEEyessubnumber \\
\Lambda_5(i) &=& \Lambda_6(i) = \eta C_{r2}(i)  \geq \Lambda_4(i).  \IEEEyessubnumber
\end{IEEEeqnarray}
The probability that the above expressions hold with equality is zero and therefore, we can set $\mathcal{I}_k(i)=0, \,\, k=1,4$. We note that if $\eta=\frac{1}{2}$ and $P_2=P_r$ hold, we also obtain $\Lambda_2(i)=\Lambda_3(i)=\Lambda_5(i)=\Lambda_6(i),\,\,\forall i$. Therefore, we distinguish the following possible cases. If $\eta \neq \frac{1}{2}$ or $P_2\neq P_r$, it is easy to see that regardless of how we select between ${\cal M}_2$ and ${\cal M}_3$, rate $\bar{R}_{2r}$ is constant and similarly, regardless of how we select between ${\cal M}_5$ and ${\cal M}_6$, rate $\bar{R}_{r2}$ is constant. Therefore, the cost function is fixed and we can use the probabilistic approach to satisfy the constraints. Thus, we set $\mathcal{I}_{3}(i)=1-\mathcal{I}_2(i)=\mathcal{C}_2(i)$ and $\mathcal{I}_{6}(i)=1-\mathcal{I}_5(i)=\mathcal{C}_4(i)$ where $p_2=\frac{E\{(1-q)C_{r2}\}}{E\{q(C_r-C_{2r})\}}$ and $p_4=\frac{E\{qC_{2r}\}}{E\{(1-q)C_{r1}\}}$ with $q(i)=q_2(i)+q_3(i)$. The considered values of $\mu_1$ and $\mu_2$  are  optimal in this case  if   $0\leq p_1\leq 1$ and $0\leq p_4 \leq 1$ hold when $\eta \neq \frac{1}{2}$ or $P_2\neq P_r$.

On the other hand, if $\eta=\frac{1}{2}$ and $P_2=P_r$, assuming that $\mathrm{C1}$ and $\mathrm{C2}$ in (\ref{OptProb}) hold, the cost function is $\eta \bar{R}_{r2}+(1-\eta)\bar{R}_{2r}= \eta E\{(q_5+q_6)C_{r2}\}+(1-\eta)E\{(q_2+q_3)C_{2r}\}=\frac{1}{2}E\{(q_2+q_3+q_5+q_6)C_{2r}\}=\frac{1}{2}E\{C_{2r}\}$ where we exploit $C_{2r}(i)=C_{r2}(i)$. Moreover, the maximum of the cost function is achieved by a probabilistic approach with two coin flips as follows
\begin{IEEEeqnarray}{rll}
    {\begin{cases}
\mathcal{I}_2(i)= \mathcal{C}_5(i) [1-\mathcal{C}_2(i)]\\
\mathcal{I}_3(i)= \mathcal{C}_5(i) \mathcal{C}_2(i) \\
\mathcal{I}_5(i)=[1-\mathcal{C}_5(i)] [1-\mathcal{C}_4(i)]\\
\mathcal{I}_6(i)=[1-\mathcal{C}_5(i)] \mathcal{C}_4(i)
\end{cases}} \IEEEyesnumber
\end{IEEEeqnarray}
where given $\omega_l= \frac{E\{C_{r2}\}}{E\{C_{r}\}}$ and $\omega_u=\frac{E\{C_{r1}\}}{E\{C_{r1}+C_{2r}\}}$, the coin flip probabilities $p_2=\frac{1-p_5}{p_5} \frac{\omega_l}{1-\omega_l}$ and $p_4=\frac{p_5}{1-p_5} \frac{1-\omega_u}{\omega_u}$ are chosen such that constraints  $\mathrm{C1}$ and $\mathrm{C2}$ in (\ref{OptProb}) are satisfied, respectively, and probability $p_5$ is chosen to maximize the cost function in (\ref{OptProb}). However, any value of $p_5$ which results in a feasible value of $p_2$ and $p_4$, i.e., $0\leq p_2,p_4\leq 1$, results in the  maximum possible value for the cost function. This leads to $\omega_l\leq p_5\leq \omega_u$.  The considered values of $\mu_1$ and $\mu_2$  are optimal in this case if  $0\leq p_1\leq 1$, $0\leq p_4 \leq 1$, and $0\leq p_5\leq 1$ hold when $\eta=\frac{1}{2}$ and $P_2=P_r$.

\noindent
\textbf{d)} If $\mu_1=0$ and $0<\mu_2<1-\eta$ hold, then only modes $\mathcal{M}_3,\mathcal{M}_4,$ and $\mathcal{M}_6$ can  be selected for the optimal strategy. Since, for a given $\mu_2$, regardless of how we choose between modes $\mathcal{M}_4$ and $\mathcal{M}_6$,  rate $\bar{R}_{r1}$ is fixed, we are able to check whether constraint $\mathrm{C2}$ in (\ref{OptProb}) holds or not. Then, if constraint $\mathrm{C2}$ in (\ref{OptProb}) holds, rate $\bar{R}_{1r}$ is fixed and consequently, the cost function is fixed, and any method of selecting between  modes $\mathcal{M}_4$ and $\mathcal{M}_6$ that satisfies $\mathrm{C1}$ in (\ref{OptProb}) is optimal. Thus, we adopt a the probabilistic approach and set $\mathcal{I}_6(i)=1-\mathcal{I}_4(i)=\mathcal{C}_3(i)$ with  $p_3=\frac{E\{q_3C_{1r}\}}{E\{qC_{r2}\}}$ and $q(i)=q_4(i)+q_6(i)$. A set of $\mu_1$ and $\mu_2$   is optimal in this case if  constraint  $\mathrm{C2}$ in (\ref{OptProb}) and $0\leq p_3 \leq 1$ hold.

\noindent
\textbf{Selection region $\mathcal{S}_2$:} For this selection region, $\eta-\mu_1 > 1-\eta-\mu_2$ has to hold which leads to $t(i)=1,\,\,\forall i$.  Following a similar procedure as for selection region $\mathcal{S}_1$, we obtain the results provided in Proposition \ref{FixProtLong}.

To summarize, the optimal values of the selection weights are obtained by a two-dimensional search over $\mu_1\in[0,\,\,\eta],$ and $\mu_2\in[0,\,\,1-\eta]$, respectively.  Moreover, for a set of $\mu_1$ and $\mu_2$ in each search step, we obtain the optimal values of the coin flip probabilities and the conditions to check whether the considered point is optimal or not.  The search process is terminated as soon as a valid point is found. This completes the proof.



\section{Proof of Optimality of Binary Relaxation}
\label{AppBinRelax}

In this appendix, we prove that the optimal solution of the problem with the relaxed constraints,
$0\leq  q_k(i)\leq 1$ and $0\leq t(i) \leq 1$, can be achieved by the boundary values of $q_k(i)$ and $t(i)$, i.e., 0 or 1.
Therefore, the binary relaxation does not change the solution of the problem. 

\subsection{Binary Relaxation of $q_k(i)$} 
We use the contradiction method for the proof. If one of the $q_k(i),\,\,k=1,\dots,6$,
assumes a non-binary value in the optimal solution, then in order to satisfy constraint $\mathrm{C3}$ in (\ref{OptProb}), there has to be at least one other non-binary selection variable in that time slot.
Assuming that the mode indices of the non-binary selection variables are $k'$ and $k''$ in the $i$-th time slot,
we obtain $\alpha_k(i)=0,\,\,k = 1,\dots,6$, from (\ref{Complementary Slackness}a), and $ \beta_{k'}(i)=0$ and
$ \beta_{k''}(i)=0$  from (\ref{Complementary Slackness}b). Then, by substituting these values into (\ref{Stationary Mode}),
we obtain
\begin{IEEEeqnarray}{lll}\label{BinRelax}
    \lambda(i) = \Lambda_{k'}(i)  \IEEEyesnumber\IEEEyessubnumber  \\
    \lambda(i)= \Lambda_{k''}(i)\IEEEyessubnumber  \\
   \lambda(i)-\beta_k(i) =  \Lambda_k(i), \quad k\neq k', k''. \IEEEyessubnumber
\end{IEEEeqnarray}
From (\ref{BinRelax}a) and  (\ref{BinRelax}b), we obtain $\Lambda_{k'}(i)=\Lambda_{k''}(i)$ and by subtracting (\ref{BinRelax}a) and  (\ref{BinRelax}b) from (\ref{BinRelax}c), we obtain
\begin{IEEEeqnarray}{rCl}
    \Lambda_{k'}(i) - \Lambda_k(i) &=& \beta_k(i), \quad \quad k\neq k', k'' \IEEEyesnumber \IEEEyessubnumber  \\
\Lambda_{k''}(i) - \Lambda_k(i) &=& \beta_k(i), \quad \quad k\neq k', k''. \IEEEyessubnumber
\end{IEEEeqnarray}
From the dual feasibility condition given in (\ref{Dual Feasibility Condition}b), we have $\beta_k(i)\geq 0$ which leads to $\Lambda_{k'}(i)=\Lambda_{k''}(i)\geq \Lambda_k(i)$.
However, because of the continuous probability density functions of the channels gains,  $\Pr\{\Lambda_{k'}(i)=\Lambda_{k''}(i)\} > 0$
holds for some transmission modes $\mathcal{M}_{k'}$ and $\mathcal{M}_{k''}$, if and only if we have
$\mu_1=0,\eta$ and $\mu_2=0,1-\eta$. However, in Appendices \ref{AppAdapLong} and \ref{AppFixLong}, we have proved that the optimal solution of the optimization problem for $\mu_1=0,\eta$ or $\mu_2=0,1-\eta$ can be achieved via a probabilistic approach, i.e., with binary values for $q_k(i)$. Therefore, binary $q_k(i)$ are optimal for all time slots.

\subsection{Binary Relaxation of $t(i)$} 

Similarly, if we assume $0<t(i)<1$, then from (\ref{Complementary Slackness}d) and (\ref{Complementary Slackness}e), we have $\phi_l(i)=0, \,\, l=0,1$. Therefore, from (\ref{Stationary t}) and $C_r(i) - C_{1r}(i) - C_{2r}(i)\leq 0$, we obtain $\eta-\mu_1=1-\eta-\mu_2$. From here, the proof of the optimality of the binary relaxation is different for fixed per-node power constraint and joint long-term power constraint. In particular, assuming $\eta-\mu_1=1-\eta-\mu_2$, we have proved in Appendix \ref{AppFixLong} that the maximum cost function in (\ref{OptProb}) with a fixed per-node power constraint, can be achieved by binary values of $t(i)$. In the following, we complete the proof  for the joint long-term power constraint by showing that assuming $\eta-\mu_1=1-\eta-\mu_2$ leads to a contradiction with the assumption $q_3(i)=0$. To this end, by assuming $q_3(i)=0$ and substituting $\eta-\mu_1=1-\eta-\mu_2$ into (\ref{Stationary Power}a) and (\ref{Stationary Power}b), we obtain
\begin{IEEEeqnarray}{rCl}\label{ContradicT}
    {\begin{cases}
     -\frac{1}{\mathrm{ln} 2} (\eta-\mu_1) \frac{S_1(i)}{1+P_1(i)S_1(i)+P_2(i)S_2(i)} +\gamma =0 \\
		 -\frac{1}{\mathrm{ln} 2} (\eta-\mu_1) \frac{S_2(i)}{1+P_1(i)S_1(i)+P_2(i)S_2(i)} +\gamma =0
    \end{cases}}
\end{IEEEeqnarray}
In Appendix \ref{AppMURegion}, we show that $\mu_1 \neq \eta$. Therefore, the above conditions can be satisfied simultaneously only if $S_1(i)=S_2(i)$, which, considering the continuous probability density functions of the channels gains, occurs with probability zero. Hence, the optimal $t(i)$ takes only the boundary values, i.e., 0 or 1, and not values in between. This completes the proof.


\section{Selection Weights Regions}
\label{AppMURegion}

In  this  appendix,  we  find the optimal intervals for the selection weights $\mu_1$ and $\mu_2$ in Theorems \ref{AdaptProt} and  \ref{FixProt}. 

\subsection{Selection Weights in Theorem \ref{AdaptProt}}

\begin{figure}
\centering
\psfrag{M1}[c][c][0.75]{$\mu_1$}
\psfrag{M2}[c][c][0.75]{$\mu_2$}
\psfrag{Mu1}[c][c][0.75]{$\mu_1 = \eta$}
\psfrag{M21}[l][l][0.75]{$\,\,\mu_2=1-\eta$}
\psfrag{S1}[c][c][0.75]{$\big\{\mathcal{M}_1,\mathcal{M}_4,\mathcal{M}_6\big\}$}
\psfrag{S2}[c][c][0.75]{\begin{tabular}{c}
   $\big\{\mathcal{M}_1,\mathcal{M}_2,\mathcal{M}_3,$\\
  $\quad\qquad\mathcal{M}_4,\mathcal{M}_6\big\}$
\end{tabular}}
\psfrag{S3}[c][c][0.75]{$\big\{\mathcal{M}_1,\mathcal{M}_2,\mathcal{M}_3\big\}$}
\psfrag{S4}[c][c][0.75]{\begin{tabular}{c}
  $\big\{\mathcal{M}_1,\mathcal{M}_2,$\\
   $\,\,\,\mathcal{M}_3,\mathcal{M}_5,$ \\
$\quad\qquad\mathcal{M}_6\big\}$
\end{tabular}}
\psfrag{S5}[c][c][0.75]{\begin{tabular}{c}
   $\big\{\mathcal{M}_1,\mathcal{M}_2,$\\
   $\,\,\,\mathcal{M}_3,\mathcal{M}_4,$ \\
$\,\,\,\mathcal{M}_5,\mathcal{M}_6\big\}$
\end{tabular}}
\psfrag{S6}[c][c][0.75]{\begin{tabular}{c}
   $\big\{\mathcal{M}_1,\mathcal{M}_4,$\\
$\,\,\,\mathcal{M}_5,\mathcal{M}_6\big\}$
\end{tabular}}
\psfrag{S7}[c][c][0.75]{$\big\{\mathcal{M}_4,\mathcal{M}_5,\mathcal{M}_6\big\}$}
\psfrag{S8}[c][c][0.75]{\begin{tabular}{c}
   $\big\{\mathcal{M}_2,\mathcal{M}_4,$\\
$\,\,\,\mathcal{M}_5,\mathcal{M}_6\big\}$
\end{tabular}}
\psfrag{S9}[c][c][0.75]{$\big\{\mathcal{M}_2,\mathcal{M}_5,\mathcal{M}_6\big\}$}
\includegraphics[width=3 in]{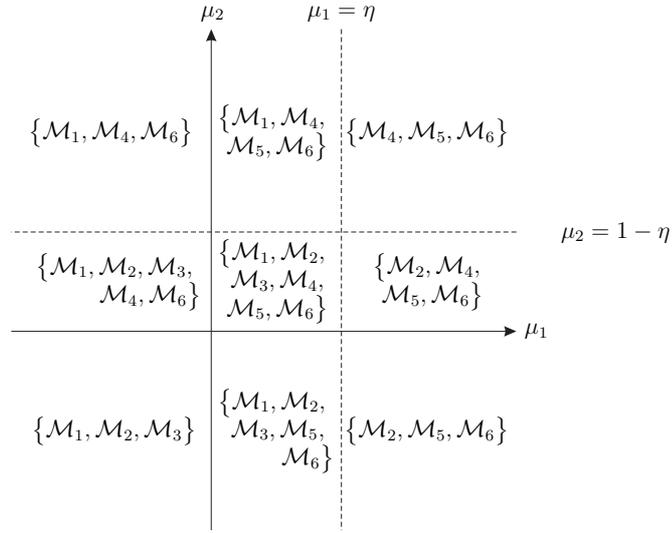}
\caption{Modes with none-zero probability for positive powers in the space of $(\mu_1,\mu_2)$.}
\label{FigMURegionAdapt}
\end{figure}

We  note  that  for  different values  of  $\mu_1$ and $\mu_2$,  some  of  the  optimal  powers  derived in (\ref{eq_11}), (\ref{P2}), (\ref{P4}), (\ref{P5}), (\ref{PM3-t0}), (\ref{PM3-t1}), and (\ref{PM6})    are  zero  for  all  channel realizations.  For  example,  if   $\mu_1\geq\eta$,  we  obtain  $P^{\mathcal{M}_1}_1(i)=0,\,\,\forall i$
from (\ref{eq_11}). Fig. \ref{FigMURegionAdapt} illustrates the set of modes that can take positive powers with non-zero probability in the space of ($\mu_1,\mu_2$). In the following, we show that values of $\mu_1$ and $\mu_2$ outside the intervals $0\leq \mu_1<\eta$ and $0\leq \mu_2<1-\eta$
cannot lead to a maximum weighted sum rate or simply lead to a violation of constraints $\mathrm{C1}$ or $\mathrm{C2}$ in (\ref{OptProb}).

\noindent
\textit{Case 1:} Sets  $\{\mathcal{M}_1,\mathcal{M}_2,\mathcal{M}_3\}$ and $\{\mathcal{M}_4,\mathcal{M}_5,\mathcal{M}_6\}$ lead to selection of either the transmission from the users to the relay or the transmission from the relay to the users, respectively,  in all time slots. This leads to violation of constraints $\mathrm{C1}$ and $\mathrm{C2}$ in (\ref{OptProb}) and thus the optimal values of $\mu_1$ and $\mu_2$ are not in this region.

\noindent
\textit{Case 2:} In set  $\{\mathcal{M}_1,\mathcal{M}_4,\mathcal{M}_6\}$, both modes $\mathcal{M}_4$ and $\mathcal{M}_6$ transmit data from the relay to user 1 and thus, need the transmission from user 2 to the relay, which cannot be realized in this set. Thus, this set leads to violation of constraint $\mathrm{C2}$ in (\ref{OptProb}). Similarly, in set $\{\mathcal{M}_2,\mathcal{M}_5,\mathcal{M}_6\}$,  both modes $\mathcal{M}_5$ and $\mathcal{M}_6$ require a transmission from user 1 to the relay which cannot be selected in this set. Thus, this region of $\mu_1$ and $\mu_2$ leads to violation of constraint $\mathrm{C1}$ in (\ref{OptProb}).

\noindent
\textit{Case 3:} In set $\{\mathcal{M}_1,\mathcal{M}_4,\mathcal{M}_5,\mathcal{M}_6\}$, there is no transmission from user 2 to
the relay. Therefore, the optimal values of $\mu_1$ and $\mu_2$ have to guarantee that modes $\mathcal{M}_4$
and $\mathcal{M}_6$ are not selected for any channel realization. However, from (\ref{OptMetAdapt}), we obtain
\begin{IEEEeqnarray}{rCl}
   \Lambda_6(i)  &=& \mu_1C_{r2}(i) + \mu_2C_{r1}(i) -\gamma P_r(i) \big |_{P_r(i)=P_r^{\mathcal{M}_6}(i)} \nonumber \\
 &\overset{(a)} {\geq}& \mu_1C_{r2}(i) + \mu_2C_{r1}(i) -\gamma P_r(i)\big |_{P_r(i)=P_r^{\mathcal{M}_5}(i)} \nonumber \\
 &\overset{(b)} {\geq}& \mu_1C_{r2}(i) -\gamma P_r(i) \big |_{P_r(i)=P_r^{\mathcal{M}_5}(i)} = \Lambda_5(i),
\end{IEEEeqnarray}
where $(a)$ follows from the fact that $P_r^{\mathcal{M}_6}(i)$ maximizes $\Lambda_6(i)$ and   $(b)$ follows
from $\mu_2C_{r1}(i)\geq 0$. Inequalities $(a)$ and $(b)$ hold with equality only if $S_1(i)=0$ which happens with  probability zero for time-continuous fading, or $\mu_2=0$ which is not included in this region, see Fig. \ref{FigMURegionAdapt}. Therefore, mode $\mathcal{M}_6$ will be selected in this region which leads to a violation of constraint $\mathrm{C2}$ in (\ref{OptProb}). A similar statement is true for set $\{\mathcal{M}_2,\mathcal{M}_4,\mathcal{M}_5,\mathcal{M}_6\}$. Thus, the optimal values of $\mu_1$ and $\mu_2$ cannot be in these two regions.

\noindent
\textit{Case 4:} For set $\{\mathcal{M}_1,\mathcal{M}_2,\mathcal{M}_3,\mathcal{M}_4,\mathcal{M}_6\}$,
we obtain
\begin{IEEEeqnarray}{rCl}
   \Lambda_6(i) &=& \mu_1C_{r2}(i) + \mu_2 C_{r1}(i)-\gamma P_r(i)\big |_{P_r(i)=P_r^{\mathcal{M}_6}(i)} \nonumber \\
 &\overset{(a)} {\leq}& \mu_2 C_{r1}(i)-\gamma P_r(i) \big |_{P_r(i)=P_r^{\mathcal{M}_6}(i)} \nonumber \\
 &\overset{(b)} {\leq}& \mu_2 C_{r1}(i)-\gamma P_r(i) \big |_{P_r(i)=P_r^{\mathcal{M}_4}(i)} = \Lambda_4(i)
\end{IEEEeqnarray}
where inequality $(a)$ comes from the fact that $\mu_1C_{r2}(i)\leq 0$ and equality holds when either $S_2(i)=0$, which occurs
with  probability zero, or $\mu_1=0$ holds. Inequality $(b)$ holds since $P_r^{\mathcal{M}_4}(i)$ maximizes $\Lambda_4(i)$ and
holds with equality only if $P_r^{\mathcal{M}_4}(i)=P_r^{\mathcal{M}_6}(i)$ and consequently $\mu_1=0$. If $\mu_1\neq0$, mode $\mathcal{M}_6$ is not selected and there is no transmission from the relay to user 2. Therefore, the optimal values of $\mu_1$ and $\mu_2$ have to guarantee that modes $\mathcal{M}_1$ and $\mathcal{M}_3$ are not selected for any channel realization. Thus, we obtain $\mu_1=\eta$ which is not contained in this region. If $\mu_1=0$,
from (\ref{OptMetAdapt}), we obtain
\begin{IEEEeqnarray}{rCl}
   \Lambda_6(i) = \Lambda_4(i) &=& \mu_2C_{r1}(i) -\gamma P_r(i)\big |_{P_r(i)=P_r^{\mathcal{M}_4}(i)} \nonumber \\
 &\overset{(a)} {\leq}& \eta C_{r1}(i) -\gamma P_r(i)\big |_{P_r(i)=P_r^{\mathcal{M}_4}(i)} \nonumber \\
 &\overset{(b)} {\leq}& \eta C_{r1}(i)-\gamma P_r(i) \big |_{P_r(i)=P_1^{\mathcal{M}_1}(i)} = \Lambda_1(i) \quad\,\,
\end{IEEEeqnarray}
where inequality $(a)$ holds only if $\mu_2\leq \eta$ and inequality $(b)$ holds with equality if $\mu_2=\eta$. If $\mu_2 < \eta$, modes $\mathcal{M}_4$ and $\mathcal{M}_6$ are not selected. Thus, there is no transmission from the relay to the users which leads to a violation of $\mathrm{C1}$ and $\mathrm{C2}$ in (\ref{OptProb}). If $\mu_2=\eta$, we obtain $P_2^{\mathcal{M}_2}(i)=0$. Hence, mode $\mathcal{M}_2$ cannot be selected and either $P_1^{\mathcal{M}_3}(i)=0$ or $P_2^{\mathcal{M}_3}(i)=0$, and thus mode $\mathcal{M}_3$ cannot be selected either. Since modes $\mathcal{M}_4$ and $\mathcal{M}_6$ require the transmission from user 2 to the relay, and modes $\mathcal{M}_2$ and $\mathcal{M}_3$ are not selected, constraint $\mathrm{C2}$ in (\ref{OptProb}) is violated and $\mu_1=0$ and $\mu_2\leq \eta$ cannot be optimal. Therefore, if $\mu_1=0$, we must have $\mu_2>\eta$, and by considering that in this region, we always have $\mu_2\leq 1-\eta$, we obtain a necessary condition for $\mu_1=0$ as $1-\eta>\eta$ or $\eta<\frac{1}{2}$. A similar statement is true for set $\{\mathcal{M}_1,\mathcal{M}_2,\mathcal{M}_3,\mathcal{M}_5,\mathcal{M}_6\}$ where we obtain that the necessary condition for $\mu_2=0$ is $\eta>\frac{1}{2}$.

We note that we have proved that except for set $\{\mathcal{M}_1,\mathcal{M}_2,\mathcal{M}_3,\mathcal{M}_4,\mathcal{M}_5,\mathcal{M}_6\}$, the other sets lead to a contradiction. Therefore, the optimal selection weights must be in the following intervals $\mu_1\in[0,\,\,\eta]$ and $\mu_2\in[0,\,\,1-\eta]$. To complete the proof, we must consider the boundary values of the selection weights. Specifically, if $\mu_1=\eta$ holds, we obtain that $P_1^{\mathcal{M}_1}=P_1^{\mathcal{M}_3}=0$ from (\ref{eq_11}), (\ref{PM3-t0}), and (\ref{PM3-t1}). Therefore, there is no transmission from user 1 to the relay, i.e., we obtain one-way relaying from user 2 to user 1. However, this leads to $\eta=0$ which is not included in  the optimization problem for obtaining the boundary surface of the rate region.  Similarly, $\mu_2=1-\eta$ holds only if $\eta=1$ which is also excluded for obtaining the boundary surface. Thus, the feasible intervals for the optimal selection weights are $0\leq\mu_1<\eta$ and $0\leq\mu_2<1-\eta$. We also note that the selection pairs cannot be simultaneously zero, i.e., $\mu_1=\mu_2=0$, since this leads to exclusion of all modes from the relay to the users which in turn leads to violation of $\mathrm{C1}$ and $\mathrm{C2}$ in (\ref{OptProb}).

\subsection{Selection Weights in Theorem \ref{FixProt}}
\begin{figure}
\centering
\psfrag{M1}[c][c][0.75]{$\mu_1$}
\psfrag{M2}[c][c][0.75]{$\mu_2$}
\psfrag{Mu1}[c][c][0.75]{$\mu_1 = \eta$}
\psfrag{M21}[c][c][0.75]{$\,\,\,\mu_2=1-\eta$}
\psfrag{S1}[c][c][0.75]{$\big\{\mathcal{M}_1,\mathcal{M}_4\big\}$}
\psfrag{S2}[c][c][0.75]{$\big\{\mathcal{M}_3,\mathcal{M}_4\big\}$}
\psfrag{S3}[c][c][0.75]{$\big\{\mathcal{M}_3\big\}$}
\psfrag{S4}[c][c][0.75]{$\big\{\mathcal{M}_3,\mathcal{M}_5\big\}$}
\psfrag{S5}[c][c][0.75]{$\big\{\mathcal{M}_3,\mathcal{M}_6\big\}$}
\psfrag{S6}[c][c][0.75]{$\big\{\mathcal{M}_1,\mathcal{M}_6\big\}$}
\psfrag{S7}[c][c][0.75]{$\big\{\mathcal{M}_6\big\}$}
\psfrag{S8}[c][c][0.75]{$\big\{\mathcal{M}_2,\mathcal{M}_6\big\}$}
\psfrag{S9}[c][c][0.75]{$\big\{\mathcal{M}_2,\mathcal{M}_5\big\}$}
\includegraphics[width=3 in]{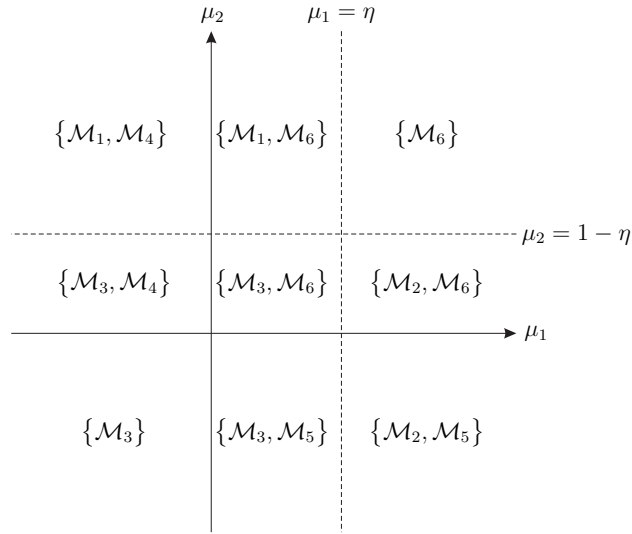}
\caption{Candidate modes for selection in the space of $(\mu_1,\mu_2)$.}
\label{FigMURegionFix}
\end{figure}

Herein, we use the fact that in some regions of $(\mu_1,\mu_2)$, some of the selection metrics are negative while the others are positive for all channel realizations. Thus, the transmission modes corresponding to the selection metrics with negative value are not selected at all. Fig. \ref{FigMURegionFix} represents the set of candidate selection modes for all channel realizations in the space of $(\mu_1,\mu_2)$.  These candidate modes are obtained based on the necessary selection condition introduced in (\ref{OptMetFix}).

\noindent
\textbf{Mode $\mathcal{M}_1$:} If $\mu_1< \eta$ and $\mu_2 \geq 1-\eta$, then $\mathcal{M}_1$ is a candidate for selection since we obtain $\Lambda_1(i)\geq \max\{\Lambda_2(i),\Lambda_3(i)\}$ from (\ref{MetFix}).

\noindent
\textbf{Mode $\mathcal{M}_2$:} If $\mu_1\geq \eta$ and $\mu_2< 1-\eta$, then $\mathcal{M}_2$ is a candidate for selection since we obtain $\Lambda_2(i)\geq \max\{\Lambda_1(i),\Lambda_3(i)\}$ from (\ref{MetFix}).

\noindent
\textbf{Mode $\mathcal{M}_3$:} If $\mu_1 \leq 1-\eta$ and $\mu_2 \leq 1-\eta$, then, based on (\ref{MU12}a) and (\ref{M25}a), $\mathcal{M}_3$ is a candidate for selection since we obtain  $\Lambda_3(i)\geq \max\{\Lambda_1(i),\Lambda_2(i)\}$ from (\ref{MetFix}).

\noindent
\textbf{Mode $\mathcal{M}_4$:} If $\mu_1 \leq 0$ and $\mu_2> 0$, then $\mathcal{M}_4$ is a candidate for selection since we obtain $\Lambda_4(i)\geq \max\{\Lambda_5(i),\Lambda_6(i)\}$ from (\ref{MetFix}).

\noindent
\textbf{Mode $\mathcal{M}_5$:} If $\mu_1> 0$ and $\mu_2 \leq 0$, then $\mathcal{M}_5$ is a candidate  for selection since we obtain $\Lambda_5(i)\geq \max\{\Lambda_4(i),\Lambda_6(i)\}$ from (\ref{MetFix}).

\noindent
\textbf{Mode $\mathcal{M}_6$:} If $\mu_1 \geq 0$ and $\mu_2 \geq 0$, then $\mathcal{M}_6$ is a candidate for selection since we obtain $\Lambda_6(i)\geq \max\{\Lambda_4(i),\Lambda_5(i)\}$ from (\ref{MetFix}).

It is straightforward to see that all sets in Fig.~\ref{FigMURegionFix} contradict one or both constraints  $\mathrm{C1}$ and $\mathrm{C2}$ in (\ref{OptProb}), except $\{\mathcal{M}_3,\mathcal{M}_6\}$. We note that $(\mu_1,\mu_2)=(0,0)$ and $(\mu_1,\mu_2)=(\eta,1-\eta)$ lead to the exclusion of the relay-to-user modes and user-to-relay modes, respectively, and thus, violate constraints $\mathrm{C1}$ and $\mathrm{C2}$ in (\ref{OptProb}). However, in general, the other boundary values of the selection weights do not lead to a direct contradiction for all channel statistics and $\eta$. In particular, if $\mu_1=0,\eta$ and $\mu_2=0,1-\eta$ hold, we obtain that point-to-point modes $\mathcal{M}_4$, $\mathcal{M}_2$, $\mathcal{M}_5$, and $\mathcal{M}_1$ are also possible candidates for mode selection, respectively. Therefore, the feasible intervals containing the optimal values of the selection weights are $0\leq \mu_1 \leq \eta$ and $0\leq \mu_2 \leq 1-\eta$. This completes the proof.

\bibliographystyle{IEEEtran}
\bibliography{refs}

\end{document}